\documentclass[prd,aps,superscriptaddress,tightenlines,nofootinbib,floatfix,preprintnumbers,11pt,longbibliography]{revtex4-1}
\usepackage{epsfig}
\usepackage{amsfonts}
\usepackage{amsmath}
\usepackage{amssymb}
\usepackage{dsfont}
\usepackage{hyperref}
\usepackage{xspace}
\usepackage{slashed}
\hypersetup{
    colorlinks=true,       
    linkcolor=blue,          
    citecolor=blue,        
    filecolor=blue,      
    urlcolor=blue           
}
\usepackage{pifont}
\usepackage{color}
\usepackage[dvipsnames]{xcolor}
\usepackage{graphicx}  %
\usepackage{bm}  %
\usepackage{comment}
\usepackage{graphics}
\usepackage{float}
\usepackage{caption}
\usepackage{ragged2e}
\usepackage{tikz} 
\usetikzlibrary{shapes,arrows,positioning,automata,backgrounds,calc,er,patterns}
\usepackage{tikz-feynman}
\tikzfeynmanset{compat=1.0.0}
\usepackage{multirow}

\newcommand{\dslash}[1]{#1 \llap{/\kern-0.5pt}}
\newcommand{\Dslash}[1]{#1 \llap{/\kern+1.5pt}}
\newcommand{\DDslash}[1]{#1 \llap{/\kern+2.3pt}}
\newcommand{\dslashh}[1]{#1 \llap{/\kern+1pt}}
\newcommand{\abs}[1]{|#1|}

\newcommand{\bea}{\begin{eqnarray}}
\newcommand{\eea}{\end{eqnarray}}
\newcommand{\be}{\begin{equation}}
\newcommand{\ee}{\end{equation}}
\newcommand{\bma}{\begin{pmatrix}}
\newcommand{\ema}{\end{pmatrix}}

\begin{document}
\preprint{	LA-UR-26-28874}

\title{
\vspace*{0.5cm}
\huge
Three-nucleon potentials and their matrix elements for $\delta_{\mathrm{NS}}$ in superallowed $\beta$ decays}

\author{Graham Chambers-Wall}
\affiliation{Department of Physics, Washington University in Saint Louis, Saint Louis, MO 63130, USA}

\author{Maria Dawid}
\affiliation{Department of Physics, Washington University in Saint Louis, Saint Louis, MO 63130, USA}

\author{Emanuele Mereghetti}
\affiliation{Theoretical Division, Los Alamos National Laboratory, Los Alamos, 
NM 87545, USA}

\author{Saori Pastore}
\affiliation{Department of Physics, Washington University in Saint Louis, Saint Louis, MO 63130, USA}
\affiliation{McDonnell Center for the Space Sciences at Washington University in St. Louis, MO 63130, USA}

\author{Maria Piarulli}
\affiliation{Department of Physics, Washington University in Saint Louis, Saint Louis, MO 63130, USA}
\affiliation{McDonnell Center for the Space Sciences at Washington University in St. Louis, MO 63130, USA}

\begin{abstract}
    We derive three-nucleon potentials for radiative corrections at order $\mathcal{O}(\alpha \epsilon_\chi^2)$ in chiral effective field theory where $\alpha$ is the electromagnetic coupling and $\epsilon_\chi$ is the chiral expansion parameter. These corrections contribute to the nuclear-structure-dependent term $\delta_\mathrm{NS}$ for determinations of the Cabibbo–Kobayashi–Maskawa matrix element $V_{ud}$ from superallowed nuclear $\beta$ decays. We compute matrix elements of three-nucleon contributions at orders $\mathcal{O}(\alpha^2)$ and $\mathcal{O}(\alpha \epsilon_\chi^2)$ using quantum Monte Carlo methods. We find the matrix elements at order $\mathcal{O}(\alpha^2)$ to be large and to have an effect on the extraction of $V_{ud}$ at the scale of $10^{-4}$. In the case of $^{10}$C, the matrix elements of $\mathcal{O}(\alpha \epsilon_\chi^2)$ three-nucleon operators
    provide a 2\% correction to $\delta_{\text{NS}}$,
    and are thus beyond the current precision goals. 
    The relative size of three-nucleon corrections to $\delta_{\text{NS}}$ in medium mass nuclei, where two-body contributions tend to be smaller, remains to be investigated.      We then use these results to determine $V_{ud}$ from the superallowed transition ${}^{10}\mathrm{C}\rightarrow{}^{10}\mathrm{B}$.
\end{abstract}

\maketitle

\tableofcontents

\newpage

\section{Introduction}

In the Standard Model (SM) of particle physics, the Cabibbo-Kobayashi-Maskawa (CKM) matrix \cite{Cabibbo_1963,Kobayashi_1973}, which describes the relative strengths of quark weak interactions, is unitary. Violation of CKM unitarity  would indicate physics beyond the Standard Model (BSM), and could be explained by a variety of possible scenarios, such as vector-like quarks and leptons \cite{Belfatto_2021,Cheung_2020,Crivellin_2021A,Branco_2021,Crivellin_2020A,Kirk_2021,Belfatto:2023tbv}, corrections to the Fermi constant \cite{Marciano_1999,Crivellin_2021B}, and violation of lepton flavor universality \cite{Crivellin_2021C,Crivellin_2021E,Marzocca_2021,Capdevila_2021}. Due to experimental precision, investigating the unitarity of the first row of the CKM matrix, namely the prediction
\begin{align}\label{ckm}
    |V_{ud}|^2+|V_{us}|^2+|V_{ub}|^2=1,
\end{align}
provides one of the most stringent CKM unitarity tests to date, with a precision that corresponds to probing BSM physics at the scale of about 10 TeV, comparable to direct LHC searches \cite{Gorchtein_2024,Cirigliano:2023nol,Gialidi_2026}. Currently
the sum of CKM elements on the left-hand side of~\eqref{ckm} deviates from 1 at
the $2-2.5\sigma$ level \cite{Seng:2018qru,Seng:2018yzq,Cirigliano:2022yyo,Hardy_2020,PDG_2024}, highlighting the need for precise determinations of each element $V_{ui}$ to make definitive statements about CKM unitarity. 

The most precise determinations of $V_{ud}$ come from superallowed $(0^+\rightarrow0^+)$ $\beta$ decays and neutron decays with uncertainties $\delta V_{ud}/V_{ud}$ around $3\times10^{-4}$ and $5\times 10^{-4}$, respectively \cite{Towner_2010,PDG_2024,Cirigliano_2023B}. Meson decays, $K\rightarrow\pi l\nu$ and $K,\pi\rightarrow\mu\nu$, provide information on $V_{us}$ and the ratio $V_{us}/V_{ud}$. The value of the last element, $V_{ub}$, is smaller than the precision of the other elements and therefore not of significance for current tests. 

Extracting $V_{ud}$ from experiment requires theory input in order to convert measurements of the half-life into a value of the CKM element.
The relation between $V_{ud}$ and the half-life $t$ has received considerable attention in the literature, see for example the discussion in Refs.~\cite{Hardy_2020,Seng_2023}. 
In an effective field theory (EFT) approach that attempts to factorize the contributions of photons with different energy and momentum, the master formula for the extraction of $V_{ud}$ is given by \cite{Cirigliano_2024A,Cirigliano_2024B}
\begin{align}
    \frac{1}{t}&=\frac{G_F^2|V_{ud}|^2m_e^5}{\pi^3\log2}\left[C_\mathrm{eff}^{(g_V)}(\mu)\right]^2\left[1+\Bar{\delta}_R^\prime(\mu)\right]\left[1+\Bar{\delta}_{\mathrm{NS}}-\Bar{\delta}_C\right]\times \Bar{f}(\mu)\label{eq:masterformula}
\end{align}
where $\Bar{f}$ is a phase-space factor including the Fermi function and $\Bar{\delta}_C$ is an isospin-symmetry-breaking (ISB) correction. 
The ``outer correction''
$\Bar{\delta}^\prime_R$  
encodes contributions from low-energy photons, with energy of order of the reaction's $\mathcal Q$ value.
$C_\mathrm{eff}^{(g_V)}$ collects contribution from integrating out ``hard'' and ``soft'' photons, 
that is  contributions from the weak scale, from the typical scale of QCD, $|{\bf q}_\gamma | \approx m_N$, where ${\bf q}_\gamma$ is the photon three-momentum and $m_N$ is the nucleon mass, and a subset of the corrections arising at the nuclear scale, $|{\bf q}_\gamma| \approx m_\pi$, with the pion mass $m_\pi$. 
Finally,  $\Bar{\delta}_\mathrm{NS}$ is the nuclear-structure-dependent correction, which, in the EFT formalism, arises from the exchange of ``potential'' and soft photons \cite{Cirigliano_2024A},
which also have $|{\bf  q}_\gamma| \approx m_\pi$.
As done in Refs.~\cite{Cirigliano_2024A,Cirigliano_2024B}, in Eq.~\eqref{eq:masterformula} we denote with a bar quantities derived in the EFT formalism. These are closely related to object appearing in the traditional factorization formula used, for example in Ref.~\cite{Hardy_2020}. The detailed mapping    
is discussed in Ref.~\cite{Cirigliano_2024A}.

In the EFT formalism, $\Bar\delta_{\text{NS}}$
is given by the matrix elements of two- and higher-body electroweak operators,
which have a double expansion in the electromagnetic coupling $\alpha$ and in the chiral expansion parameter $\epsilon_\chi = \mathcal O(m_\pi/\Lambda_\chi)$. To provide a confident estimate on the theoretical uncertainties entering Eq. \eqref{eq:masterformula}, radiative corrections must be computed beyond leading order (LO). For the nuclear structure-dependent correction $\Bar{\delta}_{\mathrm{NS}}$, electroweak operators, or ``potentials'', have been derived up to order $\mathcal{O}(\alpha\epsilon_\chi)$ and $\mathcal{O}(\alpha^2)$. Three-nucleon potentials arise at $\mathcal O(\alpha^2)$,
from the exchange of two photons between the electron/positron and two distinct nucleons in the nucleus \cite{Cirigliano_2024A}. 
Their matrix elements, however,
have been assumed to provide a relative small contribution in light nuclei and have not been explicitly computed in Refs.~\cite{Cirigliano_2024A} and \cite{King_2025}.

Additional three-nucleon contributions appear at $\mathcal{O}(\alpha\epsilon^2_\chi)$, where, in their simplest form, they can be thought of as a one-body current connected with a two-body current by a photon exchange.
Contributions of two-nucleon currents and correlations are crucial for single-$\beta$ decays calculations, resolving the need to quench $g_A$ from its physical nucleon-axial coupling value of 1.27 \cite{King:2020wmp,Gysbers:2019uyb,Pastore_2018}. Testing whether this two-nucleon physics is equally important for observables with larger values of momentum transfer has not been fully explored. The role of three-nucleon contributions to $0\nu\beta\beta$, which has momentum transfer of similar order as $\Bar{\delta}_{\mathrm{NS}}$, has a contribution in-line with power counting expectations for light nuclei \cite{Chambers-Wall_2026}. Evaluating the size of these three-nucleon contributions for superallowed beta decays should be tested for a robust determination of $\Bar{\delta}_{\mathrm{NS}}$  and its theoretical error, and therefore of $V_{ud}$. To this end, we derive the three-nucleon potentials at $\mathcal{O}(\alpha\epsilon_\chi^2)$ and compute matrix elements at $\mathcal{O}(\alpha^2)$ and $\mathcal{O}(\alpha\epsilon_\chi^2)$ in the three-nucleon sector for light nuclei using quantum Monte Carlo methods to solve the many-body problem of strongly interacting nucleons \cite{Carlson_2015, Gandolfi_2020}. 

The paper is organized as follows.
In Section \ref{sec:EFT}, we overview the EFT methods used to derive the two- and three-nucleon potentials. In Section \ref{sec:LO}, we review the two-nucleon potentials for $\Bar{\delta}_{\mathrm{NS}}$, which we will compare the three-nucleon results against, we
derive the expressions for the three-nucleon potentials,
and we examine their consistency on what expected on the basis of vector current conservation.
In Section \ref{sec:ME}, we review the quantum Monte Carlo method and show results for the three-nucleon matrix elements in $^6\mathrm{He}(0^+;1)\rightarrow{}^6\mathrm{Li}(0^+;1)$ and $^{10}\mathrm{C}(0^+;1)\rightarrow{}^{10}\mathrm{B}(0^+;1)$. We additionally discuss the impact of these contributions compared to the leading two-nucleon results and the impact on the determination of $V_{ud}$. Finally, we conclude in Section \ref{sec:conclusion}.

\section{Effective field theory}\label{sec:EFT}

Radiative corrections to nuclear $\beta$ decay involve contributions from a hierarchy of energy scales probed by the exchanged photon. In this work, we focus on electroweak radiative corrections (EWRC) that depend on nuclear structure and arise from
potential photons, namely photons with energy  $q^0_\gamma\simeq \boldsymbol{q}_\gamma^2/m_N$ and momentum $\abs{\boldsymbol{q}_\gamma}\simeq m_\pi$. In this regime, the relevant degrees of freedom are hadrons. Chiral EFT provides a framework to organize the interactions of pions and nucleons among themselves and with external currents based on the symmetries of QCD. 

In the limit of vanishing quark masses and no electromagnetic or weak interactions, the QCD Lagrangian is invariant under chiral transformations. The spontaneous and explicit breaking of this symmetry restricts the form of the interactions of pions and nucleons and can be accounted for perturbatively. In this section, we summarize the power counting expansion for few-nucleon interactions and provide the interactions among hadrons and electroweak currents in order to construct the two- and three-nucleon RC potentials.

\subsection{Power counting}

The contributions of 
potential photons to $\beta$ decays can be organized according to an expansion in $\alpha$ and in the chiral expansion parameter $\epsilon_\chi=p/\Lambda_\chi$, where $p$ is a low-momentum scale and $\Lambda_\chi\simeq1$ GeV is the chiral symmetry breaking scale. We will explicitly count powers of $\alpha$ by inspection of the relevant diagrams. The counting in $\epsilon_\chi$ can be determined using the standard arguments in chiral EFT \cite{Weinberg_1990,Weinberg_1991,Friar:1996zw}.   Using Naive Dimensional Analysis (NDA) \cite{Manohar:1983md}, the interactions in the Lagrangian are organized by chiral index $\chi=d+n/2-2$ with $d$ derivatives, insertions of $m_\pi$, or insertions of $\Delta$ and $n$ baryon fields.
The scaling of $A$-nucleon diagrams 
then depends on the  number $V_i$ of insertions of vertices with chiral index $\chi_i$, the number of irreducible loops, $L$, and the number of separately connected pieces, $C$.
To count these, Weinberg assigned  
to each disconnected nucleon line a factor of $(4\pi)^2 p^{-4}$, from the four-dimensional energy-momentum conservation delta function
\cite{Weinberg_1991}, leading to the Weinberg's power counting (WPC) chiral index
\begin{equation}\label{eq:WPC}
    \delta_{\text{WPC}} = 2 (A-1-C) + 2L + \sum V_i \chi_i,
\end{equation}
where $C$ is the number of separately connected pieces in one diagram.
Friar, on the other hand, assigned the factor 
$(4\pi) p^{-4}$ to the delta function
\cite{Friar:1996zw},
which is compatible with the non-relativistic nature of bound state loops. For diagrams with $1 \le C \le A-1$, this results in a Friar's power counting (FPC) index \cite{Hammer:2019poc} 
\begin{equation}\label{eq:FPC}
    \delta_{\, \rm FPC} =  (A-1-C) + 2L + \sum V_i \chi_i.
\end{equation}
These schemes describe the organization of irreducible diagrams. When solving the Lippmann-Schwinger (or Schr\"{o}dinger) equation, reducible contributions, where an intermediate state of a diagram is equal to the initial or final state, are accounted for iteratively by resumming the irreducible contributions with intermediate free nucleon propagators. The electroweak Hamiltonian governing single $\beta$ decays can be parameterized as
\begin{align}
    H_\beta&=\sqrt{2}G_FV_{ud}\Bar{e}_L\gamma_\mu \nu_L \mathcal{J}^\mu_{\mathrm{W}}
\end{align}
where $\mathcal{J}_\mathrm{W}^\mu$ is the weak nuclear current.
At lowest order in $\alpha$ and $\epsilon_\chi$, the current is given by the one-body contributions 
\begin{align}
   \left[ {\mathcal J}_W^\mu \right]^{(-2) }  
&= 
 \sum_{j=1}^A g_V   \Big(    \delta^{\mu 0} - \frac{g_A}{g_V}  \delta^{\mu i} \sigma^{i}_j  \Big)  \tau^{+}_j,  
\end{align}
where $g_V = 1 + \mathcal O(\alpha)$,  $g_A/g_V = 1.2753(13)$ \cite{PDG_2024}, and the superscript $(-2)$ denotes the chiral index. 
At higher orders in chiral EFT, the weak vector and axial currents receive two- and higher-body contributions
\cite{Pastore:2011ip,Pastore:2009is,Kolling:2009iq,Krebs:2016rqz}.
In the isospin limit ($\alpha =0$, $m_u = m_d$), however, the zero momentum Fermi matrix element remains protected by current conservation, so $M_F = \sqrt{2}$ at all orders in the chiral expansion. 
This relation is broken by electromagnetic corrections. In particular,
potential photon exchanges induce correction to both the temporal and spatial components of the current operator. We can write the current as 
\begin{align}
   \left[ {\mathcal J}_W^\mu \right]^{ }   =  \left[ {\mathcal J}_W^\mu \right]_\chi  + \delta^{\mu 0}  \mathcal V^0 
+ \delta^{\mu i} \mathcal V_i   
 + \dots ,
\label{eq:Hpotential}
\end{align}
where $\left[ {\mathcal J}_W^\mu \right]_\chi$ denotes the chiral currents in the isospin limit, $\mathcal V_{0 (i)}$
the time- (space-) component of photon-mediated EW operators, while $\ldots$ denote operators that depend on the lepton energy, which we will not consider in what follows.
The time component $\mathcal V^0$ is particularly relevant for Fermi decays, and can be further expanded in $\alpha$ as
\begin{equation}
    \mathcal V^0 = \alpha \sum_{\delta} \left[\mathcal V^{0} \right]_\alpha^{(\delta)} + \alpha^2 \sum_{\delta} \left[\mathcal V^{0} \right]_{\alpha^2}^{(\delta)} +\ldots,
    \end{equation}
where $\alpha = e^2/4\pi$.
Ref.~\cite{Cirigliano_2024A} showed that the construction of the
 $\mathcal O(\alpha)$ two-body potential 
requires the insertion of vertices from the subleading chiral Lagrangian, with $\chi =1$. 
 According to the power counting rules 
of Eq. \eqref{eq:WPC}, one would find a chiral index $\delta =1$,
implying that the scaling of the photon-induced two-body contribution relative to the one-body weak vector current is   
\begin{equation}
    e^2 \frac{Q^3}{(4\pi F_\pi)^3} = \alpha \epsilon_\chi^2.
\end{equation}
In FPC, $\delta =0$, the suppression is reduced to $\mathcal O(\alpha \epsilon_\chi)$. As this latter scaling is in better agreement with explicit calculations \cite{Cirigliano_2024A,King_2025}, we will denote the leading two-body contributions to $\delta_{\text{NS}}$ as $\mathcal O(\alpha \epsilon_\chi)$, or next-to-leading order (NLO). 

As we will see, in the three-body sector one can construct lepton energy-independent operators using vertices from the leading Lagrangian, $\chi=0$. 
Using the counting rules of Eq.~\eqref{eq:WPC}, we would then expect three-body 
contributions to be suppressed by one power of $\epsilon_\chi$ with respect to two-body operators, and thus to scale as $\mathcal O(\alpha \epsilon_\chi^2)$, or next-to-next-to-leading order (N$^2$LO). FPC rules in Eq.~\eqref{eq:FPC} would lead to the expectation that two- and three-body operators are equally important. Since for other  electroweak operators FPC  overestimates three-body contributions  \cite{Chambers-Wall_2026}, we will take the $\mathcal O(\alpha \epsilon_\chi^2)$ scaling
as our working hypothesis and check it explicitly by calculating  matrix elements in light nuclei.

The $\mathcal O(\alpha^2)$ potentials have been built in Ref.~\cite{Cirigliano_2024A}. These operators are connected with terms that scale as $\mathcal O(\alpha^2 Z^2 \log R)$ and $\mathcal O(\alpha^2 Z \log R)$, where $R$ is the nucleus' charge radius,
in the traditional treatment of the Fermi function \cite{Wilkinson:1993hz} and of $\mathcal O(\alpha^2 Z)$ corrections \cite{Jaus:1970tah,Jaus:1986te,Sirlin:1986cc}. Because of the $Z$ enhancement, these operator can lead to sizable effects, notwithstanding the $\alpha^2$ suppression.  

In terms of the operator $\mathcal V^0$, $\Bar{\delta}_{\text{NS}}$ is given by
\begin{equation}\label{eq:deltaNSmomentum}
    \Bar{\delta}_{\text{NS}} = \frac{2}{M^{(0)}_F} \langle f | \mathcal V^0 | i \rangle = \frac{2}{M^{(0)}_F}  \left( \alpha \,\langle f | \left\{\left[\mathcal V^0\right]_{\alpha \epsilon_\chi} + \left[\mathcal V^0\right]_{\alpha \epsilon_\chi^2}\right\}| i \rangle + \alpha^2 \langle f |  \left[\mathcal V^0\right]_{\alpha^2} | i \rangle + \ldots \right), 
\end{equation}
with $M^{(0)}_F$ the Fermi matrix element in the isospin limit, and we replaced the chiral index $\delta$ with subscripts that more explicitly denote the scaling of the corrections relative to $M^{(0)}_F$.

\subsection{Electroweak and chiral Lagrangian}

To construct the chiral interactions, we introduce the pion field $U=u^2=\exp(i\boldsymbol{\pi}\cdot\boldsymbol{\tau}/F_0)$, nucleon ($\Delta$) fields $N$ $\left(T_i^\mu\right)$, and external currents $l_\mu=\frac{e}{2}A_\mu \tau^3-2\sqrt{2}G_F\left[V_{ud}\Bar{e}_L\gamma_\mu\nu_L\tau^++\mathrm{h.c.}\right]$, $\hat{l}_\mu=\hat{r}_\mu=\frac{e}{6}A_\mu$, and $r_\mu=\frac{e}{2}A_\mu\tau^3$, where $F_0$ is the pion decay constant in the chiral limit, $A_\mu$ is the photon field, $\tau^a$ is the Pauli matrix, and $G_F$ is the Fermi constant. Weak neutral currents, which do not have a significant contribution, are omitted. At the order we are working, we identify $F_0$ with the physical pion decay constant, $F_0=F_\pi=92.3$ MeV. We can now write the LO, denoted by index 0, pion interactions as
\begin{equation}\label{Eq:LagPi}
\mathcal L^{(0)}_{\pi}  = \frac{F_0^2}{4} \mathrm{Tr}\left[(D_\mu U)^\dagger (D^\mu U)\right]+ \frac{F_0^2}{4} \mathrm{Tr}\left[U^\dagger \chi + U \chi^\dagger\right]\,,
\end{equation}
where $\chi=2B\times\mathrm{diag}(m_u,m_d)$ with quark masses $m_{u,d}$, $B\simeq2.8$ GeV, and the chiral covariant derivative is 
\begin{equation}
D_\mu U = \partial_\mu U - i (l_\mu + \hat{l}_\mu) U + i U (r_\mu + \hat{r}_\mu)\,.
\end{equation}

The LO baryon Lagrangian is
\begin{equation}\label{LagN}
    \mathcal L^{(0)}_{\pi N} = \bar N_v i v \cdot \mathcal D N_v  + g_A \bar N_v S \cdot u N_v  - \bar{T}_{v\, i}^\mu (i v \cdot \mathcal D^{i j} - \Delta \delta^{ij} + g_1 S\cdot u^{i j}) T_{v\, \mu j}  + h_A \left( \bar{T}_{v\, i}^\mu w^i_\mu N_v + \bar N_v w^i_\mu T_{v\, i}^\mu  \right)
\end{equation}
with nucleon velocity and spin $v^\mu$ and $S^\mu$ where $v^\mu=(1,\boldsymbol{0})$ and $S^\mu=(0,\boldsymbol{\sigma}/2)$ in the rest frame of the nucleon. Indices $i,j$ denote isospin in the $\Delta$ interactions. In the remainder, we neglect the velocity subscript $v$. Among the low energy constants (LECs) appearing in Eq. (\ref{LagN}), only $g_A$ will be needed when computing matrix elements, for which we use the physical value of $g_A=1.27$ \cite{PDG_2024}. The objects contracted with the $\Delta$ fields are
\begin{eqnarray}
w^i_\mu = \frac{1}{2} {\rm Tr} (\tau^i u_\mu), \qquad
u_\mu^{i j} =  \xi^{i l}_{3/2} u_\mu \, \xi^{l j}_{3/2}, \, \qquad  
\xi^{i j}_{3/2} = \frac{2}{3} \delta^{i j} - \frac{i}{3} \epsilon^{i j k}\tau^k,
\end{eqnarray}
which depend on $u_\mu$, given by
\begin{eqnarray}
u_\mu &=& -i\left[u^\dagger\left(\partial_\mu -i (l_\mu +\hat{l}_\mu)\right)u-u\left(\partial_\mu - i (r_\mu + \hat{r}_\mu)\right)u^\dagger\right]\,. \end{eqnarray}
Finally, the nucleon and $\Delta$ covariant derivatives are
\begin{eqnarray}
\mathcal D_\mu N &=& (\partial_\mu + \Gamma_\mu)N, \\
\mathcal D^{i j}_\mu T_{\nu j} &=& \left(\partial_\mu \delta^{i j} + \Gamma_\mu \delta^{i j} - i \epsilon^{i j k} \, {\rm Tr} (\tau^k \Gamma_\mu)\right)\, T_{\nu j},  \\
\Gamma_\mu &=& \frac{1}{2}\left[u^\dagger\left(\partial_\mu -i(l_\mu + 3\hat{l}_\mu)\right)u+u\left(\partial_\mu - i (r_\mu + 3\hat{r}_\mu) \right)u^\dagger\right]\,.
\end{eqnarray}
The lowest-order Lagrangians in Eqs.~\eqref{Eq:LagPi} and \eqref{LagN} are needed for the construction of the three-body potential at N$^2$LO. In addition, the two-body operators depend on two LECs, $g_{V1,\, V2}^{\text{NN}}$. These would be naively suppressed in Weinberg's power counting, but are needed for the renormalization of the NLO potential \cite{Cirigliano_2024B,Cirigliano_2024A}.
As in $0\nu\beta\beta$, these LECs do not generate three-body operators at tree level, but might contribute at one loop \cite{Chambers-Wall_2026}, beyond the accuracy of our calculation.

\section{Two- and three-nucleon radiative corrections}\label{sec:LO}

Within the EFT approach, corrections sensitive to nuclear structure, such as the exchange of potential photons, are accounted for in $\Bar{\delta}_\mathrm{NS}$. The calculation of $\Bar{\delta}_\mathrm{NS}$  then reduces to computing matrix elements of these transition operators between initial and final nuclear states. The first corrections appear at orders $\mathcal{O}(\alpha \epsilon_\chi)$, $\mathcal{O}(\alpha\mathcal{Q}/m_\pi)$, and $\mathcal{O}(\alpha^2)$ where $\mathcal{Q}$ is the $\mathcal{Q}$-value of the transition. The first contributions are independent of the outgoing lepton energy but require an insertion of both leading and subleading vertices. The next contribution requires only leading vertices but depends on the lepton energies. The last contribution results from the exchange of more than one photon. The energy dependence can be separated:
\begin{align}
    \Bar{\delta}_\mathrm{NS}&=\delta^{(0)}_\mathrm{NS}+\overline{\delta^E_\mathrm{NS}}.
\end{align}
where $\delta^{(0)}_\mathrm{NS}$ denotes the energy-independent contribution and $\overline{\delta^E_\mathrm{NS}}$ denotes the energy-dependent contribution averaged over the lepton phase space. The $\mathcal{O}(\alpha\epsilon_\chi)$,
$\mathcal{O}(\alpha\epsilon^2_\chi)$
and $\mathcal{O}(\alpha^2)$ energy-independent contributions are
\begin{align}
    \delta_{\mathrm{NS}}^{(0)}&=\frac{2}{g_V(\mu_\pi)M_F^{(0)}}\left[\alpha\left(M_{\mathrm{GT}}^{\mathrm{mag}}+M_{\mathrm{T}}^{\mathrm{mag}}+M_{\mathrm{GT}}^\mathrm{CT}+M_{\mathrm{LS}}+M_\mathrm{3N}\right)+\alpha^2 \left(M_{\mathrm{2N}}^++M_{\mathrm{3N}}^+\right)\right]\label{eq:dNS0}
\end{align}
where the vector coupling constant $g_V(m_\pi)=1.01494(12)$ \cite{Cirigliano_2023B} and Fermi matrix element $M_F^{(0)}=\sqrt{2}$. The subscripts $i=\{\mathrm{F,GT,T,LS}\}$ denote the Fermi, Gamow-Teller, Tensor, and Spin-Orbit contributions of the two-nucleon potentials. 
The superscripts $i=\{\mathrm{mag,CT},+\}$ denote additional radial dependence of the operators. The two-nucleon contributions are traditionally further decomposed into isospin structure $N=n,p$ depending on the identity of the spectator nucleon. 
We do not report the definitions of the two-body operators here, but use the conventions described in Ref.~\cite{King_2025}.
The subscript $\mathrm{3N}$ denotes the three-nucleon potential, which is the result of this work. We do not explicitly decompose the three-nucleon potential into spin and isospin operator structure in Eq. (\ref{eq:dNS0}).

The definitions of $\delta_\mathrm{NS}^{(0)}$ and $\overline{\delta_\mathrm{NS}^{E}}$ absorb factors of $\alpha=e^2/(4\pi)$ and $\alpha^2$ which would appear in matrix elements of these RC potentials. Therefore, both the momentum- and coordinate-space potentials in this work have $(e^2/(4\pi))^n$ factored out with $n=1,2$ in the case of order $\mathcal{O}(\alpha)$ and $\mathcal{O}(\alpha^2)$, respectively.

\subsection{Two- and three-nucleon operators at $\mathcal O(\alpha^2)$}\label{sec:NLO}

Two- and three-nucleon corrections at order $\mathcal{O}(\alpha^2)$ were derived in Ref. \cite{Cirigliano_2024A} within dimensional regularization and the $\overline{\mathrm{MS}}_\chi$ scheme. We report them here for convenience,

\begin{align}
    \mathcal{V}_+^{\mathrm{2N}}&=-\sum_{i\neq j}\log(\Lambda r_{ij})\tau_i^\pm P_p^j,\label{eq:log2b}
    \\
    \mathcal{V}_+^{\mathrm{3N}}&=-\frac{1}{2}\sum_{i\neq j\neq k}\left[\log\left(\Lambda\frac{r_{ij}+r_{ik}+r_{jk}}{2}\right)+\frac{1}{2}\frac{r_{ij}^2+r_{ik}^2-r_{jk}(r_{ij}+r_{ik})}{r_{ij}r_{ik}}\right]\tau_i^\pm P_p^j P_p^k,\label{eq:log3b}
\end{align}
where $P_p=\frac{1}{2}(1+\tau^3)$ is the proton projection operator and $\tau^\pm$ is the isospin raising/lowering operator. The arbitrary subtraction scale $\Lambda$, appearing in these potentials is compensated for in matching coefficient $C_{\text{eff}}^{(g_V)}$. When computing matrix elements, we take $\Lambda^{-1}=R_A=1.2A^{1/3}$. The contributions to $\delta_{\text{NS}}$ are then given by the matrix elements
\begin{equation}
    M^+_\mathrm{2N} = \langle f | \mathcal V_+^{\text{2N}} | i \rangle, \qquad     M^+_{3\text{N}} = \langle f | \mathcal V_+^{\text{3N}} | i \rangle. 
\end{equation}

Due to the combined electron and photon propagators, the momentum space kernels scale as $\boldsymbol{q}^{-3}$ and $\boldsymbol{q}^{-6}$ for the two- and three-nucleon potentials, respectively, which results in a coordinate-space expression that is not suppressed at large nucleon pair distances. The effect of this enhancement and the particular isospin structure of these operators is a scaling of the two- and three-nucleon contributions by $\mathcal{O}(\alpha^2Z)$ and $\mathcal{O}(\alpha^2Z^2)$, respectively. Therefore, these NLO three-nucleon contributions become more important than the two-nucleon contributions for heavier nuclei. We provide explicit calculations of these contributions with QMC and their impact on $V_{ud}$ determinations from ${}^{10}\mathrm{C}\rightarrow{}^{10}\mathrm{B}$ in Sec. \ref{sec:results}.

\subsection{Three-nucleon operators at $\mathcal{O}(\alpha\epsilon_\chi^2)$}\label{sec:N2LO_3N}

The three-nucleon $\Bar{\delta}_\mathrm{NS}$ potentials entering at $\mathcal{O}(\alpha\epsilon_\chi^2)$ are shown in Fig. \ref{fig:diagrams}. These radiative corrections are mediated by the exchange of a photon with the outgoing electron or vertex containing pion(s) or $\Delta$ resonance. In principle, radiative corrections mediated by pion-mass-splitting rather than the photon appear at this order. However, QMC calculations of pion-mass splitting contributions in the LO two-nucleon sector reveal a small ($\lesssim5\%$) correction relative to the photon exchange contributions \cite{King_2025,Cirigliano_2024A}. 
The matrix elements of $\mathcal{O}(\alpha\epsilon_\chi^2)$ photon-exchange three-nucleon contributions, as described in Sec. \ref{sec:ME}, will reveal a small impact on the total matrix element. For this reason, we neglect the three-nucleon pion-ranged contributions since they would be beyond the current precision goals of CKM unitarity tests. 

\begin{figure}[t]
\begin{center}
\scalebox{.75}{
   \begin{minipage}[c]{3cm}
\begin{tikzpicture}[line width=1.2 pt]
    \begin{feynman}
        \vertex (a);
        \vertex[right=.75 of a, dot, minimum size=1.2mm, Cerulean] (b) {};
        \vertex[above=.75 of b] (bu);
        \vertex[right=1.5 of b] (c);
        \vertex[right=.75 of b] (d);
        \vertex[above=.375 of d,dot, minimum size=1.2mm, Cerulean] (d1) {};
        \vertex[above=.75 of c, Cerulean] (d2) {};
        \vertex[below=1 of a] (aa);
        \vertex[below=1 of aa] (aaa);
        \vertex[below=1 of b, dot, minimum size=1.2mm, Cerulean] (bb) {};
        \vertex[below=1 of bb] (bbb);
        \vertex[below=1 of c] (cc);
        \vertex[below=1 of cc] (ccc);

        \vertex[below=2.44 of d1, dot, minimum size=1.2mm, Cerulean] (bphoton) {};

        \diagram*{

        (a) -- [Cerulean] (c);
        (bu) -- [fermion, Cerulean] (b);
        (b) -- [fermion, Cerulean] (d1);
        (d1) -- [fermion, Cerulean] (d2);
        (aa) -- [Cerulean] (cc);
        (aaa) -- [Cerulean] (ccc);
        (b) -- [dashed, Cerulean] (bb);
        (d1) -- [photon, Cerulean] (bphoton);
        };

    \end{feynman}
\end{tikzpicture}
\caption{(a)}
\end{minipage}
\hspace{0cm}
   \begin{minipage}[c]{3cm}
   \vspace{-4mm}
\begin{tikzpicture}[line width=1.2 pt]
    \begin{feynman}
        \vertex (a);
        \vertex[right=.75 of a, dot, minimum size=1.2mm] (b) {};
        
        \vertex[above=.75 of b,dot,minimum size=1.2mm] (bpion) {};
        \vertex[above=.75 of bpion] (bu);
        \vertex[right=1.5 of b] (c);
        \vertex[right=.75 of b] (d);
        \vertex[right=.75 of bpion,dot, minimum size=1.2mm] (d1);
        \vertex[right=.25 of d1] (d2) {};
        \vertex[below=1 of a] (aa);
        \vertex[below=1 of aa] (aaa);
        \vertex[below=1 of b, dot, minimum size=1.2mm] (bb);
        \vertex[below=1 of bb] (bbb);
        \vertex[below=1 of c] (cc);
        \vertex[below=1 of cc] (ccc);
	\vertex[above=.375 of b,dot,minimum size=1.2mm] (bpi) {};

        \vertex[below=1.69 of d1, dot, minimum size=1.2mm] (bphoton) {};

        \vertex[below=1. of bphoton,dot,minimum size=1.2mm] (bphoton2) {};

        \diagram*{

        (a) -- [] (c);
        (bu) -- [fermion] (bpion);
        (bpion) -- [fermion] (d2);
        (b) -- [dashed] (bpion);
        (aa) -- [] (cc);
        (aaa) -- [] (ccc);
        (bpi) -- [photon,quarter left] (bphoton);
        (bphoton) -- [dashed] (bphoton2);
        };

    \end{feynman}
\end{tikzpicture}
\caption{\textcolor{black}{(b)}}
\end{minipage}
\hspace{0cm}
   \begin{minipage}[c]{3cm}
   \vspace{-4mm}
\begin{tikzpicture}[line width=1.2 pt]
    \begin{feynman}
        \vertex (a);
        \vertex[right=.75 of a, dot, minimum size=1.2mm] (b) {};
        \vertex[above=.75 of b,dot, minimum size=1.2mm] (bpion) {};
        \vertex[above=.75 of bpion] (bu);
        \vertex[right=1.5 of b] (c);
        \vertex[right=.75 of b] (d);
        \vertex[right=.75 of bpion,dot, minimum size=1.2mm] (d1);
        \vertex[right=.25 of d1] (d2) {};
        \vertex[below=1 of bb,dot,minimum size=1.2mm] (bbb) ;

        \vertex[below=1 of a] (aa);
        \vertex[below=1 of aa] (aaa);
        \vertex[below=1 of b, dot, minimum size=1.2mm] (bb) {};
        \vertex[below=1 of bb,dot,minimum size=1.2mm] (bbb) {};
         \vertex[below=.5 of bb, dot,minimum size=1.2mm] (bbbb) {};       \vertex[below=1 of c] (cc);
        \vertex[below=1 of cc] (ccc);
	\vertex[above=.375 of b, dot, minimum size=1.2mm] (bpi) {};
        
        \diagram*{
        
        (a) -- [] (c);
        (bu) -- [fermion] (bpion);
        (bpion) -- [fermion] (d2);
        (b) -- [dashed] (bpion);
        (aa) -- [] (cc);
        (aaa) -- [] (ccc);
        (bbb) -- [dashed] (bb);
        (bpi) -- [photon,quarter left] (bbbb);
        };

    \end{feynman}
\end{tikzpicture}
\caption{\textcolor{black}{(c)}}
\end{minipage}
\hspace{0cm}
   \begin{minipage}[c]{3cm}
      \vspace{.1cm}
\begin{tikzpicture}[line width=1.2 pt]
    \begin{feynman}
        \vertex (a);
        \vertex[right=.75 of a, dot, minimum size=1.2mm, Cerulean] (b) {};
        \vertex[above=.75 of b] (bu);
        \vertex[right=1.5 of b] (c);
        \vertex[right=.75 of b] (d);
        \vertex[above=.375 of d,dot, minimum size=1.2mm, Cerulean] (d1) {};
        \vertex[above=.75 of c, Cerulean] (d2) {};
        \vertex[below=1 of a] (aa);
        \vertex[below=1 of aa] (aaa);
        \vertex[below=1 of b, Cerulean] (bb) {};
        \vertex[below=1 of bb] (bbb);
        \vertex[below=1 of c] (cc);
        \vertex[below=1 of cc] (ccc);

        \vertex[left=.5 of b, dot,minimum size=1.2mm, Cerulean] (bl) {};
        \vertex[left=.5 of bb,dot,minimum size=1.2mm, Cerulean] (bbl) {};

        \vertex[below=2.435 of d1, dot, minimum size=1.2mm, Cerulean] (bphoton) {};

        \diagram*{

        (a) -- [Cerulean] (c);
        (bu) -- [fermion, Cerulean] (b);
        (b) -- [fermion, Cerulean] (d1);
        (d1) -- [fermion, Cerulean] (d2);
        (aa) -- [Cerulean] (cc);
        (aaa) -- [Cerulean] (ccc);
        (bl) -- [dashed, Cerulean] (bbl);
        (d1) -- [photon, Cerulean] (bphoton);
        (bl) -- [double, Cerulean] (b);
        };
        \vertex[left=.5 of b, dot,minimum size=1.2mm, Cerulean] (bl) {};
        \vertex[right=0 of b,dot,minimum size=1.2mm, Cerulean] (bbl) {};

    \end{feynman}
\end{tikzpicture}
\caption{(d)}
\end{minipage}
\hspace{0cm}
   \begin{minipage}[c]{3cm}
   \vspace{8.5mm}
\begin{tikzpicture}[line width=1.2 pt]
    \begin{feynman}
        \vertex (a);
        \vertex[right=.75 of a, dot, minimum size=1.2mm, Cerulean] (b) {};
        \vertex[below=.5 of b] (bu);
        \vertex[right=1.5 of b] (c);
        \vertex[right=.75 of b] (d);
        \vertex[below=.5 of b,dot, minimum size=1.2mm] (d1);
        \vertex[right=1.5 of bu, Cerulean] (d2) {};
        \vertex[right=.75 of bu, dot, minimum size=1.2mm, Cerulean] (d1u) {};

        \vertex[below=.5 of b, dot, minimum size=1.2mm, Cerulean] (bu) {};
        \vertex[below=1 of a] (aa);
        \vertex[below=1 of aa] (aaa);
        \vertex[below=1 of b, Cerulean] (bb) {};
        \vertex[below=1 of bb] (bbb);
        \vertex[below=1 of c] (cc);
        \vertex[below=1 of cc] (ccc);

        \vertex[left=.5 of b, dot,minimum size=1.2mm, Cerulean] (bl) {};
        \vertex[left=.5 of bb,dot,minimum size=1.2mm, Cerulean] (bbl) {};

        \vertex[below=1.5 of d1u, dot, minimum size=1.2mm, Cerulean] (bphoton) {};

        \vertex[above=.75 of b] (buu);
        \vertex[above=.4 of d1u] (d1uu);
        
        \diagram*{

        (a) -- [Cerulean] (c);
        (bu) -- [fermion, Cerulean] (d1u);
        (d1u) -- [fermion, Cerulean] (d2);
        (aa) -- [Cerulean] (cc);
        (aaa) -- [Cerulean] (ccc);
        (bl) -- [dashed, Cerulean] (bbl);
        (d1u) -- [photon, Cerulean] (bphoton);
        (bl) -- [double, Cerulean] (b);
        (b) -- [dashed, Cerulean] (bu);
        (d1uu) -- [fermion, Cerulean] (bu);
        };
        \vertex[left=.5 of b, dot,minimum size=1.2mm, Cerulean] (bl) {};
        \vertex[right=0 of b,dot,minimum size=1.2mm, Cerulean] (bbl) {};

    \end{feynman}
\end{tikzpicture}
\caption{(e)}
\end{minipage}
\hspace{0cm}
   \begin{minipage}[c]{3cm}
   \vspace{8.5mm}
\begin{tikzpicture}[line width=1.2 pt]
    \begin{feynman}
        \vertex (a);
        \vertex[right=1. of a, dot, minimum size=1.2mm, Cerulean] (b) {};
        \vertex[below=.5 of b] (bu);
        \vertex[right=1.25 of b] (c);
        \vertex[right=.75 of b] (d);
        \vertex[below=.5 of b,dot, minimum size=1.2mm] (d1);
        \vertex[right=1.5 of bu, Cerulean] (d2) {};
        \vertex[right=.75 of bu, dot, minimum size=1.2mm] (d1u);

        \vertex[below=.5 of b, dot, minimum size=1.2mm, Cerulean] (bu) {};
        \vertex[below=1 of a] (aa);
        \vertex[below=1 of aa] (aaa);
        \vertex[below=1 of b, Cerulean] (bb) {};
        \vertex[below=1 of bb] (bbb);
        \vertex[below=1 of c] (cc);
        \vertex[below=1 of cc] (ccc);

        \vertex[left=.5 of b, dot,minimum size=1.2mm, Cerulean] (bl) {};
        \vertex[left=.5 of bb,dot,minimum size=1.2mm, Cerulean] (bbl) {};

        \vertex[below=1.5 of bu, dot, minimum size=1.2mm, Cerulean] (bphoton) {};

        \vertex[above=.75 of b] (buu);
        \vertex[above=.4 of d1u] (d1uu);
        
        \diagram*{

        (a) -- [Cerulean] (c);
        (bu) -- [fermion, Cerulean] (d1u);
        (aa) -- [Cerulean] (cc);
        (aaa) -- [Cerulean] (ccc);
        (bl) -- [dashed, Cerulean] (bbl);
        (bu) -- [photon, Cerulean] (bphoton);
        (bl) -- [double, Cerulean] (b);
        (b) -- [dashed, Cerulean] (bu);
        (d1uu) -- [fermion,Cerulean] (bu);
        };
        \vertex[left=.5 of b, dot,minimum size=1.2mm, Cerulean] (bl) {};
        \vertex[right=0 of b,dot,minimum size=1.2mm, Cerulean] (bbl) {};

    \end{feynman}
\end{tikzpicture}
\caption{(f)}
\end{minipage}
   \begin{minipage}[c]{3cm}
\begin{tikzpicture}[line width=1.2 pt]
    \begin{feynman}
        \vertex (a);
        \vertex[right=.75 of a, dot, minimum size=1.2mm,Cerulean] (b) {};
        \vertex[above=.75 of b] (bu);
        \vertex[right=1.5 of b] (c);
        \vertex[right=.75 of b] (d);
        \vertex[above=.375 of d,dot, minimum size=1.2mm,Cerulean] (d1) {};
        \vertex[above=.75 of c,Cerulean] (d2) {};
        \vertex[below=1 of bb,dot,minimum size=1.2mm] (bbb) ;

        \vertex[below=1 of a] (aa);
        \vertex[below=1 of aa] (aaa);
        \vertex[below=1 of b, dot, minimum size=1.2mm,Cerulean] (bb) {};
        \vertex[below=1 of bb,dot,minimum size=1.2mm,Cerulean] (bbb) {};
         \vertex[below=.5 of bb, dot,minimum size=1.2mm,Cerulean] (bbbb) {};       \vertex[below=1 of c] (cc);
        \vertex[below=1 of cc] (ccc);
        
        \diagram*{

        (a) -- [Cerulean] (c);
        (bu) -- [fermion,Cerulean] (b);
        (b) -- [fermion,Cerulean] (d1);
        (d1) -- [fermion,Cerulean] (d2);
        (aa) -- [Cerulean] (cc);
        (aaa) -- [Cerulean] (ccc);
        (bbb) -- [dashed,Cerulean] (bb);
        (d1) -- [photon,quarter left,Cerulean] (bbbb);
        };

    \end{feynman}
\end{tikzpicture}
\caption{(g)}
\end{minipage}}

\end{center}
\setcounter{figure}{0} 
\captionsetup{labelformat=empty}

\begin{center}
\scalebox{.75}{
   \begin{minipage}[c]{3cm}
   \vspace{13.5mm}
\begin{tikzpicture}[line width=1.2 pt]
    \begin{feynman}
        \vertex (a);
        \vertex[right=.75 of a,Cerulean] (b) {};
        \vertex[above=.75 of b] (bu);
        \vertex[right=1.5 of b] (c);

        \vertex[below=1 of a] (aa);
        \vertex[below=1 of aa] (aaa);
        \vertex[below=1 of b, dot, minimum size=1.2mm,Cerulean] (bb) {};
        \vertex[below=1 of bb,dot,minimum size=1.2mm,Cerulean] (bbb) {};
         \vertex[above=.5 of bbb, dot,minimum size=1.2mm,Cerulean] (bbbb) {}; 
         \vertex[right=.75 of bbbb,dot,minimum size=1.2mm,Cerulean] (b4) {};
         \vertex[right=.75 of b4] (b5);
         \vertex[below=1 of c] (cc);
        \vertex[below=1 of cc] (ccc);

        \vertex[above=1.5 of b4, dot, minimum size=1.2mm,Cerulean] (b6) {};

        \vertex[below=.4 of b4] (b7);
        
        \diagram*{

        (a) -- [Cerulean] (c);
  
        (aa) -- [Cerulean] (cc);
        (aaa) -- [Cerulean] (ccc);
        (bbb) -- [dashed,Cerulean] (bb);
        (bbbb) -- [fermion,Cerulean] (b4);
        (b4) -- [fermion,Cerulean] (b5);
        (b4) -- [photon,Cerulean] (b6);
        (b7) -- [fermion,Cerulean] (bbbb);
        };

    \end{feynman}
\end{tikzpicture}
\caption{(h)}
\end{minipage}
\hspace{0cm}
   \begin{minipage}[c]{3cm}
   \vspace{1.4cm}
\begin{tikzpicture}[line width=1.2 pt]
    \begin{feynman}
        \vertex (a);
        \vertex[right=.75 of a,dot, minimum size=1.2mm,Cerulean] (b) {};
        \vertex[above=.75 of b] (bu);
        \vertex[right=1.5 of b] (c);

        \vertex[below=1 of a] (aa);
        \vertex[below=1 of aa] (aaa);
        \vertex[below=1 of b, dot, minimum size=1.2mm,Cerulean] (bb) {};
        \vertex[below=1 of bb,dot,minimum size=1.2mm,Cerulean] (bbb) {};
         \vertex[above=.5 of bbb, dot,minimum size=1.2mm,Cerulean] (bbbb) {}; 
         \vertex[right=.75 of bbbb,dot,minimum size=1.2mm] (b4);
         \vertex[right=.25 of b4] (b5);
         \vertex[below=1 of c] (cc);
        \vertex[below=1 of cc] (ccc);

        \vertex[above=1.5 of b4, dot, minimum size=1.2mm] (b6) ;

        \vertex[below=.4 of b4] (b7);
        
        \diagram*{

        (a) -- [Cerulean] (c);
  
        (aa) -- [Cerulean] (cc);
        (aaa) -- [Cerulean] (ccc);
        (bbb) -- [dashed,Cerulean] (bb);
        (bbbb) -- [fermion,Cerulean] (b5);
        (b) -- [photon,quarter left,Cerulean] (bbbb);
        (b7) -- [fermion,Cerulean] (bbbb);
        };

    \end{feynman}
\end{tikzpicture}
\caption{(i)}
\end{minipage}
\hspace{0cm}
   \begin{minipage}[c]{3cm}
   \vspace{.5cm}
\begin{tikzpicture}[line width=1.2 pt]
    \begin{feynman}
        \vertex (a);
        \vertex[right=.75 of a, dot, minimum size=1.2mm,Cerulean] (b) {};
        \vertex[above=.75 of b] (bu);
        \vertex[right=1.5 of b] (c);
        \vertex[right=.75 of b] (d);
        \vertex[above=.375 of d,dot, minimum size=1.2mm,Cerulean] (d1) {};
        \vertex[above=.75 of c,Cerulean] (d2) {};
        \vertex[below=1 of a] (aa);
        \vertex[below=1 of aa] (aaa);
        \vertex[below=1 of b, dot, minimum size=1.2mm] (bb);
        \vertex[below=1 of bb] (bbb);
        \vertex[below=1 of c] (cc);
        \vertex[below=1 of cc] (ccc);

        \vertex[below=1.44 of d1, dot, minimum size=1.2mm,Cerulean] (bphoton) {};

        \vertex[below=1 of bphoton,dot,minimum size=1.2mm,Cerulean] (bphoton2) {};
        
        \diagram*{

        (a) -- [Cerulean] (c);
        (bu) -- [fermion,Cerulean] (b);
        (b) -- [fermion,Cerulean] (d1);
        (d1) -- [fermion,Cerulean] (d2);
        (aa) -- [Cerulean] (cc);
        (aaa) -- [Cerulean] (ccc);
        (d1) -- [photon,Cerulean] (bphoton);
        (bphoton) -- [dashed,Cerulean] (bphoton2);
        };

    \end{feynman}
\end{tikzpicture}
\caption{(j)}
\end{minipage}
\hspace{0cm}
   \begin{minipage}[c]{3cm}
\begin{tikzpicture}[line width=1.2 pt]
    \begin{feynman}
        \vertex (a);
        \vertex[right=.75 of a, dot, minimum size=1.2mm] (b) {};
	\vertex[above=.75 of b,dot, minimum size=1.2mm] (bpion) {};
        \vertex[above=.75 of bpion] (bu);
        \vertex[right=1.5 of b] (c);
        \vertex[right=.75 of b] (d);
        \vertex[right=.75 of bpion,dot, minimum size=1.2mm] (d1) {};
        \vertex[right=.875 of d1] (d2) {};
        \vertex[below=1 of bb,dot,minimum size=1.2mm] (bbb) ;

        \vertex[below=1 of a] (aa);
        \vertex[below=1 of aa] (aaa);
        \vertex[below=1 of b, dot, minimum size=1.2mm] (bb) {};
        \vertex[below=1 of bb,dot,minimum size=1.2mm] (bbb) {};
         \vertex[below=.5 of bb, dot,minimum size=1.2mm] (bbbb) {};       \vertex[below=1 of c] (cc);
        \vertex[below=1 of cc] (ccc);

        \diagram*{

        (a) -- [] (c);
        (bu) -- [fermion] (bpion);
        (bpion) -- [fermion] (d1);
        (d1) -- [fermion] (d2);
	(b) -- [dashed] (bpion);
        (aa) -- [] (cc);
        (aaa) -- [] (ccc);
        (bbb) -- [dashed] (bb);
        (d1) -- [photon,quarter left] (bbbb);
        };

    \end{feynman}
\end{tikzpicture}
\caption{\textcolor{black}{(k)}}
\end{minipage}
\hspace{0cm}
   \begin{minipage}[c]{3cm}
\begin{tikzpicture}[line width=1.2 pt]
    \begin{feynman}
        \vertex (a);
        \vertex[right=.75 of a, dot, minimum size=1.2mm] (b) {};
        \vertex[above=.75 of bpion] (bu);
	\vertex[above=.75 of b,dot,minimum size=1.2mm] (bpion) {};
        \vertex[right=1.5 of b] (c);
        \vertex[right=.75 of b] (d);
        \vertex[right=.75 of bpion,dot, minimum size=1.2mm] (d1) {};
        \vertex[right=.85 of d1] (d2) {};
        \vertex[below=1 of a] (aa);
        \vertex[below=1 of aa] (aaa);
        \vertex[below=1 of b, dot, minimum size=1.2mm] (bb);
        \vertex[below=1 of bb] (bbb);
        \vertex[below=1 of c] (cc);
        \vertex[below=1 of cc] (ccc);

        \vertex[below=1.75 of d1, dot, minimum size=1.2mm] (bphoton) {};

        \vertex[below=1. of bphoton,dot,minimum size=1.2mm] (bphoton2) {};
        
        \diagram*{

        (a) -- [] (c);
        (bu) -- [fermion] (bpion);
        (bpion) -- [fermion] (d1);
        (d1) -- [fermion] (d2);
	(b) -- [dashed] (bpion);
        (aa) -- [] (cc);
        (aaa) -- [] (ccc);
        (d1) -- [photon] (bphoton);
        (bphoton) -- [dashed] (bphoton2);
        };

    \end{feynman}
\end{tikzpicture}
\caption{\textcolor{black}{(l)}}
\end{minipage}
\hspace{0cm}
   \begin{minipage}[c]{3cm}
\begin{tikzpicture}[line width=1.2 pt]
    \begin{feynman}
        \vertex (a);
        \vertex[right=.75 of a, dot, minimum size=1.2mm] (b) {};
        \vertex[above=.75 of bpion] (bu);
        \vertex[above=.75 of b,dot,minimum size=1.2mm] (bpion) {};
        \vertex[right=1.5 of b] (c);
        \vertex[right=.75 of b] (d);
        \vertex[right=.75 of bpion,dot, minimum size=1.2mm] (d1);
        \vertex[right=.25 of d1] (d2) {};
        \vertex[below=1 of a] (aa);
        \vertex[below=1 of aa] (aaa);
        \vertex[below=1 of b, dot, minimum size=1.2mm] (bb);
        \vertex[below=1 of bb] (bbb);
        \vertex[below=1 of c] (cc);
        \vertex[below=1 of cc] (ccc);

        \vertex[below=1.69 of d1, dot, minimum size=1.2mm] (bphoton) {};

        \vertex[below=1. of bphoton,dot,minimum size=1.2mm] (bphoton2) {};

        \diagram*{

        (a) -- [] (c);
        (bu) -- [fermion] (bpion);
        (bpion) -- [fermion] (d2);
        (b) -- [dashed] (bpion);
        (aa) -- [] (cc);
        (aaa) -- [] (ccc);
        (bpion) -- [photon,quarter left] (bphoton);
        (bphoton) -- [dashed] (bphoton2);
        };

    \end{feynman}
\end{tikzpicture}
\caption{\textcolor{black}{(m)}}
\end{minipage}
\hspace{0cm}
   \begin{minipage}[c]{3cm}
\begin{tikzpicture}[line width=1.2 pt]
    \begin{feynman}
        \vertex (a);
        \vertex[right=.75 of a, dot, minimum size=1.2mm] (b) {};
        \vertex[above=.75 of b,dot, minimum size=1.2mm] (bpion) {};
        \vertex[above=.75 of bpion] (bu);
        \vertex[right=1.5 of b] (c);
        \vertex[right=.75 of b] (d);
        \vertex[right=.75 of bpion,dot, minimum size=1.2mm] (d1);
        \vertex[right=.25 of d1] (d2) {};
        \vertex[below=1 of bb,dot,minimum size=1.2mm] (bbb) ;

        \vertex[below=1 of a] (aa);
        \vertex[below=1 of aa] (aaa);
        \vertex[below=1 of b, dot, minimum size=1.2mm] (bb) {};
        \vertex[below=1 of bb,dot,minimum size=1.2mm] (bbb) {};
         \vertex[below=.5 of bb, dot,minimum size=1.2mm] (bbbb) {};       \vertex[below=1 of c] (cc);
        \vertex[below=1 of cc] (ccc);
        
        \diagram*{
        
        (a) -- [] (c);
        (bu) -- [fermion] (bpion);
        (bpion) -- [fermion] (d2);
        (b) -- [dashed] (bpion);
        (aa) -- [] (cc);
        (aaa) -- [] (ccc);
        (bbb) -- [dashed] (bb);
        (bpion) -- [photon,quarter left] (bbbb);
        };

    \end{feynman}
\end{tikzpicture}
\caption{\textcolor{black}{(n)}}
\end{minipage}}
\end{center}
\setcounter{figure}{1} 
\captionsetup{labelformat=empty}

\begin{center}
\scalebox{.75}{
   \begin{minipage}[c]{3cm}
      \vspace{0.2cm}
\begin{tikzpicture}[line width=1.2 pt]
    \begin{feynman}
        \vertex (a);
        \vertex[right=.75 of a] (b);
        \vertex[right=1.5 of b] (c);
        \vertex[right=.75 of b] (d);
        \vertex[below=1 of bb] (bbb);
	\vertex[right=1.05 of a,dot,minimum size=1.2mm] (d) {};
	\vertex[above=.75 of d] (du);
	\vertex[below=1 of d,dot,minimum size=1.2mm] (dd) {};
	\vertex[below=1 of dd,dot,minimum size=1.2mm] (ddd) {};
	\vertex[right=1 of bb] (b3);
	\vertex[above=1.4 of b3] (b4);

        \vertex[below=1 of a] (aa);
        \vertex[below=1 of aa] (aaa);
        \vertex[below=1 of b] (bb);
        \vertex[below=1 of c] (cc);
        \vertex[below=1 of cc] (ccc);
        
        \diagram*{
        
        (a) -- [] (c);
        (aa) -- [] (cc);
        (aaa) -- [] (ccc);
	(du) -- [fermion] (d);
	(d) -- [fermion] (b4);
	(d) -- [dashed] (dd);
	(dd) -- [photon] (ddd);
        };

    \end{feynman}
\end{tikzpicture}
\caption{\textcolor{black}{(o)}}
\end{minipage}
\hspace{0cm}
   \begin{minipage}[c]{3cm}
      \vspace{-.05cm}
\begin{tikzpicture}[line width=1.2 pt]
    \begin{feynman}
        \vertex (a);
        \vertex[right=.75 of a, dot, minimum size=1.2mm] (b) {};
        \vertex[above=.75 of b] (bu);
        \vertex[right=1.5 of b] (c);
        \vertex[right=.75 of b] (d);
        \vertex[above=.375 of d] (d1);
        \vertex[above=.75 of c] (d2) {};
        \vertex[below=1 of a] (aa);
        \vertex[below=1 of aa] (aaa);
        \vertex[below=1 of b, dot, minimum size=1.2mm] (bb) {};
        \vertex[below=1 of bb] (bbb);
        \vertex[below=1 of c] (cc);
        \vertex[below=1 of cc] (ccc);
	\vertex[below=.5 of b,dot,minimum size=1.2mm] (bd) {};

        \vertex[below=2.31 of d1, dot, minimum size=1.2mm] (bphoton) {};

        \diagram*{

        (a) -- [] (c);
        (bu) -- [fermion] (b);
        (b) -- [fermion] (d1);
        (aa) -- [] (cc);
        (aaa) -- [] (ccc);
        (b) -- [dashed] (bb);
        (bd) -- [photon,quarter left] (bphoton);
        };

    \end{feynman}
\end{tikzpicture}
\caption{\textcolor{black}{(p)}}
\end{minipage}
\hspace{0cm}
   \begin{minipage}[c]{3cm}
\begin{tikzpicture}[line width=1.2 pt]
    \begin{feynman}
        \vertex (a);
        \vertex[right=.75 of a, dot, minimum size=1.2mm] (b) {};
        \vertex[above=.75 of b] (bu);
	\vertex[right=.75 of b,dot,minimum size=1.2mm] (delta) {};
        \vertex[right=1.5 of b] (c);
        \vertex[right=.75 of delta] (d);
        \vertex[above=.375 of d] (d1);
        \vertex[above=.75 of c] (d2) {};
        \vertex[below=1 of a] (aa);
        \vertex[below=1 of aa] (aaa);
        \vertex[below=1 of b, dot, minimum size=1.2mm] (bb) {};
        \vertex[below=1 of bb] (bbb);
        \vertex[below=1 of c] (cc);
        \vertex[below=1 of cc] (ccc);
	\vertex[below=.5 of delta,dot,minimum size=1.2mm] (dpi) {};
	\vertex[right=.75 of dpi] (dfermi1);
	\vertex[below=.35 of dfermi1] (dfermi2);

        \vertex[below=1 of bb, dot, minimum size=1.2mm] (bphoton) {};

        \diagram*{

        (a) -- [] (b);
        (delta) -- [] (c);
        (b) -- [double] (delta);
        (aa) -- [] (cc);
        (aaa) -- [] (ccc);
        (b) -- [dashed] (bb);
        (bb) -- [photon] (bphoton);
	(delta) -- [dashed] (dpi);
	(dpi) -- [fermion] (dfermi1);
	(dfermi2) -- [fermion] (dpi);
        };
    \end{feynman}
\end{tikzpicture}
\caption{\textcolor{black}{(q)}}
\end{minipage}
\hspace{0cm}
   \begin{minipage}[c]{3cm}
\begin{tikzpicture}[line width=1.2 pt]
    \begin{feynman}
        \vertex (a);
        \vertex[right=.75 of a, dot, minimum size=1.2mm] (b) {};
        \vertex[above=.75 of b] (bu);
        \vertex[right=1.5 of b] (c);
        \vertex[right=.75 of b] (d);
        \vertex[above=.375 of d] (d1);
        \vertex[above=.75 of c] (d2) {};
	\vertex[right=.75 of b,dot,minimum size=1.2mm] (delta) {};
        \vertex[below=1 of a] (aa);
        \vertex[below=1 of aa] (aaa);
        \vertex[below=1 of b, dot, minimum size=1.2mm] (bb) {};
        \vertex[below=1 of bb] (bbb);
        \vertex[below=1 of c] (cc);
        \vertex[below=1 of cc] (ccc);
        \vertex[below=.5 of b,dot,minimum size=1.2mm] (bd) {};
	\vertex[below=.5 of delta,dot,minimum size=1.2mm] (dpi) {};
        \vertex[right=.75 of dpi] (dfermi1);
        \vertex[below=.35 of dfermi1] (dfermi2);
        \vertex[below=2.31 of d1, dot, minimum size=1.2mm] (bphoton) {};

        \diagram*{

        (a) -- [] (b);
        (delta) -- [] (c);
        (delta) -- [double] (b);
        (aa) -- [] (cc);
        (aaa) -- [] (ccc);
        (b) -- [dashed] (bb);
        (bd) -- [photon,quarter left] (bphoton);
        (delta) -- [dashed] (dpi);
        (dpi) -- [fermion] (dfermi1);
        (dfermi2) -- [fermion] (dpi);
        };

    \end{feynman}
\end{tikzpicture}
\caption{\textcolor{black}{(r)}}
\end{minipage}
\hspace{0cm}
   \begin{minipage}[c]{3cm}
\begin{tikzpicture}[line width=1.2 pt]
    \begin{feynman}
        \vertex (a);
        \vertex[right=.75 of a, dot, minimum size=1.2mm] (b) {};
        \vertex[above=.75 of b] (bu);
        \vertex[right=.75 of b,dot,minimum size=1.2mm] (delta) {};
        \vertex[right=1.5 of b] (c);
        \vertex[right=.75 of delta] (d);
        \vertex[above=.375 of d] (d1);
        \vertex[above=.75 of c] (d2) {};
        \vertex[below=1 of a] (aa);
        \vertex[below=1 of aa] (aaa);
        \vertex[below=1 of b, dot, minimum size=1.2mm] (bb) {};
        \vertex[below=1 of bb] (bbb);
        \vertex[below=1 of c] (cc);
        \vertex[below=1 of cc] (ccc);
        \vertex[below=.5 of delta,dot,minimum size=1.2mm] (dpi);
        \vertex[above=.75 of delta] (dfermi1);
        \vertex[above=.35 of c] (dfermi2);

        \vertex[below=1 of bb, dot, minimum size=1.2mm] (bphoton) {};

        \diagram*{

        (a) -- [] (b);
        (delta) -- [] (c);
        (b) -- [double] (delta);
        (aa) -- [] (cc);
        (aaa) -- [] (ccc);
        (b) -- [dashed] (bb);
        (bb) -- [photon] (bphoton);
        (dfermi1) -- [fermion] (delta);
        (delta) -- [fermion] (dfermi2);
        };
    \end{feynman}
\end{tikzpicture}
\caption{\textcolor{black}{(s)}}
\end{minipage}
\hspace{0cm}
   \begin{minipage}[c]{3cm}
\begin{tikzpicture}[line width=1.2 pt]
    \begin{feynman}
        \vertex (a);
        \vertex[right=.75 of a, dot, minimum size=1.2mm] (b) {};
        \vertex[above=.75 of b] (bu);
        \vertex[right=1.5 of b] (c);
        \vertex[right=.75 of b] (d);
        \vertex[above=.375 of d] (d1);
        \vertex[above=.75 of c] (d2) {};
        \vertex[right=.75 of b,dot,minimum size=1.2mm] (delta) {};
        \vertex[below=1 of a] (aa);
        \vertex[below=1 of aa] (aaa);
        \vertex[below=1 of b, dot, minimum size=1.2mm] (bb) {};
        \vertex[below=1 of bb] (bbb);
        \vertex[below=1 of c] (cc);
        \vertex[below=1 of cc] (ccc);
        \vertex[below=.5 of b,dot,minimum size=1.2mm] (bd) {};
        \vertex[below=.5 of delta,dot,minimum size=1.2mm] (dpi);
        \vertex[above=.75 of delta] (dfermi1);
        \vertex[above=.35 of c] (dfermi2);
        \vertex[below=2.31 of d1, dot, minimum size=1.2mm] (bphoton) {};

        \diagram*{

        (a) -- [] (b);
        (delta) -- [] (c);
        (delta) -- [double] (b);
        (aa) -- [] (cc);
        (aaa) -- [] (ccc);
        (b) -- [dashed] (bb);
        (bd) -- [photon,quarter left] (bphoton);
        (dfermi1) -- [fermion] (delta);
        (delta) -- [fermion] (dfermi2);
        };

    \end{feynman}
\end{tikzpicture}
\caption{\textcolor{black}{(t)}}
\end{minipage}
\hspace{0cm}
   \begin{minipage}[c]{3cm}
\begin{tikzpicture}[line width=1.2 pt]
    \begin{feynman}
        \vertex (a);
        \vertex[right=.75 of a, dot, minimum size=1.2mm] (b) {};
        \vertex[above=.75 of b] (bu);
        \vertex[right=1.5 of b] (c);
        \vertex[right=.75 of b] (d);
        \vertex[above=.375 of d] (d1);
        \vertex[above=.75 of c] (d2) {};
        \vertex[right=.75 of b,dot,minimum size=1.2mm] (delta) {};
        \vertex[below=1 of a] (aa);
        \vertex[below=1 of aa] (aaa);
        \vertex[below=1 of b, dot, minimum size=1.2mm] (bb) {};
        \vertex[below=1 of bb] (bbb);
        \vertex[below=1 of c] (cc);
        \vertex[below=1 of cc] (ccc);
        \vertex[below=.5 of b,dot,minimum size=1.2mm] (bd);
        \vertex[below=.7 of delta,dot,minimum size=1.2mm] (dpi) {};
        \vertex[right=.75 of dpi] (dfermi1);
        \vertex[above=.35 of dfermi1] (dfermi2);
        \vertex[below=2.31 of d1, dot, minimum size=1.2mm] (bphoton) {};
	\vertex[below=.35 of delta,dot,minimum size=1.2mm] (dpi2) {};

        \diagram*{
        
        (a) -- [] (b);
        (delta) -- [] (c);
        (delta) -- [double] (b); 
        (aa) -- [] (cc);
        (aaa) -- [] (ccc);
        (b) -- [dashed] (bb); 
        (dpi2) -- [photon,quarter right] (bphoton);
        (delta) -- [dashed] (dpi); 
        (dpi) -- [fermion] (dfermi1);
        (dfermi2) -- [fermion] (dpi);
        };

    \end{feynman}
\end{tikzpicture}
\caption{\textcolor{black}{(u)}}
\end{minipage}}

\end{center}
\setcounter{figure}{0} 
\captionsetup{labelformat=default}
\caption{Diagrams contributing to the three-nucleon potentials at $\mathcal{O}(\alpha\epsilon_\chi^2)$. Solid, dashed, wavy, arrowed, and double lines denote nucleons, pions, photons, leptons, and deltas, respectively. Dots denote LO vertices. Black diagrams are suppressed, being proportional to the outgoing lepton energy, or do not contribute to Fermi $\beta$ decays.}\label{fig:diagrams}
\end{figure}

There are 21 diagrams given in Fig. \ref{fig:diagrams}. However, most of these are suppressed, being proportional to the energy of the outgoing lepton or not contributing to $0^+\rightarrow0^+$ decays. The momentum space potentials for the 8 remaining diagrams, shown in blue in Fig. \ref{fig:diagrams}, are given by
\begin{align}
    \mathcal{V}_0^{(a)}&=2\pi\frac{g_A^2}{F_\pi^2}\sum_{i\neq j\neq k}\frac{\boldsymbol{\sigma}_i\cdot\boldsymbol{q}_k}{\boldsymbol{q}_k^4}\frac{\boldsymbol{\sigma}_j\cdot\boldsymbol{q}_j}{\boldsymbol{q}_j^2+m_\pi^2}P_p^{(k)}(\tau_i^\pm\tau_j^3-\tau_i^3\tau_j^\pm) \label{eq:potfirst}
    \\
    \mathcal{V}_0^{(d)}&=\frac{8\pi}{9}\frac{g_Ah_A^2}{F_\pi^2\Delta}\sum_{i\neq j\neq k}\frac{1}{\boldsymbol{q}_k^4}\frac{\boldsymbol{\sigma}_j\cdot\boldsymbol{q}_j}{\boldsymbol{q}_j^2+m_\pi^2}P_p^{(k)}\left(4\boldsymbol{q}_k\cdot\boldsymbol{q}_j\tau_j^\pm+\boldsymbol{\sigma}_i\cdot\boldsymbol{q}_k\times\boldsymbol{q}_j(\boldsymbol{\tau}_i\times\boldsymbol{\tau}_j)^\pm\right)
    \\
    \mathcal{V}_0^{(e)}&=-\frac{8\pi}{9}\frac{g_Ah_A^2}{F_\pi^2\Delta}\sum_{i\neq j\neq k}\frac{1}{\boldsymbol{q}_k^2}\frac{1}{\boldsymbol{q}_k^2+m_\pi^2}\frac{\boldsymbol{\sigma}_j\cdot\boldsymbol{q}_j}{\boldsymbol{q}_j^2+m_\pi^2}P_p^{(k)}\left(4\boldsymbol{q}_k\cdot\boldsymbol{q}_j\tau_j^\pm+\boldsymbol{\sigma}_i\cdot\boldsymbol{q}_k\times\boldsymbol{q}_j(\boldsymbol{\tau}_i\times\boldsymbol{\tau}_j)^\pm\right)
    \\ 
    \mathcal{V}_0^{(f)}&=\frac{8\pi}{9}\frac{g_Ah_A^2}{F_\pi^2\Delta}\sum_{i\neq j\neq k}\frac{1}{\boldsymbol{q}_k^2}\frac{1}{\boldsymbol{q}_k^2+m_\pi^2}\frac{\boldsymbol{\sigma}_j\cdot\boldsymbol{q}_j}{\boldsymbol{q}_j^2+m_\pi^2}P_p^{(k)}\left(4\boldsymbol{q}_k\cdot\boldsymbol{q}_j\tau_j^\pm+\boldsymbol{\sigma}_i\cdot\boldsymbol{q}_k\times\boldsymbol{q}_j(\boldsymbol{\tau}_i\times\boldsymbol{\tau}_j)^\pm\right)
\\
\mathcal{V}_0^{(g)}&=2\pi\frac{g_A^2}{F_\pi^2}\sum_{i\neq j\neq k}\frac{1}{\boldsymbol{q}_i^4}\frac{\boldsymbol{\sigma}_k\cdot\boldsymbol{q}_k}{\boldsymbol{q}_k^2+m_\pi^2}\left(\boldsymbol{\sigma}_j\cdot\boldsymbol{q}_j-ig_A\boldsymbol{\sigma}_i\cdot\boldsymbol{q}_k\times\boldsymbol{q}_i\frac{\boldsymbol{\sigma}_j\cdot\boldsymbol{q}_j}{\boldsymbol{q}_j^2+m_\pi^2}\right)\tau_i^\pm(\tau_j^+\tau_k^--\tau_j^-\tau_k^+)
    \\
    \mathcal{V}_0^{(h)}&=-2\pi\frac{g_A^2}{F_\pi^2}\sum_{i\neq j\neq k}\frac{1}{\boldsymbol{q}_k^4}\frac{\boldsymbol{\sigma}_i\cdot\boldsymbol{q}_i}{\boldsymbol{q}_i^2+m_\pi^2}\boldsymbol{\sigma}_j\cdot\boldsymbol{q}_jP_p^{(k)}(\tau_i^\pm\tau_j^3-\tau_i^3\tau_j^\pm)
\\
\mathcal{V}_0^{(i)}&=\pi\frac{g_A^2}{F_\pi^2}\sum_{i\neq j\neq k}\frac{1}{\boldsymbol{q}_i^2}\frac{\boldsymbol{\sigma}_j\cdot\boldsymbol{q}_j}{\boldsymbol{q}_j^2+m_\pi^2}\frac{\boldsymbol{\sigma}_k\cdot\boldsymbol{q}_k}{\boldsymbol{q}_k^2+m_\pi^2}P_p^{(i)}(\tau_j^\pm\tau_k^3+\tau_j^3\tau_k^\pm)
\\
\mathcal{V}_0^{(j)}&=2\pi\frac{g_A^2}{F_\pi^2}\sum_{i\neq j\neq k}\frac{1}{\boldsymbol{q}_i^4}\frac{\boldsymbol{\sigma}_k\cdot\boldsymbol{q}_k}{\boldsymbol{q}_k^2+m_\pi^2}\left(-\boldsymbol{\sigma}_j\cdot\boldsymbol{q}_i+ig_A\boldsymbol{\sigma}_i\cdot\boldsymbol{\sigma}_j\times\boldsymbol{q}_i\right)\tau_i^\pm(\tau_k^+\tau_j^--\tau_k^-\tau_j^+). \label{eq:potlast}
\end{align}

\noindent where $\boldsymbol{\sigma}$ ($\boldsymbol{\tau}$) is the Pauli spin (isospin) matrix, $\boldsymbol{q}_i$ is the momentum transferred to nucleon $i$, and $\tau^\pm$ is the isospin raising or lowering operator.

As we discussed earlier, we factored out a power of $\alpha$, which explains the appearance of factors of $\pi$ in Eqs.~\eqref{eq:potfirst}--\eqref{eq:potlast}.
The three diagrams with an intermediate $\Delta$ correspond to $\mathcal{V}_0^{(d)}$, $\mathcal{V}_0^{(e)}$, and $\mathcal{V}_0^{(f)}$. In these potentials, $\Delta$ should be understood as the mass splitting $\Delta = m_\Delta-m_N$. However, these potentials are parity odd and therefore do not contribute to superallowed beta decay. Two of the $\Delta$ diagrams feature an additional cancellation: $\mathcal{V}_0^{(e)}=-\mathcal{V}_0^{(f)}$. This can be understood at the level of vertices, where, upon neglecting the electron mass, the combination leads to a vanishing contribution independent of the precense of a $\Delta$:

\begin{figure}[H]
\begin{center}
\scalebox{.75}{
   \begin{minipage}[c]{2.5cm}
\begin{tikzpicture}[line width=1.2 pt]
    \begin{feynman}
        \vertex (a); 
        \vertex[right=1 of a, dot,minimum size=1.2mm] (b) {};
        \vertex[right=1 of b, dot,minimum size=1.2mm] (bb) {};
        \vertex[right=1 of bb,dot] (c);
        \vertex[right=1 of c] (d);
        \vertex[above left=1 of b] (e);
        \vertex[above right=1 of bb] (f);
    
        \diagram*{
        (a) -- [dashed] (b);
        (bb) -- [photon] (c);
        (b) -- [fermion] (bb);
        (e) -- [fermion] (b); 
        (bb) -- [fermion] (f);
        };
    \end{feynman}
\end{tikzpicture} 
\end{minipage}%
\hspace{.6cm}
   \begin{minipage}[c]{1cm}
    \[+\vspace{.2cm}\]
\end{minipage}
   \begin{minipage}[c]{2.5cm}
\begin{tikzpicture}[line width=1.2 pt]
    \begin{feynman}
        \vertex (a);
        \vertex[right=1 of a, dot,minimum size=1.2mm] (b) {};
        \vertex[right=1 of b,dot] (c);
        \vertex[right=1 of c] (d);
        \vertex[above left=1 of b] (e);
        \vertex[above right=1 of b] (f);

        \diagram*{
        (a) -- [dashed] (b);
        (b) -- [photon] (c);
        (e) -- [fermion] (b);
        (b) -- [fermion] (f);
        };
    \end{feynman}
\end{tikzpicture}
\end{minipage}%
}
   \begin{minipage}[c]{8cm}
    \[=-\sqrt{2}eF_\pi G_F V_{ud}(\delta_{1a}+i\delta_{2a})\bar{u}_L(p_e)\gamma^\mu\left(\frac{q^2+\slashed{q}m_e}{q^2-m_e^2}-1\right)v_L(p_\nu)\hspace{10cm}\vspace{.2cm}\]
\end{minipage}

\end{center}
\end{figure}

\noindent with the pion having an isospin index $a$ and photon having momentum $q$ and Lorentz index $\mu$. Therefore, there are no contributions from the $\Delta$ in these $\mathcal{O}(\alpha\epsilon_\chi^2)$ potentials. In a $\Delta$-less interaction, the tree-level contributions of the $\Delta$ would be absorbed by the LECs $c_{3,4}$, which would appear at one order higher in the expansion. However, the $2\pi2N$ vertex containing $c_{3,4}$ is momentum-dependent, and the $\Delta$-less version of $\mathcal{V}_0^{(d)}$ would also be parity odd and therefore not contribute to $0^+\rightarrow0^+$. The $\Delta$-less version of $\mathcal{V}_{0}^{(e)}$ and $\mathcal{V}_0^{(f)}$ would feature the same cancellation because the previous vertex-level argument is independent of what vertex the photon is contracted with.

\subsection{The role of vector and axial currents at order $\mathcal{O}(\alpha)$}\label{sec:VA}

The foundational work formulating radiative corrections for semileptonic decays, such as the exchange of $\gamma$ and $Z$ between hadrons and leptons for pion and Fermi $\beta$ decays, was performed exploiting current algebra and Ward identities \cite{Sirlin_1978}. A modern approach is the dispersive representation (DR) for the case of the simultaneous exchange of the $W$ and $\gamma$ (so-called $\gamma W$-box diagram) \cite{Seng:2018yzq,Seng:2018qru,Gorchtein_2019,Seng_2023}. A central result of these calculations is the role of vector-axial (VA) contributions as opposed to vector-vector (VV) for the calculation of corrections to Fermi decays at order $\mathcal{O}(\alpha)$. However, in the three-nucleon potentials derived within the EFT approach, there appear to be both VV and VA contributions, i.e. terms proportional to $g_A^2$ and $g_A^3$, respectively. To reconcile these differences, we first review the central argument in the notation of the DR approach and then show how this maps into the language of Heavy Baryon Chiral Perturbation Theory (HB$\chi$PT).

\subsubsection{Argument in the dispersive approach}

The essential object of the EWRC is the generalized Compton tensor given by the time-order product,
\begin{align}
    T^{\mu\nu}(q;p^\prime,p)\equiv \frac{1}{2}\int \mathrm{d}^4x\;e^{iq\cdot x}\langle \Psi_f(p^\prime)|T\{J_{\mathrm{EM}}^\mu(x) J_\mathrm{W}^\nu(0)\}|\Psi_i(p)\rangle,
\end{align}
with initial (final) hadronic states $\Psi_i(p)$ ($\Psi_f(p^\prime)$) and photon momentum $q$. For a more thorough overview, see Ref. \cite{Seng_2023}. The EM current $J_{\mathrm{EM}}^\mu$ is purely vector whereas the charged weak current has both vector and axial components. This motivates decomposing the tensor into two parts: $T^{\mu\nu}=T_{\mathrm{VV}}^{\mu\nu}+T_{\mathrm{VA}}^{\mu\nu}$. 
In the forward limit ($p=p^\prime$) the non-Born part of the Compton tensor take the general form
\begin{align}
    T^{\mu\nu}_{\mathrm{VV}}&=\left(-g^{\mu\nu}+\frac{q^\mu q^\nu}{q^2}\right)T_1(\nu,Q^2)+\frac{\hat{p}^\mu\hat{p}^\nu}{m_N\nu}T_2(\nu,Q^2), \quad T^{\mu\nu}_{\mathrm{VA}}=\frac{i\epsilon^{\mu\nu\alpha\beta}p_\alpha q_\beta}{2m_N\nu}T_3(\nu,Q^2)
\end{align}
where $\nu=q^0$, $Q^2=-q^2$, $\hat{p}=p^\mu-p\cdot q q^\mu/q^2$, and $T_{1,2,3}(\nu,Q^2)$ are Lorentz scalars. This Compton tensor is contracted with the leptonic tensor, which, after expanding the Dirac matrices and taking $p_e\rightarrow0$, becomes
\begin{align}
    L^{\mu\nu}=\Bar{u}(p_e)\gamma^\mu\slashed{q}\gamma^\nu P_Lv(p_\nu)&=\bar u(p_e) \left[ q^\nu g^{\mu \beta} + q^\mu g^{\nu\beta}  - g^{\mu \nu} q^\beta + i \varepsilon^{\mu \alpha \nu \beta} q_\alpha \right]\gamma_\beta  P_L v(p_\nu)
   \nonumber  \\
    &=\Bar{u}(p_e)\,l^{\mu\nu\beta}(q)\gamma_\beta P_Lv(p_\nu)
\end{align}
with $\epsilon^{0123}=-1$ and we have defined $l^{\mu\nu\beta}=\left[q^\nu g^{\mu\beta}+q^\mu g^{\nu\beta}-g^{\mu\nu}q^\beta+i\epsilon^{\mu\alpha\nu\beta}q_\alpha\right]$ stripped of the spinors, which will be useful in the discussion later. The first three terms are symmetric with respect to the exchange of $\mu\leftrightarrow\nu$, and the last term is antisymmetric.  Because $T_\mathrm{VV}^{\mu\nu}$ is symmetric in the forward limit, the antisymmetric term does not contribute in the contraction with the leptonic tensor, and the contraction reduces to
\begin{align}
    L_{\mu\nu}T_\mathrm{VV}^{\mu\nu}=2q_\mu T_\mathrm{VV}^{\mu\nu}\Bar{u}(p_e)\gamma_\nu P_L v(p_\nu)-T_{\mathrm{VV},\mu}^\mu\bar{u}(p_e)\slashed{q}P_Lv(p_\nu).
\end{align}

\noindent The first term vanishes with $q_\mu T_\mathrm{VV}^{\mu\nu}=0$. The trace $q_\beta T^\mu_{\mathrm{VV},\mu}$ remains a nonvanishing nuclear-structure-dependent correction but, in Sirlin's representation \cite{Sirlin_1978}, is absorbed by a similar term appearing in a hadronic vertex correction (see Ref. \cite{Seng_2021} for more detail). Therefore, while the VV contribution to the Compton tensor does not vanish, it has no contribution to $\Bar{\delta}_\mathrm{NS}$ at this order. The VA component survives the contraction with the antisymmetric part of the lepton tensor, which is retained in the $\gamma W$-box contribution.

\subsubsection{Argument in the HB$\chi$PT approach}

Now we will demonstrate the same argument in the HB$\chi$PT formalism. Matching onto nucleon and pion fields, the single-nucleon vector and axial currents are

\begin{align}
    \mathcal{J}_\mathrm{W,V}^\mu&=\Bar{N}\left[g_V(\mathbf{q}^2)\left(v^\mu+\frac{p^\mu+p^{\prime \mu}}{2m_N}\right)+ig_M(\mathbf{q}^2)\epsilon^{\mu\nu\alpha\beta}\frac{v_\alpha S_\beta q_\nu}{m_N}\right]\tau^+N
    \\
    \mathcal{J}_\mathrm{W,A}^\mu&=\Bar{N}\left[-2g_A(\mathbf{q}^2)\left(S^\mu-\frac{S\cdot(p+p^\prime)}{2m_N}v^\mu+\frac{S\cdot q}{\mathbf{q}^2+m_\pi^2}q^\mu\right)\right]\tau^+N,
\end{align}
where the divergence of each current can be written,
\begin{align}
    q_\mu \mathcal{J}^\mu_\mathrm{W,V}&=g_V\Bar{N}\left(v\cdot q+\frac{p^{\prime2}-p^2}{2m_N}\right)\tau^+N\xrightarrow[\text{}]{\text{on-shell}}0\label{eq:divV}
    \\
    q_\mu\mathcal{J}_\mathrm{W,A}^\mu&=-2g_A\frac{m_\pi^2}{\mathbf{q}^2+m_\pi^2}\Bar{N}S\cdot q\tau^+N\neq 0.
\end{align}
We emphasize that Eq. (\ref{eq:divV}) vanishes only when the nucleon is free and on-shell, which is not the case in a bound nuclear system. This subtlety complicates the argument regarding VV contributions, which we begin with by constructing the leading two- and three-nucleon operators explicitly, mirroring the dispersive approach. For the $\mathcal{O}(\alpha)$ diagrams with the photon attaching to an electron, the operator can be factorized into a product of EM and W currents. Retaining the temporal component relevant for Fermi decays, the operator takes the form
\begin{align}
    \mathcal{V}_0=-e\frac{l_{\mu\nu0}(q)}{q^4}\mathcal{J}_\mathrm{EM}^\mu(q)\mathcal{J}_\mathrm{W}^\nu(-q)
\end{align}
with photon momentum $q$ and $l^{\mu\nu\beta}$ the same as in the previous section. Each of these currents can be expanded in many-nucleon terms,
\begin{align}
    \mathcal{J}_\mathrm{X}^\mu=\sum_i\mathcal{J}_{\mathrm{X},i}^\mu+\sum_{i\neq j}\mathcal{J}_{\mathrm{X},ij}^\mu+\ldots=\mathcal{J}_\mathrm{X,\,1N}^\mu+\mathcal{J}_\mathrm{X,\,2N}^\mu+\ldots,\quad\quad\mathrm{X}\in\{\mathrm{EM,W}\}.\label{eq:curr_expansion}
\end{align}
Performing the contractions of the leptonic tensor with the currents, there are five structures:
\begin{align}
    \mathcal{V}_0&=-\frac{e}{q^4}\left[\mathcal{J}_\mathrm{EM}^0\left(q\cdot\mathcal{J}_\mathrm{W}\right)+\left(q\cdot\mathcal{J}_\mathrm{EM}\right)\mathcal{J}_\mathrm{W}^0-q^0\mathcal{J}_\mathrm{EM}^0\mathcal{J}_\mathrm{W}^0+q^0\bm{\mathcal{J}}_\mathrm{EM}\cdot\bm{\mathcal{J}}_\mathrm{W}+i\boldsymbol{q}\cdot\left(\bm{\mathcal{J}}_\mathrm{EM}\times\bm{\mathcal{J}}_\mathrm{W}\right)\right]\label{eq:V0}
\end{align}
with the explicit momentum-dependence of the currents dropped. The first four terms come from the symmetric part of $l^{\mu\nu\beta}$, and the last term comes from the antisymmetric part. The parity structure of $\mathcal{J}_\mathrm{EM}^\mu$ and $\mathcal{J}_\mathrm{W}^\mu$ will restrict which of these five terms can contribute to the VV and VA structure. Since $\mathcal{J}_\mathrm{EM}^0$ ($\bm{\mathcal{J}}_\mathrm{EM}$) and $\mathcal{J}_\mathrm{W,V}^0$ ($\bm{\mathcal{J}}_\mathrm{W,V}$) are scalar (polar vector) and $\mathcal{J}_\mathrm{W,A}$ ($\bm{\mathcal{J}}_\mathrm{W,A}$) is a pseudoscalar (axial vector), the first four terms are even (odd) and the last term is odd (even) for the VV (VA) components. Therefore, $0^+\rightarrow0^+$ transitions select only the first four terms for VV operators and only the last term for VA operators. This is the equivalent argument in the dispersive approach with the symmetric parts of the leptonic tensor contributing to VV and antisymmetric contributing to VA. With the fourth term of Eq. (\ref{eq:V0}) suppressed at LO, the VV component is
\begin{align}
    \mathcal{V}_{0,\mathrm{VV}}=-\frac{e}{q^4}\left[\mathcal{J}_\mathrm{EM}^0\left(q\cdot\mathcal{J}_\mathrm{W}\right)+\left(q\cdot\mathcal{J}_\mathrm{EM}\right)\mathcal{J}_\mathrm{W}^0-q^0\mathcal{J}_\mathrm{EM}^0\mathcal{J}_\mathrm{W}^0\right].\label{eq:V_VV}
\end{align}
Inserting the current expansion of eq. (\ref{eq:curr_expansion}) into eq. (\ref{eq:V_VV}) produces the two- and three-nucleon operators. The two-nucleon piece, built from single-nucleon currents, is
\begin{align}
    \mathcal{V}_{0,\mathrm{VV}}^\mathrm{2N}&=-e\sum_{i\neq j}\frac{1}{q^4}\left[\mathcal{J}_{\mathrm{EM},i}^0\left(q\cdot\mathcal{J}_{\mathrm{W,V},j}\right)+\left(q\cdot\mathcal{J}_{\mathrm{EM},i}\right)\mathcal{J}_{\mathrm{W,V},j}^0-q^0\mathcal{J}_{\mathrm{EM},i}^0\mathcal{J}_{\mathrm{W,V},j}^0\right].
\end{align}
The three-nucleon operators can be constructed from a product of one- and two-nucleon currents. Since the two-nucleon charges are higher order, the possibilities are
\begin{align}
    \left.\mathcal{V}_{0,\mathrm{VV}}^\mathrm{3N}\right|_{(a),(h)}=-e\sum_{i\neq j\neq k}\frac{1}{q_k^4}\mathcal{J}_{\mathrm{EM},k}^0\left(q\cdot\mathcal{J}_{\mathrm{W,V},ij}\right),\quad\quad\left.\mathcal{V}_{0,\mathrm{VV}}^\mathrm{3N}\right|_{(g),(j)}=-e\sum_{i\neq j\neq k}\frac{1}{q_i^4}\left(q\cdot\mathcal{J}_{\mathrm{EM},jk}\right)\mathcal{J}_{\mathrm{W,V},i}^0,
\end{align}
which can be identified with the potentials of diagrams $(a)$ and $(h)$ or $(g)$ and $(j)$. Utilizing the Heisenberg equation of motion, the many-body embodiment of current conservation is
\begin{align}
\mathbf{q}\cdot\bm{\mathcal{J}}_\mathrm{X}(q)=[H,\mathcal{J}_\mathrm{X}^0(q)],\quad\quad H=T+v_\pi+v_\mathrm{CT}+\ldots.
\end{align}

Expanding the Hamiltonian $H$ up to LO and matching $A$-nucleon terms, the commutation relations become
\begin{align}
\mathbf{q}\cdot\bm{\mathcal{J}}_{\mathrm{X,1N}}(q)=[T,\mathcal{J}_{\mathrm{X,1N}}^0(q)],\quad\quad\mathbf{q}\cdot\bm{\mathcal{J}}_{\mathrm{X,2N}}(q)=[v_\pi,\mathcal{J}_\mathrm{X,1N}^0(q)]\label{eq:cc}
\end{align}
\noindent where we have restored the explicit momentum-dependence of the currents to emphasize that Eq. (\ref{eq:cc}) would have a sign change when the current is instead a function of $-q$. Additionally, the LO contact $v_\mathrm{CT}$ commutes with the LO EM and W charges, so it is absent from eq. (\ref{eq:cc}). This encapsulates the relationship between the one- and two-nucleon currents: the divergence of one is fixed by the commutation of the other with the one-pion exchange potential at this order. Using $T=\sum_a T_a$, the potentials  $\mathcal{V}_{0,\mathrm{VV}}^\mathrm{2N}$ and $\mathcal{V}_{0,\mathrm{VV}}^\mathrm{3N}$ can now be expressed in terms of the Hamiltonian as

\begin{align}
    \mathcal{V}_{0,\mathrm{VV}}^\mathrm{2N}&=-\sum_{a\neq b}\frac{e}{q^4}\left(q^0\mathcal{J}_{\mathrm{EM},a}^0\mathcal{J}_{\mathrm{W,V},b}^0+[T_b-T_a,\mathcal{J}_{\mathrm{EM},a}^0\mathcal{J}_{\mathrm{W,V},b}^0]\right),\label{eq:V2N}
    \\
    \mathcal{V}_{0,\mathrm{VV}}^\mathrm{3N}&=\sum_{a\neq b\neq c}\frac{e}{q^4}\left([v_{\pi}^{ab},\mathcal{J}_{\mathrm{EM},a}^0\mathcal{J}_{\mathrm{W,V},c}^0]-[v_{\pi}^{ab},\mathcal{J}_{\mathrm{EM},c}^0\mathcal{J}_{\mathrm{W,V},a}^0]\right).\label{eq:V3N}
\end{align}

We can now use these results to show that the VV components of the two- and three-nucleon contributions cancel with each other. Let $|\phi_i\rangle$ ($\langle\phi_f|$) be an eigenstate of $T$ with energy $E_i$ ($E_f$), then when the double charge operator acts on the free state, we get
\begin{align}
\begin{split}    \mathcal{J}_{\mathrm{EM},a}^0\mathcal{J}_{\mathrm{W,V},b}^0|\phi_i\rangle&=-G_0^{(f)}[T_a-T_b,\mathcal{J}_{\mathrm{EM},a}^0\mathcal{J}_{\mathrm{W,V},b}^0]|\phi_i\rangle-2G_0^{(f)}[T_b,\mathcal{J}_{\mathrm{EM},a}^0\mathcal{J}_{\mathrm{W,V},b}^0]|\phi_i\rangle
    \\
    &+G_0^{(f)}\mathcal{J}_{\mathrm{EM},a}^0 \mathcal{J}_{\mathrm{W,V},b}^0G_0^{-1(f)}|\phi_i\rangle\label{eq:doublecharge}
\end{split}
\end{align}

\noindent where  propagators are given by $G_0^{(i)}=(E_i-T)^{-1}$ and $G_0^{(f)}=(E_f-T)^{-1}$. This can be shown by acting $G_0^{-1(f)}$ on the left side of Eq. (\ref{eq:doublecharge}) and rewriting the result in terms of the commutator of $T$ with the double charges. The last term of Eq. (\ref{eq:doublecharge}) vanishes when $E_i=E_f$. This result can be used to expand Eq. (\ref{eq:V3N}) in this limit to show the connection between the two- and three-nucleon potentials,
\begin{align}
    \begin{split}
        \langle\phi_f|\mathcal{V}_{0,\mathrm{VV}}^\mathrm{3N}|\phi_i\rangle&=\sum_{a\neq b\neq c}\frac{e}{q^4}\left[\langle\phi_f|\left(v_\pi^{ab}G_0^{(f)}\left(\mathcal{V}_{0,\mathrm{VV},ac}^\mathrm{2N}+\mathcal{V}_{0,\mathrm{VV},ca}^\mathrm{2N}\right)+\left(\mathcal{V}_{0,\mathrm{VV},ac}^\mathrm{2N}+\mathcal{V}_{0,\mathrm{VV},ca}^\mathrm{2N}\right)G_0^{(i)}v_\pi^{ab}\right)|\phi_i\rangle\right.
        \\
        &+\left.\langle\phi_f|v_\pi^{ab}G_0^{(f)}q^0(\mathcal{J}_{\mathrm{EM},a}^0\mathcal{J}_{\mathrm{W,V},c}^0+\mathcal{J}_{\mathrm{EM},c}^0\mathcal{J}_{\mathrm{W,V},a}^0)\right.
        \\
        &\left.+q^0(\mathcal{J}_{\mathrm{EM},a}^0\mathcal{J}_{\mathrm{W,V},c}^0+\mathcal{J}_{\mathrm{EM},c}^0\mathcal{J}_{\mathrm{W,V},a}^0)G_0^{(i)}v_\pi^{ab}|\phi_i\rangle\right]\label{eq:free_states}
    \end{split}
\end{align}
with $\mathcal{V}_{0,\mathrm{VV}}^\mathrm{2N}=-e\sum_{a\neq b}\frac{1}{q^4}V_{0,\mathrm{VV},ab}^\mathrm{2N}$ and $q^0\mathcal{J}^0_{\mathrm{EM},a}\mathcal{J}^0_{\mathrm{W,V},c}\equiv-[T_c,\mathcal{J}^0_{\mathrm{EM},a}\mathcal{J}^0_{\mathrm{W,V},c}]$. Therefore, the two-nucleon VV potential, when iterated with the one-pion exchange and nucleon propagator, cancels with the VV component of the three-nucleon potential. In the limit $q_0=0$, the matrix element of $\mathcal{V}_{\mathrm{0,VV}}^\mathrm{2N}$ between free states vanishes. However, for interacting systems, the contribution is nonzero:

\begin{align}
    \langle\phi_f|v_\pi^{ab}G_0^{(f)}\mathcal{V}_{\mathrm{0,VV},ac}^\mathrm{2N}|\phi_i\rangle\neq0.
\end{align}

\noindent Fig. \ref{fig:free_cancellation} represents Eq. (\ref{eq:free_states}) diagrammatically. In the next section, we will demonstrate this with explicit expressions of these potentials formed from the iteration of the two-nucleon potential with $v_\pi$.

\newcommand{\cfscale}{0.7}       
\colorlet{cffree}{white}                
\tikzset{
  cfpic/.style={line width=1.2pt, baseline={([yshift=-.5ex]0,-1)}},
  cfchg/.style={circle, draw=black, fill=white, line width=0.9pt, inner sep=0pt, minimum size=2.0mm},
}
 
\newcommand{\cfLines}{\draw (-0.45,0)--(2.55,0); \draw (-0.45,-1)--(2.55,-1); \draw (-0.45,-2)--(2.55,-2);}

\newcommand{\cfBlobs}[3]{%
  \draw[line width=0.8pt, fill=#3] (-0.45,-1) ellipse (0.4 and 1.35); \node[font=\Large] at (-0.45,-1) {$#1$};
  \draw[line width=0.8pt, fill=#3] (2.55,-1) ellipse (0.4 and 1.35); \node[font=\Large] at (2.55,-1) {$#2$};}
\newcommand{\cfDiag}[4]{%
  \begin{tikzpicture}[cfpic]
    \cfLines
    \begin{feynman} #4 \end{feynman}
    \cfBlobs{#1}{#2}{#3}
  \end{tikzpicture}}

\newcommand{\cfLep}[2]{%
  \vertex[#2] (w) at (#1,0) {};
  \vertex (nu) at (#1,0.75);
  \vertex[dot, minimum size=1.2mm] (e1) at ({#1+0.75},0.375) {};
  \vertex (e2) at ({#1+1.5},0.75);
  \diagram*{(nu) -- [fermion] (w) -- [fermion] (e1) -- [fermion] (e2)};}

\newcommand{\cfPion}[3]{%
  \vertex[dot, minimum size=1.2mm] (pa) at (#1,#2) {};
  \vertex[dot, minimum size=1.2mm] (pb) at (#1,#3) {};
  \diagram*{(pa) -- [dashed] (pb)};}

\newcommand{\cfDa}[3]{\cfDiag{#1}{#2}{#3}{%
  \cfLep{0.75}{dot, minimum size=1.2mm}
  \vertex[dot, minimum size=1.2mm] (p) at (0.75,-1) {};
  \vertex[dot, minimum size=1.2mm] (g) at (1.5,-2) {};
  \diagram*{(w) -- [dashed] (p), (e1) -- [photon] (g)};}}

\newcommand{\cfDg}[3]{\cfDiag{#1}{#2}{#3}{%
  \cfLep{0.75}{dot, minimum size=1.2mm}
  \vertex[dot, minimum size=1.2mm] (p1) at (0.75,-1) {};
  \vertex[dot, minimum size=1.2mm] (p2) at (0.75,-2) {};
  \vertex[dot, minimum size=1.2mm] (pm) at (0.75,-1.5) {};
  \diagram*{(p2) -- [dashed] (p1), (e1) -- [photon, quarter left] (pm)};}}

\newcommand{\cfDh}[3]{\cfDiag{#1}{#2}{#3}{%
  \vertex[dot, minimum size=1.2mm] (p1) at (0.75,0) {};
  \vertex[dot, minimum size=1.2mm] (p2) at (0.75,-1) {};
  \vertex[dot, minimum size=1.2mm] (wv) at (0.75,-0.5) {};
  \vertex (nu) at (1.4,-0.15);
  \vertex[dot, minimum size=1.2mm] (e1) at (1.5,-0.5) {};
  \vertex (e2) at (2.1,-0.5);
  \vertex[dot, minimum size=1.2mm] (g) at (1.5,-2) {};
  \diagram*{(p1) -- [dashed] (p2), (nu) -- [fermion] (wv) -- [fermion] (e1) -- [fermion] (e2),
            (e1) -- [photon] (g)};}}

\newcommand{\cfDj}[3]{\cfDiag{#1}{#2}{#3}{%
  \cfLep{0.75}{dot, minimum size=1.2mm}
  \vertex[dot, minimum size=1.2mm] (g) at (1.5,-1) {};
  \vertex[dot, minimum size=1.2mm] (p2) at (1.5,-2) {};
  \diagram*{(e1) -- [photon] (g), (g) -- [dashed] (p2)};}}
 
\newcommand{\cfRv}[5]{\cfDiag{#1}{#2}{#3}{%
  \cfLep{0.35}{#4}
  \vertex[#5] (g) at (1.1,-2) {};
  \diagram*{(e1) -- [photon] (g)};
  \cfPion{1.9}{0}{-1}}}

\newcommand{\cfRw}[5]{\cfDiag{#1}{#2}{#3}{%
  \cfLep{0.35}{#4}
  \vertex[#5] (g) at (1.1,-2) {};
  \diagram*{(e1) -- [photon] (g)};
  \cfPion{1.9}{-1}{-2}}}

\begin{figure}[t]
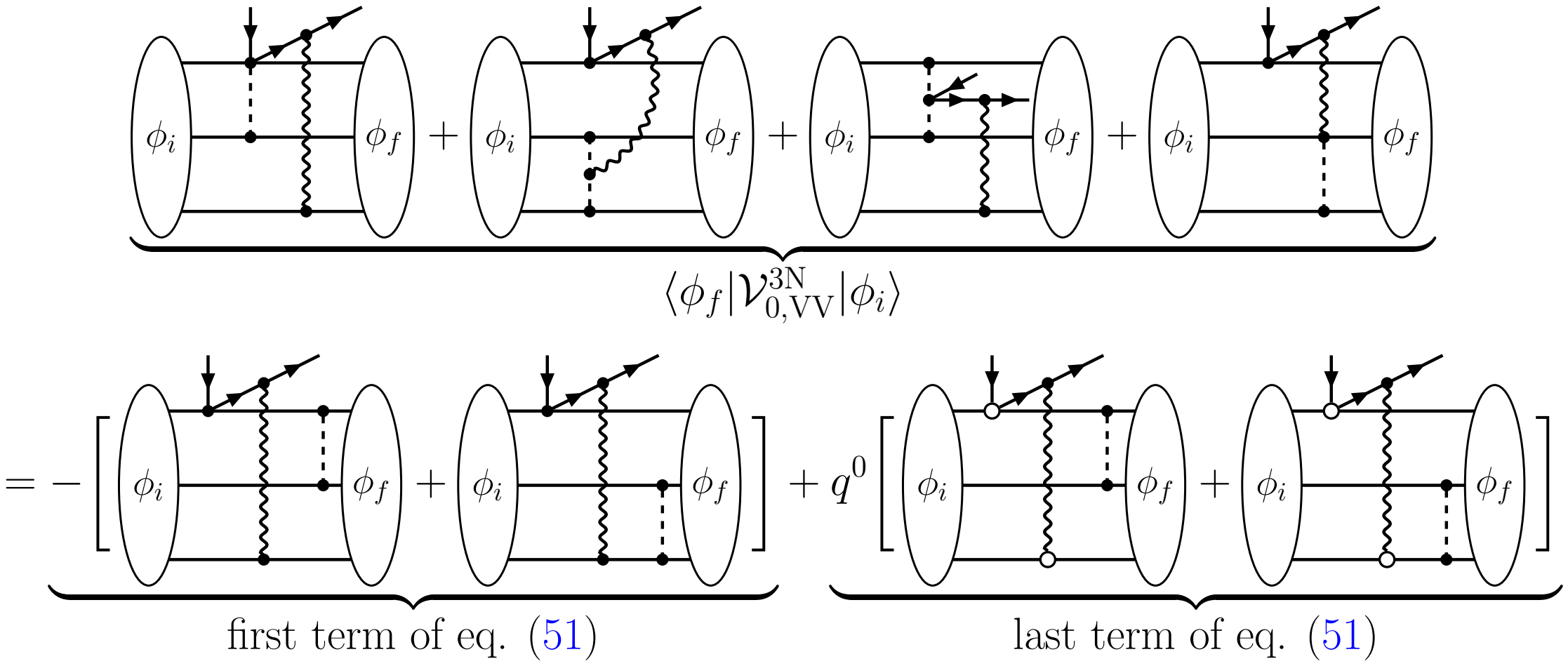

\centering
\scalebox{\cfscale}{\LARGE$\displaystyle
\underbrace{\cfDa{\phi_i}{\phi_f}{cffree}
+\cfDg{\phi_i}{\phi_f}{cffree}
+\cfDh{\phi_i}{\phi_f}{cffree}
+\cfDj{\phi_i}{\phi_f}{cffree}}_{\textstyle\langle\phi_f|\mathcal{V}^\mathrm{3N}_\mathrm{0,VV}|\phi_i\rangle}$}\\[6pt]
\scalebox{\cfscale}{\LARGE$\displaystyle
={\underbrace{-\Bigg[\,
\cfRv{\phi_i}{\phi_f}{cffree}{dot, minimum size=1.2mm}{dot, minimum size=1.2mm}
+\cfRw{\phi_i}{\phi_f}{cffree}{dot, minimum size=1.2mm}{dot, minimum size=1.2mm}
\,\Bigg]}_{\textstyle\text{first term of eq. (\ref{eq:free_states})}}}
+{\underbrace{q^0\Bigg[\,
\cfRv{\phi_i}{\phi_f}{cffree}{cfchg}{cfchg}
+\cfRw{\phi_i}{\phi_f}{cffree}{cfchg}{cfchg}
\,\Bigg]}_{\textstyle\text{last term of eq. (\ref{eq:free_states})}}}$}
\captionsetup{labelformat=default}
\caption{A diagrammatic representation of eq. (\ref{eq:free_states}) showing the cancellation of the matrix element using free states of the three-nucleon potential with the two-nucleon potential and the double charge operator iterated with $v_\pi$. Open vertices represent the charge. Only one iteration of each diagram is shown as a representative with all iterations implied in the cancellation.}
\label{fig:free_cancellation}
\end{figure}

This argument can be generalized beyond free states to bound states constructed from a Lippmann-Schwinger iteration with the LO potential,

\begin{align}
    |\Psi_i\rangle&=G_0^{(i)}(v_\pi+v_\mathrm{CT})|\Psi_i\rangle.\label{eq:BLS}
\end{align}

The bound-state scenario complicates the derivation. In Eq. (\ref{eq:BLS}), the LO potential is summed over all pairs in the nucleus. Therefore, a useful relation is 

\begin{align}
    \sum_{a\neq b\neq c}[v_\pi^{ab},\mathcal{J}_{\mathrm{EM},a}^0]\mathcal{J}_{\mathrm{W,V},c}^0&=\sum_{a\neq c}\left([v_\pi+v_\mathrm{CT},\mathcal{J}_{\mathrm{EM},a}^0]-[v_\pi^{ac},\mathcal{J}_{\mathrm{EM},a}^0]\right)\mathcal{J}_{\mathrm{W,V},c}^0
\end{align}

\noindent  The matrix element of the three-nucleon potential between bound states is then

\begin{align}
\begin{split}
\langle\Psi_f|\mathcal{V}_{0,\mathrm{VV}}^\mathrm{3N}|\Psi_i\rangle&=\langle\Psi_f|\sum_{a\neq b\neq c}\frac{e}{q^4}\left([v_\pi^{ab},\mathcal{J}_{\mathrm{EM},a}^0\mathcal{J}_{\mathrm{W,V},c}^0]-[v_\pi^{ab},\mathcal{J}_{\mathrm{EM},c}^0\mathcal{J}_{\mathrm{W,V},a}^0]\right)|\Psi_i\rangle      
\end{split}
\\
\begin{split}
&=\langle\Psi_f|\sum_{a\neq c}\frac{e}{q^4}\left([v_\pi+v_\mathrm{CT},\mathcal{J}_{\mathrm{EM},a}^0]-[v_\pi^{ac},\mathcal{J}_{\mathrm{EM},a}^0]\right)\mathcal{J}_{\mathrm{W,V},c}^0|\Psi_i\rangle
\\
&-\langle\Psi_f|\sum_{a\neq c}\frac{e}{q^4}\mathcal{J}_{\mathrm{EM},c}^0\left([v_\pi+v_\mathrm{CT},\mathcal{J}_{\mathrm{W,V},a}^0]-[v_\pi^{ac},\mathcal{J}_{\mathrm{W,V},a}^0]\right)|\Psi_i\rangle\label{eq:half_final}
\end{split}
\\
\begin{split}
&=-\langle\Psi_f|\left(\mathcal{V}_{0,\mathrm{VV}}^\mathrm{2N}-q^0\rho_\mathrm{VV}^\mathrm{2N}\right)|\Psi_i\rangle
\\
&+\langle\Psi_f|\sum_{a\neq c}\frac{e}{q^4}\left(\mathcal{J}_{\mathrm{EM},c}^0[v_\pi^{ac},\mathcal{J}_{\mathrm{W,V},a}^0]-[v_\pi^{ac},\mathcal{J}_{\mathrm{EM},a}^0]\mathcal{J}_{\mathrm{W,V},c}^0\right)|\Psi_i\rangle
\label{eq:full_final}
\end{split}
\end{align}

\noindent where we have defined a double charge operator $\rho^\mathrm{2N}_\mathrm{VV}$ as
\begin{align}
    \rho^\mathrm{2N}_\mathrm{VV}\equiv e\sum_{a\neq b}\frac{1}{q^4}\mathcal{J}_{\mathrm{EM},a}^0\,\mathcal{J}_{\mathrm{W,V},b}^0.
\end{align}

\colorlet{cffull}{gray!25}

\newcommand{\cfXsize}{0.08}
\newcommand{\cfX}[2]{%
  \draw[red, line width=1pt]
    (#1,#2) +(-\cfXsize,-\cfXsize) -- +(\cfXsize,\cfXsize)
            +(-\cfXsize,\cfXsize)  -- +(\cfXsize,-\cfXsize);}
 
\newcommand{\cfLepLong}[3]{%
  \vertex[#2] (w) at (#1,0) {};
  \vertex (nu) at (#1,0.75);
  \vertex[dot, minimum size=1.2mm] (e1) at (#3,{0.5*(#3-#1)}) {};
  \vertex (e2) at (2.2,{0.5*(2.2-#1)});
  \diagram*{(nu) -- [fermion] (w) -- [fermion] (e1) -- [fermion] (e2)};}
 
\newcommand{\cfGW}[5]{\cfDiag{#1}{#2}{#3}{%
  \cfLep{0.75}{#4}
  \vertex[#5] (g) at (1.5,-1) {};
  \diagram*{(e1) -- [photon] (g)};}}
 
\newcommand{\cfLwpg}[3]{\cfDiag{#1}{#2}{#3}{%
  \cfLepLong{0.3}{cfchg}{1.55}
  \cfPion{0.95}{0}{-1}
  \vertex[cfchg] (g) at (1.55,-1) {};
  \diagram*{(e1) -- [photon] (g)};
  \cfX{0.625}{0}
  \cfX{1.25}{-1}}}
 
\newcommand{\cfLpwg}[3]{\cfDiag{#1}{#2}{#3}{%
  \cfPion{0.3}{0}{-1}
  \cfLepLong{0.75}{cfchg}{1.65}
  \vertex[cfchg] (g) at (1.65,-1) {};
  \diagram*{(e1) -- [photon] (g)};
  \cfX{0.525}{0}
  \cfX{0.975}{-1}}}
 
\newcommand{\cfLwgp}[3]{\cfDiag{#1}{#2}{#3}{%
  \cfLepLong{0.3}{cfchg}{1.0}
  \vertex[cfchg] (g) at (1.0,-1) {};
  \diagram*{(e1) -- [photon] (g)};
  \cfPion{1.7}{0}{-1}
  \cfX{1.35}{0}
  \cfX{1.35}{-1}}}

\begin{figure}[t]
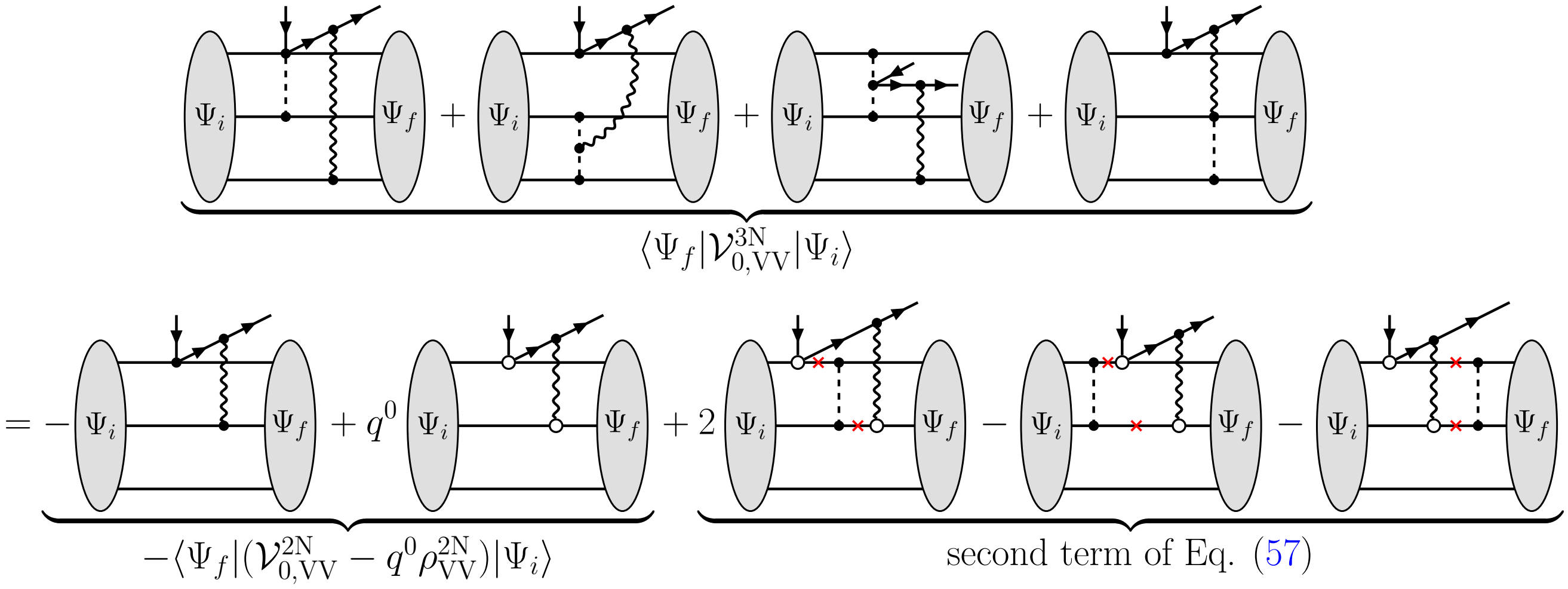

\centering
\scalebox{\cfscale}{\LARGE$\displaystyle
\underbrace{\cfDa{\Psi_i}{\Psi_f}{cffull}
+\cfDg{\Psi_i}{\Psi_f}{cffull}
+\cfDh{\Psi_i}{\Psi_f}{cffull}
+\cfDj{\Psi_i}{\Psi_f}{cffull}}_{\textstyle\langle\Psi_f|\mathcal{V}^\mathrm{3N}_\mathrm{0,VV}|\Psi_i\rangle}$}\\[6pt]
\scalebox{\cfscale}{\LARGE$\displaystyle
={\underbrace{-\cfGW{\Psi_i}{\Psi_f}{cffull}{dot, minimum size=1.2mm}{dot, minimum size=1.2mm}
+q^0\,\cfGW{\Psi_i}{\Psi_f}{cffull}{cfchg}{cfchg}}_{\textstyle -\langle\Psi_f|(\mathcal{V}^\mathrm{2N}_{0,\mathrm{VV}}-q^0\rho^\mathrm{2N}_\mathrm{VV})|\Psi_i\rangle}}
+{\underbrace{2\,\cfLwpg{\Psi_i}{\Psi_f}{cffull}
-\cfLpwg{\Psi_i}{\Psi_f}{cffull}
-\cfLwgp{\Psi_i}{\Psi_f}{cffull}}_{\textstyle\text{second term of Eq. }(\ref{eq:full_final})}}$}
\setcounter{figure}{2} 
\captionsetup{labelformat=default}
\caption{A diagrammatic representation of Eq. (\ref{eq:full_final}), relating the three-nucleon potential evaluated between bound states to two-nucleon contributions. The last three diagrams appear to have intermediate nucleon propagators; however, these diagrams should be understood as products of the charge operators with the one-pion-exchange potential, as written in Eq. (\ref{eq:full_final}), and meant as a qualitative picture only. We place a red $\times$ over the apparent nucleon propagators to emphasize this interpretation. As before, only one iteration of each diagram is shown.}
\label{fig:full_cancellation}
\end{figure}

In going from Eq. (\ref{eq:half_final}) to Eq. (\ref{eq:full_final}), we have replaced $v_\pi+v_\mathrm{CT}=H-T$ and identified  $q^0\mathcal{J}_{\mathrm{EM},a}^0\mathcal{J}_{\mathrm{W,V},c}^0\equiv-\mathcal{J}_{\mathrm{EM},a}^0[H,\mathcal{J}_{\mathrm{W,V},c}^0]$ in analogy to the free state but now with the Hamiltonian instead of $T$. When acting on a bound state ($-\mathcal{J}_{\mathrm{EM},a}^0[H,\mathcal{J}_{\mathrm{W,V},c}^0]|\Psi_i\rangle=\mathcal{J}_{\mathrm{EM},a}^0(E_i-H)\mathcal{J}_{\mathrm{W,V},c}^0|\Psi_i\rangle$), the weak charge $\mathcal{J}_{\mathrm{W,V},c}^0(-q)$ transfers momentum $-q$ and changes the charge of nucleon $c$. Then $E_i-H$ acts on the resulting state, measuring the energy removed by the weak charge, which is the photon energy.

Eq. (\ref{eq:full_final}) is qualitatively the same as the free state one: the matrix element of the three-nucleon potential cancels with the matrix element of the two-nucleon potential when iterated with $v_\pi$. The difference is the additional contribution of the second term of Eq. (\ref{eq:full_final}) formed from $v_\pi$ acting on the same nucleon pair as the electromagnetic and weak currents.
This is a two-nucleon contribution that needs to be addressed by deriving  the two-body $\Bar{\delta}_{\text{NS}}$ operators at $\mathcal{O}(\alpha \epsilon_\chi^2)$.
For both the free and bound state scenarios, the three-nucleon VV potentials should not be included in the matrix element when the corresponding two-nucleon potentials are not included to ensure consistency of the calculation. We have demonstrated this argument under a number of approximations including the outgoing leptons carrying negligible energy ($E_i=E_f$), isospin symmetry, and working at LO in the currents and potential. The cancellations shown in this section will not hold when these assumptions are relaxed but remain a subleading correction and motivate excluding VV contributions to $\Bar{\delta}_\mathrm{NS}$ in the many-body calculation at this order.

\subsubsection{Diagrammatic realization}

Consider the diagrams in Fig. \ref{fig:reducible} and their permutations, which are constructed from the LO two-nucleon contribution to $\Bar{\delta}_\mathrm{NS}$ and the one-pion-exchange potential. If both the energy-dependent and recoil terms of Eq. (\ref{eq:divV}) are kept, these
terms cancel the intermediate nucleon propagator, implying that these seemingly-reducible topologies leave behind an irreducible component. The potentials for these diagrams are given in Eqs. (\ref{eq:v}) and (\ref{eq:w}). 

\begin{align}
    \mathcal{V}_{0}^{(v)}=-\mathcal{V}_0^{(a)}-\mathcal{V}_0^{(h)}&=2\pi\frac{g_A^2}{F_\pi^2}\sum_{i\neq j\neq k}\frac{1}{\boldsymbol{q}_k^4}\frac{\boldsymbol{\sigma}_j\cdot\boldsymbol{q}_j}{\boldsymbol{q}_j^2+m_\pi^2}\boldsymbol{\sigma}_i\cdot\boldsymbol{q}_jP_p^{(k)}(\tau_i^+\tau_j^3-\tau_i^3\tau_j^+)\label{eq:v}
    \\
    \mathcal{V}_{0}^{(w)}=-\mathcal{V}_{0,\mathrm{VV}}^{(g)}-\mathcal{V}_{0,\mathrm{VV}}^{(j)}&=2\pi\frac{g_A^2}{F_\pi^2}\sum_{i\neq j\neq k}\frac{1}{\boldsymbol{q}_i^4}\frac{\boldsymbol{\sigma}_k\cdot\boldsymbol{q}_k}{\boldsymbol{q}_k^2+m_\pi^2}\boldsymbol{\sigma}_j\cdot\boldsymbol{q}_k\tau_i^+(\tau_j^+\tau_k^--\tau_j^-\tau_k^+)\label{eq:w}
\end{align}

\begin{figure}[t]
\begin{center}
\scalebox{.75}{
   \begin{minipage}[c]{3cm}
\begin{tikzpicture}[line width=1.2 pt]
    \begin{feynman}
        \vertex (a);
        \vertex[right=.75 of a, dot, minimum size=1.2mm] (b) {};
        \vertex[above=.75 of b] (bu);
        \vertex[right=1.5 of b] (c);
        \vertex[right=.75 of b] (d);
        \vertex[above=.375 of d,dot, minimum size=1.2mm] (d1) {};
        \vertex[above=.75 of c] (d2) {};
        \vertex[below=1 of a] (aa);
        \vertex[below=1 of aa] (aaa);
        \vertex[below=1 of b] (bb) {};
        \vertex[below=1 of bb] (bbb);
        \vertex[below=1 of c] (cc);
        \vertex[below=1 of cc] (ccc);

        \vertex[left=.5 of b] (bl);
        \vertex[left=.5 of bb] (bbl);

        \vertex[below=2.435 of d1, dot, minimum size=1.2mm] (bphoton) {};
        
        \vertex[right=1.25 of b, dot, minimum size=1.2mm] (pion1) {};
        \vertex[below=1 of pion1, dot, minimum size=1.2mm] (pion2) {};

        \diagram*{

        (a) -- [] (c);
        (bu) -- [fermion] (b);
        (b) -- [fermion] (d1);
        (d1) -- [fermion] (d2);
        (aa) -- [] (cc);
        (aaa) -- [] (ccc);
        (pion1) -- [dashed] (pion2);
        (d1) -- [photon] (bphoton);
        (bl) -- [] (b);
        };
        \vertex[left=.5 of b, dot,minimum size=1.2mm] (bl) ;
        \vertex[right=0 of b,dot,minimum size=1.2mm] (bbl) ;

    \end{feynman}
\end{tikzpicture}
\caption{(v)}
\end{minipage}}
\hspace{.5cm}
\scalebox{.75}{
   \begin{minipage}[c]{3cm}
\begin{tikzpicture}[line width=1.2 pt]
    \begin{feynman}
        \vertex (a);
        \vertex[right=.75 of a, dot, minimum size=1.2mm] (b) {};
        \vertex[above=.75 of b] (bu);
        \vertex[right=1.5 of b] (c);
        \vertex[right=.75 of b] (d);
        \vertex[above=.375 of d,dot, minimum size=1.2mm] (d1) {};
        \vertex[above=.75 of c] (d2) {};
        \vertex[below=1 of a] (aa);
        \vertex[below=1 of aa] (aaa);
        \vertex[below=1 of b] (bb) {};
        \vertex[below=1 of bb] (bbb);
        \vertex[below=1 of c] (cc);
        \vertex[below=1 of cc] (ccc);

        \vertex[left=.5 of b] (bl);
        \vertex[left=.5 of bb] (bbl);

        \vertex[below=2.435 of d1, dot, minimum size=1.2mm] (bphoton) {};
        
        \vertex[right=1.25 of b] (pion1) {};
        \vertex[below=1 of pion1, dot, minimum size=1.2mm] (pion2) {};
        \vertex[below=1 of pion2, dot, minimum size=1.2mm] (pion3) {};

        \diagram*{

        (a) -- [] (c);
        (bu) -- [fermion] (b);
        (b) -- [fermion] (d1);
        (d1) -- [fermion] (d2);
        (aa) -- [] (cc);
        (aaa) -- [] (ccc);
        (pion2) -- [dashed] (pion3);
        (d1) -- [photon] (bphoton);
        (bl) -- [] (b);
        };
        \vertex[left=.5 of b, dot,minimum size=1.2mm] (bl) ;
        \vertex[right=0 of b,dot,minimum size=1.2mm] (bbl) ;

    \end{feynman}
\end{tikzpicture}
\caption{(w)}
\end{minipage}}
\hspace{.5cm}
\scalebox{.75}{
   \begin{minipage}[c]{3cm}
   \vspace{7pt}
\begin{tikzpicture}[line width=1.2 pt]
    \begin{feynman}
        \vertex (a);
        \vertex[right=.75 of a, dot, minimum size=1.2mm] (b);
        \vertex[above=.75 of b] (bu);
        \vertex[right=1.5 of b] (c);
        \vertex[right=.75 of b] (d);
        \vertex[above=.75 of c] (d2);
        \vertex[below=1 of a] (aa);
        \vertex[below=1 of aa] (aaa);
        \vertex[below=1 of b, dot, minimum size=1.2mm] (bb);
        \vertex[below=1 of bb] (bbb);
        \vertex[below=1 of c] (cc);
        \vertex[below=1 of cc] (ccc);
        \vertex[left=1 of ccc, dot, minimum size=1.2mm] (c4) {};
        \vertex[right=.75 of bb, dot, minimum size=1.2mm] (b4) {};
        \vertex[above=.75 of b4] (b5);
        \vertex[above=.5 of cc] (c5);
        \vertex[above=1 of b4, dot, minimum size=1.2mm] (b6) {};
        \vertex[below=.5 of b6, dot, minimum size=1.2mm] (hb) {};
        \vertex[left=.75 of b4, dot, minimum size=1.2mm] (b7);
        \vertex[above=.94 of b7, dot, minimum size=1.2mm] (b7) {};
        \vertex[above=.75 of b7, dot, minimum size=1.2mm] (b8);
        \vertex[above=.5 of b6] (b9);

        \diagram*{

        (a) -- [] (c);
        (aa) -- [] (cc);
        (aaa) -- [] (ccc);
        (bl) -- [] (b);
        (b4) -- [dashed] (b6);
        (hb) -- [photon, quarter right] (c4);
        (b8) -- [fermion] (b7);
        (b7) -- [fermion] (b9);
        };
        
    \end{feynman}
\end{tikzpicture}
\caption{(x)}
\end{minipage}}
\end{center}
\setcounter{figure}{3} 
\captionsetup{labelformat=default}
\caption{Iterated topologies built from the leading component of $\mathcal{V}_0^\mathrm{2N}$ and $v_\pi$ or $\mathcal{J}_\mathrm{W}^0$ and $v_\gamma^\mathrm{3N}$.}\label{fig:reducible}
\end{figure}

These contributions exactly cancel the VV components of potentials $\mathcal{V}_0^{(a)}$, $\mathcal{V}_0^{(g)}$, $\mathcal{V}_0^{(h)}$, and $\mathcal{V}_0^{(j)}$. These potentials are of a similar structure as the $\gamma W$-box diagram and the LO two-nucleon EFT contribution in which the photon is attached to the electron, generating an electron propagator, and the leptonic current is contracted via the $W^\pm$ (or directly) to the hadronic state. The remaining VV contribution, $\mathcal{V}_0^{(i)}$, is a fundamentally different contribution. The $2\pi e^-\Bar{\nu}_e\gamma$ vertex featured in diagram $(i)$ is a contact interaction with the electromagnetic and weak currents acting at the same point rather than the photon attaching to the electron. This is closer to a radiative correction to the $\Psi W$ vertex rather than a true $\Bar{\delta}_{\mathrm{NS}}$ contribution, reminiscent of the trace term of the DR. Other diagrams at this order, although they are suppressed due to kinematics or other factors, are also of this type, and provide a mixing between the traditional $\delta_\mathrm{NS}$ and $\delta_\mathrm{C}$ contributions. The reducible contribution shown in diagram ($x$) is an iteration of the single-nucleon weak current with a three-nucleon ISB potential. When including the recoil correction on the $\gamma \Bar{N}N$ vertex, the potential, whose expression is given in Eq. (\ref{eq:x}), has both irreducible (no nucleon propagator) and reducible terms. The irreducible content of $\mathcal{V}_0^{(x)}$ cancels with $\mathcal{V}_0^{(i)}$.
\begin{align}
    \mathcal{V}^{(x)}_{0}&=-2\pi\frac{g_A^2}{F_\pi^2}\sum_{i\neq j\neq k}\frac{1}{\boldsymbol{q}_i^2}\frac{\boldsymbol{\sigma}_j\cdot\boldsymbol{q}_j}{\boldsymbol{q}_j^2+m_\pi^2}\frac{\boldsymbol{\sigma}_k\cdot\boldsymbol{q}_k}{\boldsymbol{q}_k^2+m_\pi^2}P_p^{(i)}\tau_j^+\tau_k^3+\mathrm{(reducible\;terms)}\label{eq:x}
\end{align}

\subsubsection{Summary of the role of VV contributions at order $\mathcal{O}(\alpha)$}

To conclude this section, we summarize the argument so far. We have now demonstrated that although VV contributions appear in the potentials of various three-nucleon $\Bar{\delta}_\mathrm{NS}$ contributions, they do not have an impact on the EWRC to superallowed Fermi decays in the EFT approach at order $\mathcal{O}(\alpha)$, as is the case in the DR. In particular, the cancellation of the VV components arises from the many-body version of CVC. 

For a many-body calculation, there are two practical consequences:

\begin{itemize}
    \item The VV components of the three-nucleon potentials of this work should only be included in a many-body calculation when the corresponding 2N potential is also included. Otherwise, their inclusion would introduce a spurious VV component. In Refs. \cite{King_2025,Cirigliano_2024A,Cirigliano_2024B} the LO two-nucleon potential $\mathcal{V}_0^\mathrm{2N}\propto q^0$ is neglected from matrix element calculations, and so we compute only the VA 3N contributions in this work.
    \item Similarly, $\mathcal{V}_0^{(i)}$ should only be included if the ISB potential $v_\gamma^\mathrm{3N}$ is also included in the calculation of $\delta_\mathrm{C}$. As shown in $\mathcal{V}_0^{(x)}$, there are additional (reducible) terms that do not cancel with the three-nucleon contributions of this work. This may be unsurprising since the conservation of the vector current holds only in the isospin-symmetric limit, so introducing ISB effects will no longer protect the decay from VV contributions. Since $v_\gamma^\mathrm{3N}$ was not included in the calculation of $\Bar{\delta}_\mathrm{C}$ in Ref. \cite{Piarulli_2026}, we exclude $\mathcal{V}_0^{(i)}$ from our matrix element calculations.
\end{itemize}

Having only operators in the VA sector survive, the three-nucleon potential that enters the calculation of $\Bar{\delta}_\mathrm{NS}$ is given by
\begin{align}
    \mathcal{V}_{0,\mathrm{VA}}^\mathrm{3N}&=\mathcal{V}_{0,\mathrm{VA}}^{(g)}+\mathcal{V}_{0,\mathrm{VA}}^{(j)}
   \nonumber \\
    &=-2\pi i\frac{g_A^3}{F_\pi^2}\sum_{i\neq j\neq k}\frac{1}{\boldsymbol{q}_i^4}\frac{\boldsymbol{\sigma}_k\cdot\boldsymbol{q}_k}{\boldsymbol{q}_k^2+m_\pi^2}\left[\boldsymbol{\sigma}_i\cdot\boldsymbol{q}_k\times\boldsymbol{q}_i\frac{\boldsymbol{\sigma}_j\cdot\boldsymbol{q}_j}{\boldsymbol{q}_j^2+m_\pi^2}+\boldsymbol{\sigma}_i\cdot\boldsymbol{\sigma}_j\times\boldsymbol{q}_i\right]\tau_i^\pm(\tau_j^+\tau_k^--\tau_j^-\tau_k^+),\label{eq:3summary}
\end{align}
and its contribution to $\bar \delta_{\text{NS}}$ is
\begin{equation}
    M_{3\text{N}} \equiv \langle f | \mathcal V^{3\text{N}}_{0, \text{VA}} | i \rangle.  \label{eq:3summary2}
\end{equation}

\section{Matrix Elements in Light Nuclei}\label{sec:ME}
\subsection{Quantum Monte Carlo Methods}

Matrix elements of $\beta$ decay transition operators can be evaluated using many-body wave functions describing the initial and final nuclei. We use the Norfolk class of two- and three-nucleon (NV2+3) potentials and quantum Monte Carlo (QMC) methods to construct these wave functions and then use them to perform matrix element calculations. We provide an overview of the principle features of these methods, but for a more detailed description, see Refs. \cite{Carlson_2015,Baroni_2016,Baroni_2018,Piarulli_2015,Piarulli_2016,Piarulli_2018,Piarulli_2020}. The first step in this approach is Variational Monte Carlo (VMC), in which a wave function $\Psi_T$ is optimized via a minimization of the expectation value of 

\begin{align}
    E_T=\frac{\langle \Psi_T|H|\Psi_T\rangle}{\langle \Psi_T|\Psi_T\rangle}\geq E_0, 
\end{align}

\noindent where $E_0$ is the true ground state energy and the Hamiltonian $H$ contains the single nucleon nonrelativistic kinetic energies and two- and three-nucleon Norfolk (NV2+3) potentials. We take as a variational ansatz

\begin{align} |\Psi_T\rangle=\mathcal{S}\prod_{i<j}\left[1+U_{ij}+\sum_{i<j\neq k}\tilde{U}_{ijk}^{\mathrm{TNI}}\right]|\Psi_J\rangle,
\end{align}

\noindent where $\mathcal{S}$ is the symmetrization operator, $\Psi_J$ is a Jastrow-like antisymmetric wave function containing the long-ranged cluster structure of the nucleus, and $U_{ij}$ and $\tilde{U}_{ijk}^\mathrm{TNI}$ are the two- and three-nucleon correlations operators, respectively, which contain the variational parameters to carry out the VMC optimization. 

The next step, in order to refine the VMC result, is Green's Function Monte Carlo (GFMC). The central idea of GFMC is that the real time ($t$) Schr\"odinger equation can be transformed into a diffusion equation via a rotation to imaginary time $\tau=it$. By expanding the variational wave function in terms of its exact eigenstates of the Hamiltonian, $|\Psi_T\rangle=\sum_i c_i|\psi_i\rangle$, the true ground state can by reached by an imaginary time propagation,

\begin{align}
    \lim_{\tau\rightarrow\infty}e^{-(H-E_0)\tau}|\Psi_T\rangle \propto c_0\psi_0.
\end{align}

\noindent This propagation is computed in $n$ small imaginary time steps $\Delta\tau$,

\begin{align}
    |\Psi(\tau)\rangle=\left(e^{-(H-E_0)\Delta\tau}\right)^n|\Psi_T\rangle,
\end{align}

\noindent until convergence is reached. Off-diagonal matrix elements, such as those of this work for $\beta$ decays, are computed with mixed estimates. Assuming $\Psi(\tau)=\Psi_T+\delta\Psi$, with small correction $\delta\Psi$, the matrix element becomes

\begin{align}
    \langle \mathcal{O}(\tau)\rangle=2\frac{\langle\Psi(\tau)|\mathcal{O}|\Psi_V\rangle}{\langle\Psi(\tau)|\Psi_V\rangle}-\frac{\langle\Psi_V|\mathcal{O}|\Psi_V\rangle}{\langle\Psi_V|\Psi_V\rangle}.
\end{align}

In the case of $\beta$-decays, the transition mixed estimate matrix element becomes

\begin{align}
    \langle\mathcal{O}(\tau)\rangle_{i\rightarrow f}\approx\sqrt{\frac{\langle \Psi_f|\Psi_f\rangle}{\langle \Psi_i|\Psi_i\rangle}}\frac{\langle\Psi_f(\tau)|\mathcal{O}|\Psi_i\rangle}{\langle \Psi_f(\tau)|\Psi_f\rangle}+\sqrt{\frac{\langle \Psi_i|\Psi_i\rangle}{\langle \Psi_f|\Psi_f\rangle}}\frac{\langle\Psi_i(\tau)|\mathcal{O}^\dagger|\Psi_f\rangle}{\langle\Psi_i(\tau)|\Psi_i\rangle}-\frac{\langle\Psi_f|\mathcal{O}|\Psi_i\rangle}{\sqrt{\langle\Psi_f|\Psi_f\rangle}\sqrt{\langle\Psi_i|\Psi_i\rangle}}.
\end{align}

The Norfolk two-nucleon (NV2) potentials include long- and
intermediate-range pion-exchange contributions through N$^2$LO and short-range contact interactions through N$^3$LO. The contact sector is parameterized by 26 LECs fitted to $NN$ scattering data from the Granada
database \cite{Navarro_2013,Navarro_2014A,Navarro_2014B}
and the deuteron binding energy.
The contact interactions are regularized by replacing the
delta-functions with a normalized Gaussian,

\begin{align}
    \delta(r)&=\frac{e^{-(r/R_S)^2}}{\pi^{3/2}R_S^3},
\end{align}

\noindent with a short-ranged cutoff $R_S$. A long-ranged regulator,

\begin{align}
    F(r)&=1-\frac{1}{(r/R_L)^6 e^{2(r-R_L)/R_L}+1},
\end{align}

\noindent with cutoff $R_L$ is required to regulate divergences appearing in pion-exchange potentials which behave like $1/r^n$ with $n\leq 6$ for small internucleon distances. The NV2 model used in these calculations, denoted with ``I" and ``a", use the same fitting of $NN$ scattering data up to $125$ MeV and regulator parameters $[R_L,R_S]=[1.2\;\mathrm{fm},0.8\;\mathrm{fm}]$. The difference in the models we use is in the fitting of the three-nucleon forces and values of their LECs, denoted $c_D$ and $c_E$, which are given in Table \ref{tab:table_cD}. Model Ia (Ia$^\star$) is fit to triton ground state energy and $nd$ scattering (Gamow-Teller $\beta$ decay matrix element) \cite{Piarulli_2018,Baroni_2018}. We focus on these models because there are already GFMC calculations with these interactions of $\Bar{\delta}_\mathrm{C}$ and two-nucleon $\Bar{\delta}_\mathrm{NS}$ matrix elements, which can be combined for a determination of $V_{ud}$ \cite{King_2025,Piarulli_2026}.

\begin{table}[H]
    \centering
    \begin{tabular}{c | c c c c c}
    \hline
    \hline
      & Ia & Ia$^\star$\\
     \hline
     $c_D$ & 3.666 & -0.635 \\
     $c_E$& -1.638 & -0.090 \\
     \hline
     \hline
    \end{tabular}
    \caption{Adimensional values of LECs $c_D$ and $c_E$ given in the NV3 interactions used in this work.}
    \label{tab:table_cD}
\end{table}

\subsubsection{Transition densities}

To study the behavior of matrix elements in greater detail, we compute their densities in coordinate-space. We define the density $C(r,R)$, given in Eq. (\ref{eq:density1}), for single $\beta$ decay in terms of the distance from the decaying nucleon to the center of mass of the two spectator nucleons as $r$ and the distance between the spectator nucleons as $R$. Explicitly, letting nucleon $i$ be the nucleon which undergoes the decay, we define $r$ and $R$ as 

\begin{align}
    r=\frac{1}{2}\sqrt{2r_{ij}^2+2r_{ik}^2-r_{jk}^2}, \quad \quad R=r_{jk},
\end{align}

\noindent which implicitly depend on indices $ijk$ but are omitted. 
The matrix element between initial and final nuclear states can be recovered from the density by integrating over these variables, as shown in Eq. (\ref{eq:density2}).

\begin{align}
    C(r,R)&=\langle \Psi_f|\sum_{i\neq j\neq k}\mathcal{O}_{ijk}\delta(r-r^\prime)\delta(R-R^\prime)|\Psi_i\rangle\label{eq:density1}
    \\
    M&=\langle\Psi_f|\tilde{\mathcal{V}}|\Psi_i\rangle=\int_0^\infty \mathrm{d}r\; C(r)=\int_0^\infty \mathrm{d}r\int_0^\infty \mathrm{d}R \;C(r,R)\label{eq:density2}
\end{align}

\noindent with coordinate space potential $\tilde{\mathcal{V}}$.

\begin{figure}
    \centering
    \includegraphics[width=.7\linewidth]{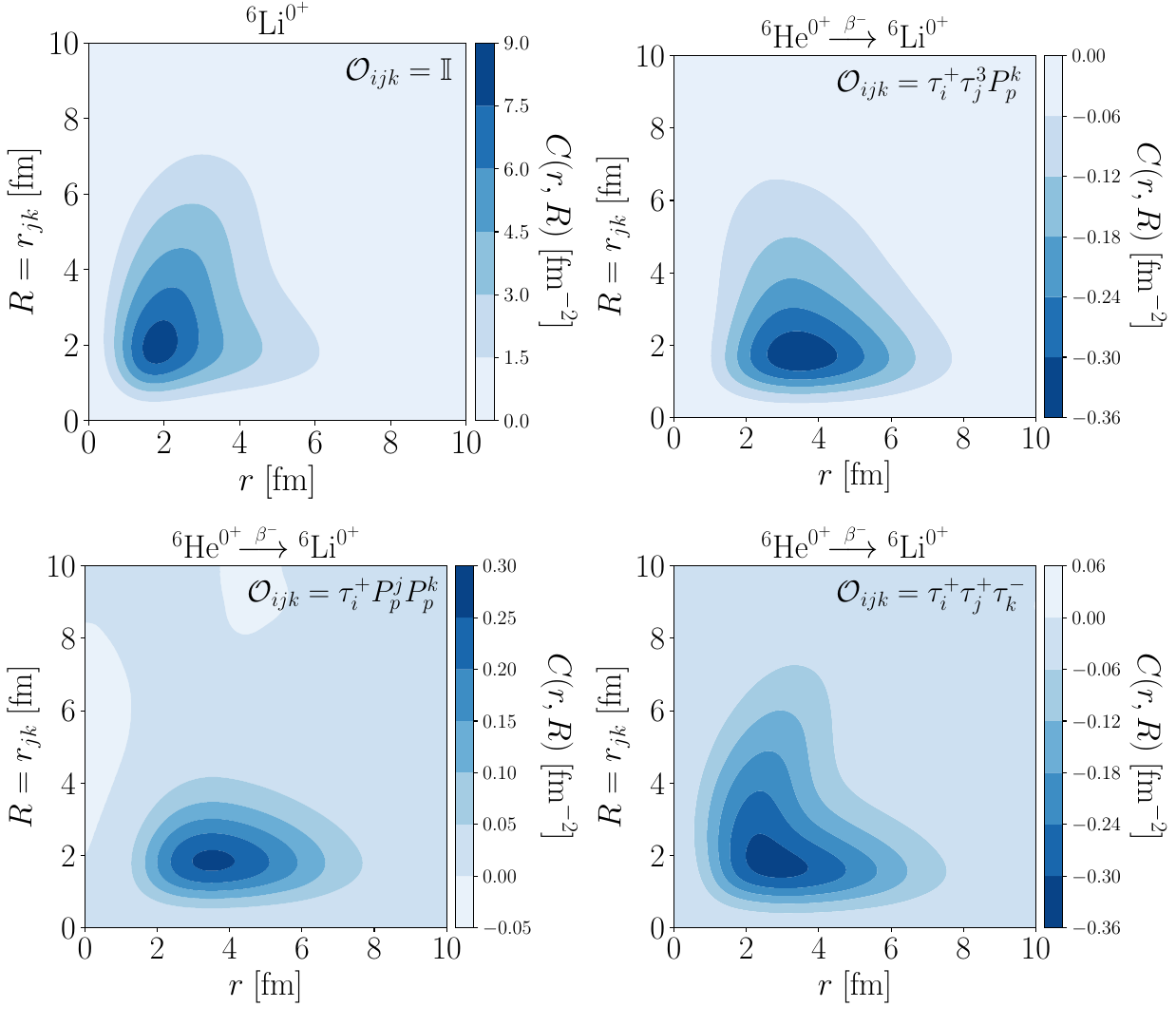}
    \captionsetup{labelformat=default}
    \caption{Densities $C(r,R)$ for the different isospin operators which appear in the three-nucleon potentials for $^6\mathrm{He}\rightarrow{}^6\mathrm{Li}$.}
    \label{fig:6HeIsospin}
\end{figure}

\begin{figure}
    \centering
    \includegraphics[width=.7\linewidth]{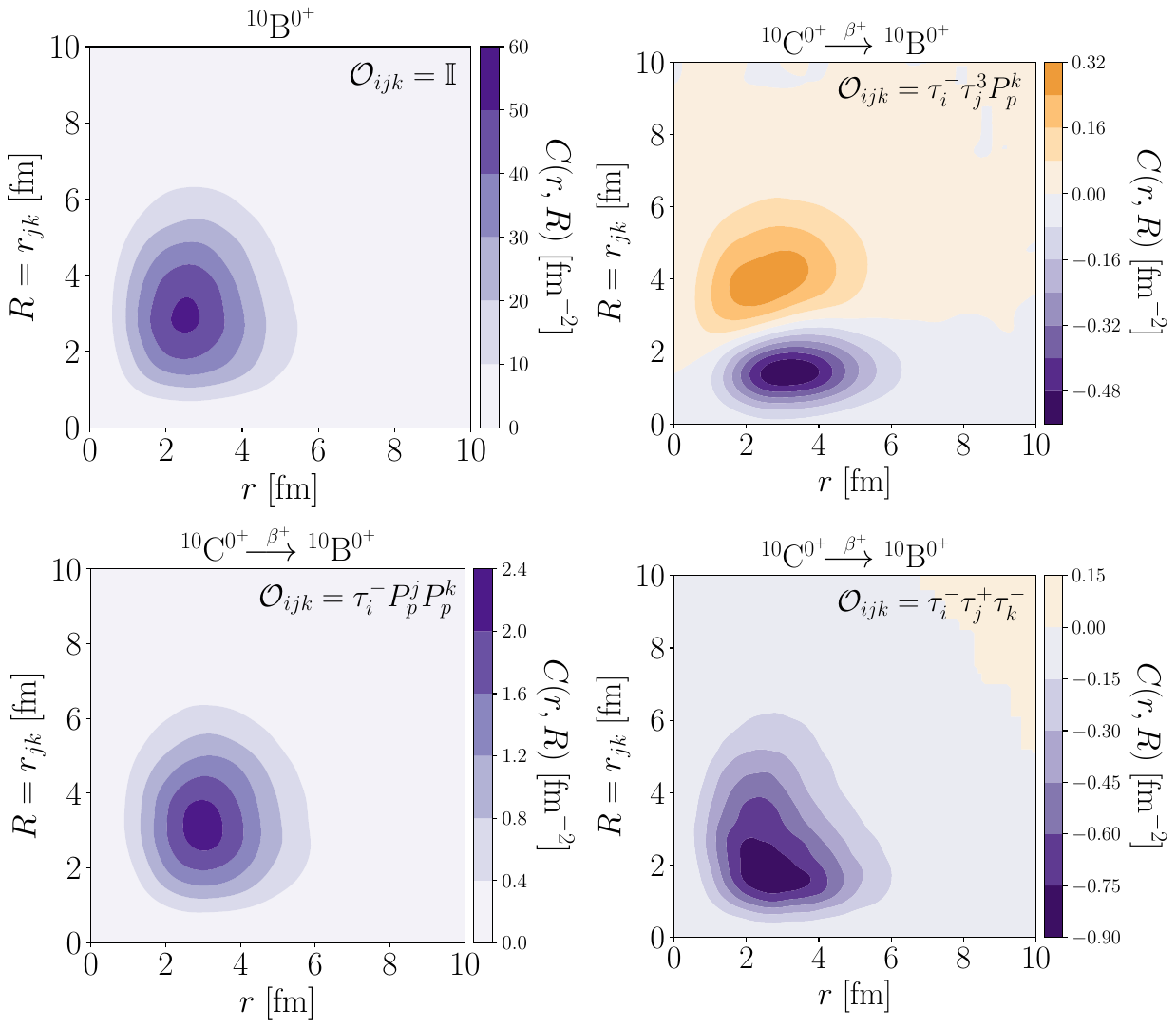}
    \captionsetup{labelformat=default}
    \caption{Densities $C(r,R)$ for the different isospin operators which appear in the three-nucleon potentials for $^{10}\mathrm{C}\rightarrow{}^{10}\mathrm{B}$.}
    \label{fig:10CIsospin}
\end{figure}

 Different operators can lead to very different matrix elements due to nuclear structure effects. As an example, we compute the densities $C(r,R)$ for ${}^{6}\mathrm{He}\rightarrow{}^{6}\mathrm{Li}$ and ${}^{10}\mathrm{C}\rightarrow{}^{10}\mathrm{B}$ for the three types of three-nucleon isospin operators that appear in the potentials studied in this work as well as the identity operator. These densities are shown in Figs. \ref{fig:6HeIsospin} and \ref{fig:10CIsospin}. For example, the densities for operators $\mathcal{O}_{ijk}\in\{\mathds{1},\tau_i^\pm P_p^j P_p^k,\tau_i^\pm\tau_j^+\tau_k^-\}$ are similar for the two transitions. However, the operator $\mathcal{O}_{ijk}=\tau_i^\pm\tau_j^3 P_p^k$ has a very different density structure for the two transitions, which can be understood in terms of the cluster structure of the nuclei involved. 
${}^{6}\mathrm{He}$ has the predominant spatial symmetry of an $\alpha$ with two valence neutrons ($[\alpha,2n]$). When the isospin operator $\tau_i^+\tau_j^3P_p^k$ acts on each permutation of each triplet in ${}^{6}\mathrm{He}$, the $\tau_i^+$ can only act on one of the two valence neutrons, and the proton projection operator $P_p^k$ (due to the photon in the diagrams) must act on one of the protons in the $\alpha$-core. The $\tau_j^3$ operator then acts on the remaining three neutrons and one proton. Due to there being more neutrons, the density becomes negative everywhere.

The same density for the $\beta^+$ decay of ${}^{10}\mathrm{C}\rightarrow{}^{10}\mathrm{B}$ has a profile with a positive and negative region. ${}^{10}\mathrm{C}$ has the spatial symmetry $[2\alpha,2p]$. The operator $\tau_i^-$ acts on one of the two valence protons. When the two spectator nucleons are close to each other ($R=r_{jk}\lesssim2$ fm), they belong to the same $\alpha$. Then $P_p^k$ can act on one of the 2 protons in the $\alpha$, and $\tau_j^3$ acts on the remaining one proton and two neutrons, resulting in a net negative contribution in this region of parameter space. However, when $R\gtrsim 2$ fm, there are two possibilities. If $P_p^k$ acts on the valence proton that does not undergo the decay, then there are four neutrons and four protons in the $2\alpha$. This results in no net contribution to the density. In the other scenario, $P_p^k$ acts on a proton in one of the $\alpha$'s. Then there are three protons or two neutrons for $\tau_j^3$ to act on (the remaining nucleons in the chosen $\alpha$ are excluded because $R$ is large), which gives a net positive contribution.

These examples show the impact of nuclear structure on the resulting densities and matrix elements. The potentials of interest for $\Bar{\delta}_\mathrm{NS}$ are more complicated, having additional spin- and radial-dependence, so a simple interpretation of the densities becomes more challenging. However, the general features of the densities provide insight into the calculation of the matrix element, as we will show in the results.

There are two natural one-dimensional densities that could be constructed from $C(r,R)$: $C(r)$ or $C(R)$. The three-nucleon density $C(r)$ is equivalent to the density of the two-nucleon potentials ($r$ is the distance from the decaying nucleon to the remaining nucleon(s) in both cases). Although information is lost when performing this reduction of the three-nucleon density, it provides a natural way to compare the two- and three-nucleon densities in terms of their relatives size and behavior. The density of $\mathcal{O}_{ijk}=\tau_i^-\tau_j^3P_p^k$ for ${}^{10}\mathrm{C}\rightarrow{}^{10}\mathrm{B}$ demonstrates a loss of information. If one were to examine $C(r)$ and $C(R)$ for this density, the same matrix element could be recovered for both. However, $C(R)$ would feature a nodal structure around $R=2.5$ fm while $C(r)$ would have no such structure with a relatively much smaller amplitude. In the remaining sections, we will show the two- and three-nucleon matrix elements and their densities for the $\Bar{\delta}_\mathrm{NS}$ potentials.

\subsection{Results and discussion}\label{sec:results}

In this section, we provide QMC calculations of matrix elements of the three-nucleon potentials entering $\Bar{\delta}_\mathrm{NS}$ at $\mathcal{O}(\alpha^2)$ and $\mathcal{O}(\alpha\epsilon_\chi^2)$. The primary interest is the superallowed decay of $^{10}\mathrm{C}\rightarrow{}^{10}\mathrm{B}$, for which there is data that can be used in conjunction with theory calculations of radiative corrections to determine $V_{ud}$. However, we also report matrix element calculations for $^{6}\mathrm{He}\rightarrow{}^{6}\mathrm{Li}$, a less computationally demanding system that serves as a benchmark and a means to demonstrate the $Z$ versus $Z^2$ scaling of the $\mathcal{O}(\alpha^2)$ potentials. We begin with the results at $\mathcal{O}(\alpha^2)$ and $\mathcal{O}(\alpha\epsilon_\chi^2)$ in Secs. \ref{sec:NLO} and \ref{sec:N2LO}, respectively, and then present the impact on $V_{ud}$ in Sec. \ref{sec:Vud}.

\subsubsection{Order $\mathcal O(\alpha^2)$ contributions}\label{sec:NLO}

We first compute matrix elements of the two- and three-nucleon potentials entering Eqs. (\ref{eq:log2b}) and (\ref{eq:log3b}). The QMC results are shown in Table~\ref{tab:MatrixElementsLog} for the transitions of $^{6}\mathrm{He}$ and ${}^{10}\mathrm{C}$ for two Norfolk interactions, denoted by $\mathrm{Ia}$ and $\mathrm{Ia^\star}$. The matrix elements $M^+_\mathrm{2N}$ and $M^+_\mathrm{3N}$ shown in the table are exact results. Some many-body methods, which can perform calculations of the heavier systems of interest for CKM unitarity tests, may face computational challenges in implementing three-nucleon transition operators. Due to its relative simplicity, having no spin-dependence, the three-nucleon potential at this order is amenable to approximation and can be constrained by the two-nucleon potential. The three nucleon potential has the geometric property of being symmetric with respect to the exchange of indices $j\leftrightarrow k$. Assuming $r_{ij}\approx r_{ik}\approx r_{jk}$ reduces the operator to the following form:
\begin{align}
    \tilde{\mathcal{V}}_{+,\mathrm{approx}}^\mathrm{3N}=-\frac{Z-1}{2}\sum_{i\neq j}\log\left(\frac{3}{2}\Lambda r_{ij}\right)\tau^+_iP_p^j\label{eq:approx3N},
\end{align}
 where the factor of $(Z-1)$ results from the sum over $P_p^k$ and would be replaced with $(Z-2)$ in the case of $\beta^+$ decay. Notice that
 here $Z$ denotes the charge of the parent nucleus.
 Eq. (\ref{eq:approx3N}) has a similar form to the two-nucleon potential. After computing the two-nucleon matrix element $M_\mathrm{2N}^+$, an effective pair distance $\langle r\rangle$ can be extracted by setting $r_{ij}$ to a constant and performing the explicit sum over the remaining isospin operator. This effective distance can then be inserted into Eq. (\ref{eq:approx3N}), providing the matrix element approximation
\begin{align}
    M_{\mathrm{3N,approx}}^+=\frac{Z-1}{2}\left[M_\mathrm{2N}^+-Z\log\left(\frac{3}{2}\right)M_\mathrm{F}^{(0)}\right],
\end{align}
 where again each $Z$ becomes $(Z-1)$ for $\beta^+$ decays.

\begin{table}[]
    \centering
    \begin{tabular}{c|c|c|c|c}
    \hline
    \hline
    $T_i\rightarrow T_f$ & Model & $M_{\mathrm{2N}}^{+}$ &  $M_{\mathrm{3N}}^{+}$ & $M_{\mathrm{3N},\mathrm{approx}}^+$
    \\
        \hline\hline
        \multirow{2}{8em}{\centering$^6\mathrm{He}\rightarrow{}^6\mathrm{Li}$}
         & Ia$\phantom{^\star}$ & -1.68 & -0.91 & -1.41\\
        & Ia$^\star$ & -1.79 & -0.94 & -1.47\\
        \hline
        \multirow{2}{8em}{\centering$^{10}\mathrm{C}\rightarrow{}^{10}\mathrm{B}$}
        & Ia$\phantom{^\star}$ & -1.64(2) & -9.10(1) & -9.01(4)\\
        & Ia$^\star$ & -2.13(4)\;[-2.45(7)] & -9.87(2)\;[-10.14(9)] & -9.99(8)\;[-10.63(14)]\\
     \hline
     \hline
    \end{tabular}
\captionsetup{justification=RaggedRight}
    \captionsetup{labelformat=default}
    \caption{VMC [GFMC] two- and three-nucleon matrix elements for the $\mathcal{O}(\alpha^2)$ potentials. }
    \label{tab:MatrixElementsLog}
\end{table}

The matrix elements using this approximation are shown in the last column in Table \ref{tab:MatrixElementsLog}. In the case of ${}^{10}\mathrm{C}$, the approximation reproduces the exact three-nucleon matrix element within $\lesssim 5\%$. However, the approximation severely overestimates the magnitude of the three-body matrix element in the case of $^6\mathrm{He}$. This can be understood in terms of the cluster structure of these transitions. The distance $r_{ij}$ is the distance from the decaying nucleon (either a proton or neutron) to another proton in the nucleus. In the case of $^{10}\mathrm{C}$ ($^{6}\mathrm{He}$), the proton (neutron) that decays is a valence nucleon. For $\beta^-$ decays in light systems, the distance between two protons, $r_{jk}$, will on average be smaller than $r_{ij}$ due to there being two protons bound closely within $\alpha$-clusters. For $\beta^+$ decays, the average $r_{ij}$ will be close to the average $pp$ pair distance because a proton is undergoing the decay. For these reasons, the approximation works better for $\beta^+$ decays. Explicit calculations of $\langle r_{pp}\rangle$ and $\langle r_{np}\rangle$, inserted appropriately into Eq. (\ref{eq:approx3N}), would improve the results for $\beta^-$ decay.

\begin{figure}[t]
    \centering    \includegraphics[width=.95\linewidth]{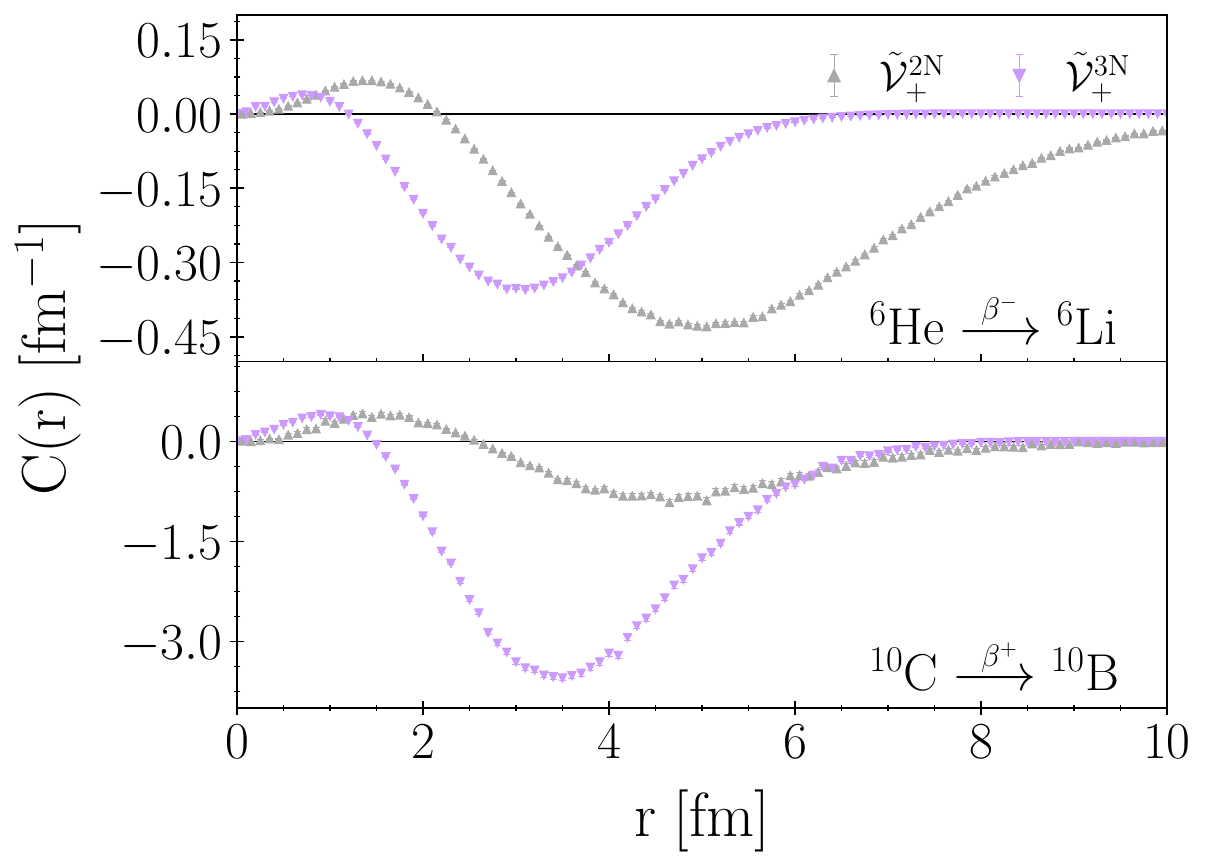}
       \captionsetup{labelformat=default}
    \caption{Density $C(r)$ for matrix element of $\tilde{\mathcal{V}}_+^{\mathrm{2N}}$ and $\tilde{\mathcal{V}}_+^{\mathrm{3N}}$ for $^6\mathrm{He}\rightarrow{}^6\mathrm{Li}$ (top) and $^{10}\mathrm{C}\rightarrow{}^{10}\mathrm{B}$ (bottom).}\label{fig:logplot}
\end{figure}

Fig. \ref{fig:logplot} shows the densities as a function on $r$, the distance from the decaying nucleon and the center of mass of the spectator nucleon(s), for the two- (gray) and three-nucleon (purple) potentials at $\mathcal{O}(\alpha^2)$. The node in each density is a result of the logarithm in the potentials, which changes signs when $r_{ij}=R_A$ or $(r_{ij}+r_{ik}+r_{jk})=2R_A$. As can be seen in Fig. \ref{fig:logplot} or the values in Table \ref{tab:MatrixElementsLog}, the two-nucleon contributions are larger than the three-nucleon contributions at $\mathcal{O}(\alpha^2)$ for $^6\mathrm{He}$. However, due to the $Z$ and $Z^2$ scaling of these contributions, the three-nucleon matrix element becomes much larger than the two-nucleon contribution ($\sim5\;\times$ greater) for $^{10}\mathrm{C}$.

\begin{figure}[t]
  \centering
  \includegraphics[width=\linewidth]{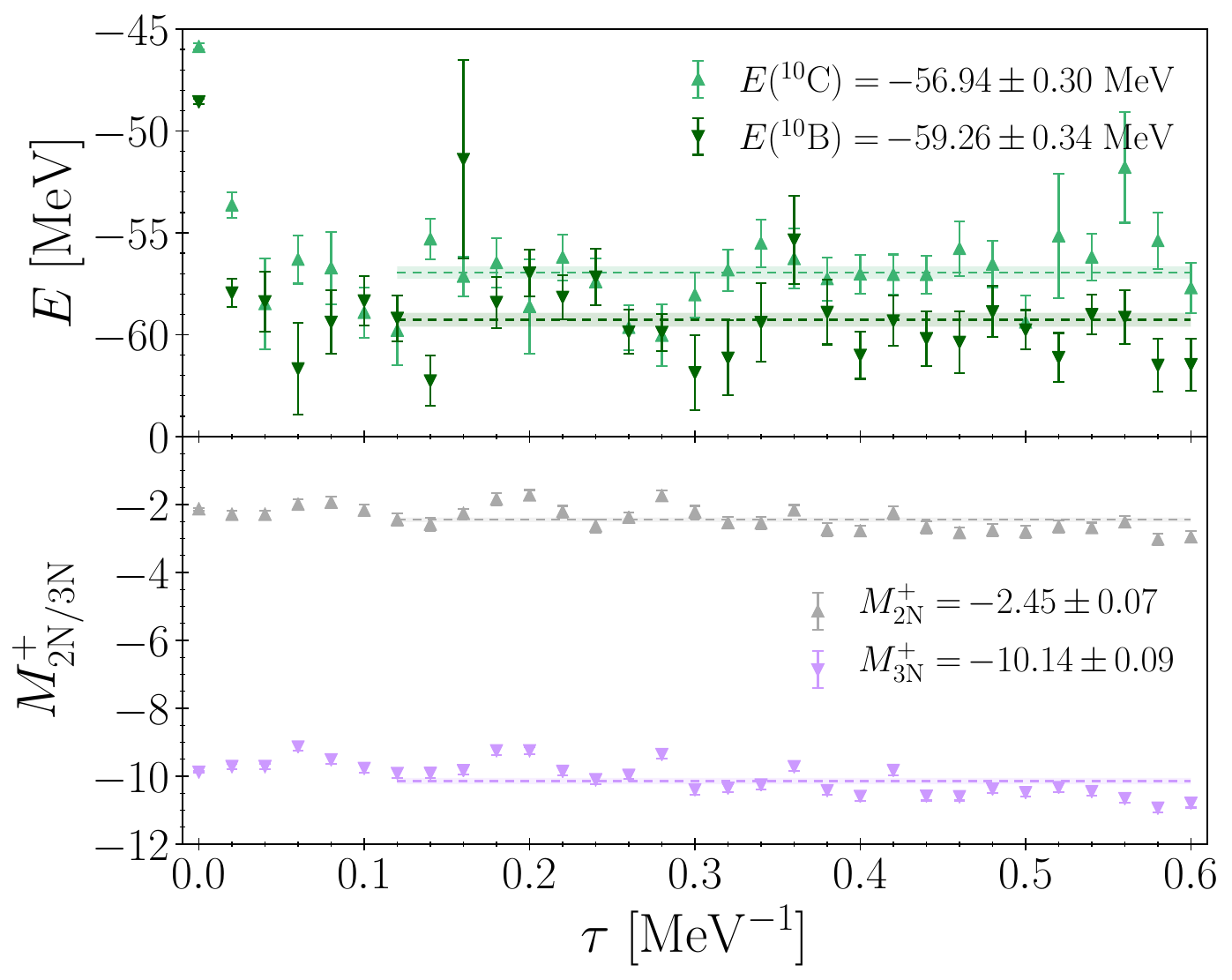}
  
\captionsetup{labelformat=default}
  \caption{GFMC energy propagation for ${}^{10}$C($0^+$;1) and ${}^{10}$B($0^+$;1) (top) and $\mathcal{O}(\alpha^2)$ matrix elements (bottom) as a function of imaginary time $\tau$ using model Ia$^\star$. The dashed lines and band denote the average energy and error estimate over the range $\tau=0.12-0.6\;\mathrm{MeV}^{-1}$.}\label{fig:GFMC_log}
\end{figure}

Due to the importance of these three-nucleon contributions, we compute their matrix elements at this order using GFMC.  Fig. \ref{fig:GFMC_log} shows the energy propagation of $^{10}\mathrm{C}$ and $^{10}\mathrm{B}$ (top panel), which converge to energies of $E({}^{10}\mathrm{C})=-56.94(30)$ MeV and $E({}^{10}\mathrm{B})=-59.26(34)$ MeV, respectively, for model Ia$^\star$. The bottom panel shows the propagation of the two- (gray) and three-nucleon (purple) matrix elements for the $^{10}\mathrm{C}\rightarrow{}^{10}\mathrm{B}$ transition. The two-nucleon matrix element changes by approximately 15\% during GFMC propagation, whereas the three-nucleon contribution changes by only 3\%. The 3\% is comparable to the typical 2–3\% GFMC systematic uncertainty, indicating that the three-nucleon contribution is relatively stable under propagation. Due to computational cost and the stability of the 3N matrix elements, we compute only the VMC matrix elements for model $\mathrm{Ia}$. 

These three-nucleon contributions will lead to a nonnegligible impact on the extraction of $V_{ud}$ at the precision of $10^{-4}$. For example, for the $^{10}$C to $^{10}$B transition, using the GFMC results one gets
\begin{equation}
    \left[\bar{\delta}_{\text{NS}}\right]_{\alpha^2}  = -1.8 \cdot 10^{-4} - 7.6 \cdot 10^{-4},
\end{equation}
where the first (second) contribution comes from the two-body (three-body) matrix element. While smaller than the $\mathcal O(\alpha \epsilon_\chi)$ correction to $\delta_{\text{NS}}$, discussed in Section \ref{sec:N2LO}, these contributions are important, and we quantify their impact  in Sec. \ref{sec:Vud}. 
Furthermore, due to the scaling of these $\mathcal{O}(\alpha^2)$ two- and three-nucleon operators, the inclusion of three-nucleon contributions is even more important for heavier nuclei.

\subsubsection{Order $\mathcal{O}(\alpha\epsilon_\chi^2)$ contributions}\label{sec:N2LO}

We now report the matrix element calculation of the three-nucleon potentials entering at $\mathcal{O}(\alpha\epsilon_\chi^2)$. As discussed in Sec. \ref{sec:VA}, only the VA part of these three-nucleon potentials should be included in a many-body calculation when the LO two-nucleon potential $\mathcal{V}_0^\mathrm{2N}\propto q^0$ and three-nucleon ISB correction $v_\gamma^\mathrm{3N}$ are not included. Potentials $\mathcal{V}_0^{(g)}$ and $\mathcal{V}_0^{(j)}$ are the only contributions at this order having a VA component, whose combined matrix element we denote by $M_\mathrm{3N}$ in Table \ref{tab:MatrixElements}.~$M_\mathrm{3N}$ is defined in Eqs.~\eqref{eq:3summary} and \eqref{eq:3summary2}, and the coordinate-space expressions for these contributions are given in Appendix \ref{app:VA}. 

\begin{figure}[t]
    \centering
    \includegraphics[width=.95\linewidth]{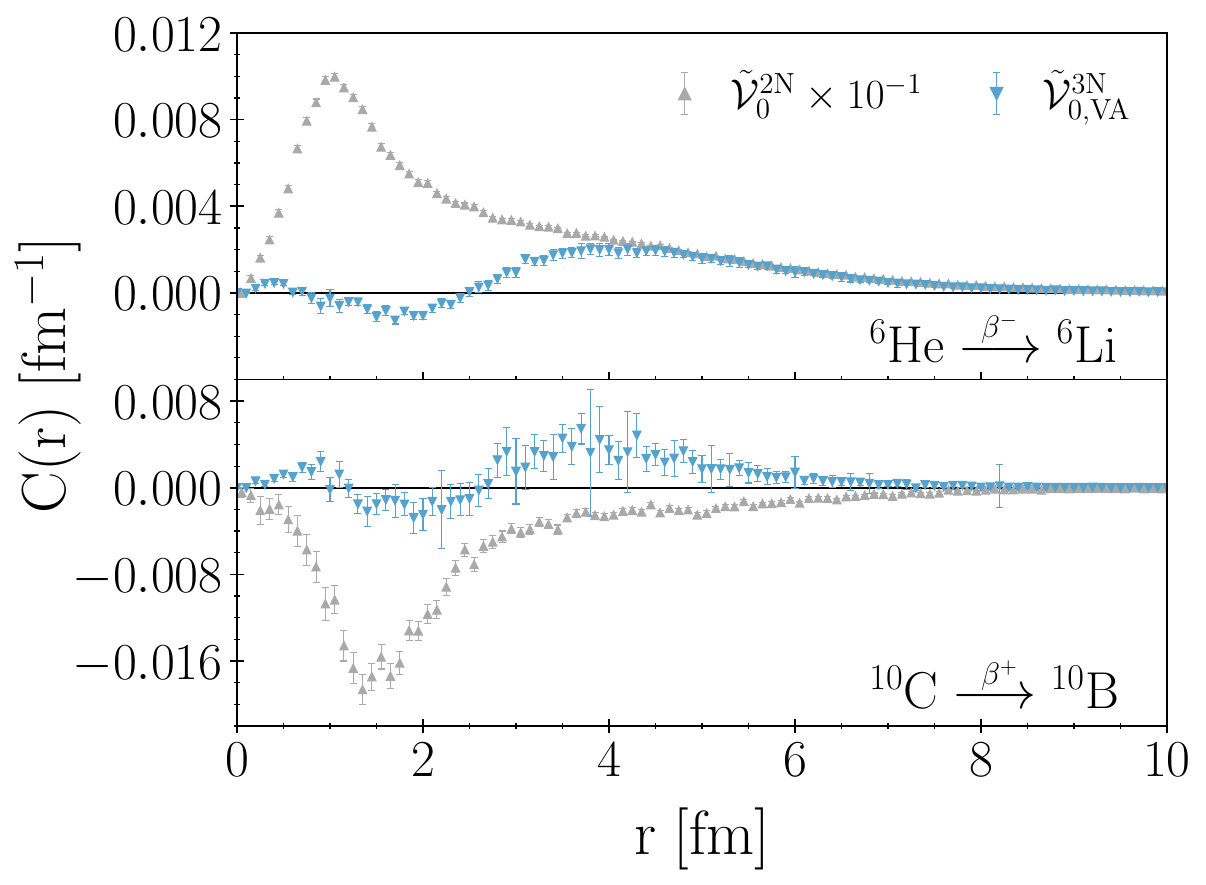}

    \captionsetup{labelformat=default}
    \caption{Density $C(r)$ for matrix element of leading two-nucleon contributions divided by 10 (gray) and the $\mathcal{O}(\alpha\epsilon_\chi^2)$ VA contributions (blue) for ${}^6\mathrm{He}\rightarrow{}^6\mathrm{Li}$ (top) and ${}^{10}\mathrm{C}\rightarrow{}^{10}\mathrm{B}$ (bottom).}    \label{fig:combinedN2LO}
\end{figure}

These three-nucleon contributions ($\mathcal{O}(\alpha\epsilon_\chi^2)$) can be interpreted as a correction to the LO two-nucleon contributions ($\mathcal{O}(\alpha\epsilon_\chi)$), whose matrix elements are also included in Table \ref{tab:MatrixElements} for the same two transitions and chiral models for comparison. As can be seen from these matrix elements or the densities in Fig. \ref{fig:combinedN2LO}, the three-nucleon contribution (blue) is roughly two orders of magnitude smaller than the LO two-nucleon contributions (gray). In both transitions, most of the contribution of $\tilde{\mathcal{V}}_{0,\mathrm{VA}}^{\mathrm{3N}}$ comes from long-ranged distance ($r\gtrsim3$ fm), due to cancellations between terms with different spin-operators at short distance, whereas the two-nucleon contribution is dominated by short-distance behavior. For the ${}^{10}\mathrm{C}\rightarrow{}^{10}\mathrm{B}$ transition, the three-nucleon contribution has opposite sign relative to the LO two-nucleon matrix element.

By naively averaging the two models,  we get that, in the case of $^{10}$C, the three-body correction to $\bar{\delta}_{\text{NS}}$ is
\begin{equation}
    \left[\bar\delta_{\text{NS}} \right]_{\text{3N}} = 0.99(12) \cdot 10^{-4}.
\end{equation}
This correction is at the level of the error on the single-nucleon vector coupling $g_V$, and, as we will discuss in Section~\ref{sec:Vud},
it does not yet affect the extraction of $V_{ud}$. No-core shell model calculations of $\delta_\mathrm{NS}$ in this same transition estimate an uncertainty from the truncation of the chiral current expansion, finding a percent-level impact as well \cite{Gennari_2025}. 
It is however possible that the three-nucleon operator will have a more significant impact in medium mass nuclei. 
In particular, for most of the $0^+ \rightarrow 0^+$ transitions in which the decaying nucleus has $T_z = 0$, where $T$ is the nucleus isospin,
the two-body piece of $\bar{\delta}_{\text{NS}}$ is suppressed, because of a cancellation between the contributions of the isoscalar and isovector nucleon magnetic moments \cite{Hardy_2020,Novario}. 
If not affected by similar cancellations, the three-nucleon operator could then have a relatively larger effect. Furthermore, the calculation of $M_{3\text{N}}$ and of
 $\mathcal{O}(\alpha\epsilon_\chi^2)$ corrections in the two-nucleon sector is necessary to provide an estimate of the EFT truncation error on $\delta_{\text{NS}}$.


\begin{table}[]
    \centering
    \begin{tabular}{c|c|cc|c}
    \hline
    \hline
    $T_i\rightarrow T_f$ & Model & $M_{\mathrm{GT}}^{\mathrm{mag}}+M_{\mathrm{T}}^{\mathrm{mag}}+M_{\mathrm{LS}}$ & & $M_{\mathrm{3N}}\;(10^{-2})$ \\
        \hline\hline
        \multirow{2}{8em}{\centering$^{6}\mathrm{He}\rightarrow{}^{6}\mathrm{Li}$}
         & Ia$\phantom{^\star}$ & 0.214 &  & 0.60(3)\\
        & Ia$^\star$ & 0.209 & & 0.55(3) \\
        \hline
        \multirow{2}{8em}{\centering$^{10}\mathrm{C}\rightarrow{}^{10}\mathrm{B}$}
        & Ia$\phantom{^\star}$ & -0.399&  & 1.05(17)\\
        & Ia$^\star$ & -0.410 & & 0.87(17) \\
     \hline
     \hline
    \end{tabular}
\captionsetup{justification=RaggedRight}
\captionsetup{labelformat=default}
\caption{VMC matrix elements for $\Bar{\delta}_{\mathrm{NS}}$ with two Norfolk interactions. The two-nucleon contributions given by $M_{\mathrm{GT}}^{\mathrm{mag}}+M_{\mathrm{T}}^{\mathrm{mag}}+M_{\mathrm{LS}}$ are order $\mathcal{O}(\alpha\epsilon_\chi)$ while the three-nucleon contribution $M_\mathrm{3N}$ is order $\mathcal{O}(\alpha\epsilon_\chi^2)$.}
\label{tab:MatrixElements}
\end{table}

\subsection{Impact on $V_{ud}$}\label{sec:Vud}

Having computed the $\Bar{\delta}_\mathrm{NS}$ contributions up to $\mathcal{O}(\alpha\epsilon_\chi^2)$ ($\mathcal{O}(\alpha^2)$) in the three-nucleon (two-nucleon) sector, we can determine $V_{ud}$ from ${}^{10}\mathrm{C}\rightarrow{}^{10}\mathrm{B}$ in conjunction with QMC calculations of $\Bar{\delta}_\mathrm{C}$ for the same transition and interaction \cite{Piarulli_2026}. The factorization formula of Eq. (\ref{eq:masterformula}) provides the inputs needed for this determination. The experimental half-life is given by Ref. \cite{Hardy_2020},
\begin{align}
    t=1321.8\pm1.8\;\mathrm{s},
\end{align}
the first factor on the rhs is known with negligible uncertainty,
\begin{align}
    \frac{G_F^2m_e^5}{\pi^3\log 2}=3.350722(3)\times10^{-4}\;\mathrm{s}^{-1}.
\end{align}
The phase-space factor and Coulomb corrections, including $\mathcal{O}(\alpha^2Z)$ corrections, are given by
\begin{align}\label{phasespace}
    \left[C_\mathrm{eff}^{(g_V)}(\mu)\right]^2\left[1+\Bar{\delta}_R^\prime(\mu)\right]\times\Bar{f}(\mu)&=2.39668(57)_{g_V}(0)_\mu. 
\end{align}

These value uses the matching coefficients $C_\mathrm{eff}^{(g_V)}$
from Ref.~\cite{Cirigliano_2024A},
with a few corrections identified in Ref.~\cite{Novario2,Novario3}, and $\mathcal O(\alpha^2 Z)$ corrections from Refs.~\cite{Crosas:2025xyv,Cao:2025zxs}. The treatment of finite size and atomic corrections is described in Ref.~\cite{Cirigliano_2024A,Novario2}, with input taken from Ref.~\cite{Towner_2010}.
The quoted errors in Eq.~\eqref{phasespace}
come from the nonperturbative part of the $W\gamma$ box, which enters in the single-nucleon vector coupling $g_V$ \cite{Seng:2018qru,Seng:2018qru}. 
The scale variation error, denoted by $\mu$, is
  negligible. The phase space $\Bar{f}$ and outer correction $\Bar{\delta}^\prime_R$ are evaluated with a variation in $\mu$ from $E_0$ to $4E_0$, and the coefficient $C_\mathrm{eff}^{(g_V)}$ is evaluated at the matching scale
\begin{align}
    \mu=\sqrt{3/5}R_\mathrm{ch}^{-1}\exp\left(\frac{1}{2}-\gamma_E\right), 
\end{align}
with charge radius $R_\mathrm{ch}$, and then evolved down to the scale $\mu=2E_0$. 
The value of the charge radius is taken from Refs.~\cite{Angeli_2013,Ohayon_2025}.

\begin{figure}[t]
\centering
\begin{minipage}{.65\textwidth}
  \centering \includegraphics[width=\linewidth]{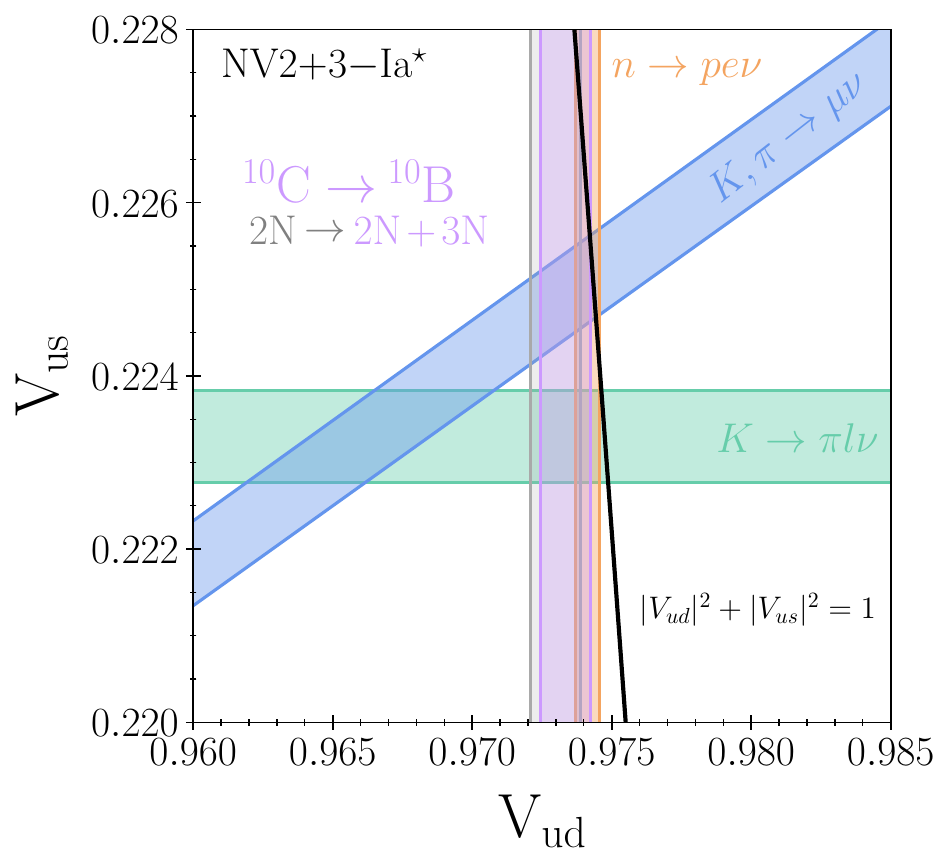}
\end{minipage}
\begin{minipage}{.34\textwidth}
  \centering
\includegraphics[width=\linewidth]{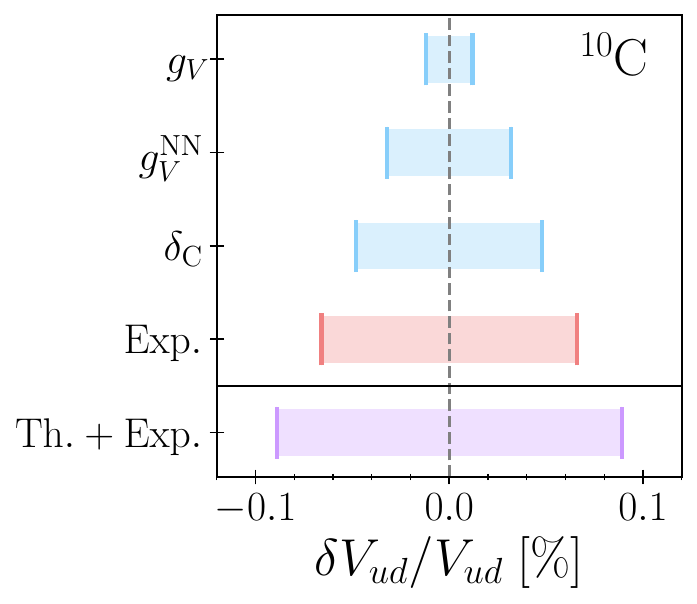}
\end{minipage}
\captionsetup{labelformat=default}
  \caption{The $V_{ud}$ vs $V_{us}$ parameter space plot (left) showing the constraint on $V_{ud}$ with only 2N matrix elements (gray) or 2N and 3N matrix elements (purple) up to NLO from the superallowed transition of ${}^{10}\mathrm{C}\rightarrow{}^{10}\mathrm{B}$. The breakdown of the uncertainty entering the purple band is shown in the right figure with theory error in blue, experiment in red, and combined in purple. Other bands are taken from Ref. \cite{Cirigliano_2023}.}
  \label{fig:VudVus}
\end{figure}

We take values of $\Bar{\delta}_\mathrm{C}$ in $\%$ from Ref. \cite{Piarulli_2026}:
\begin{align}
    \mathrm{NV2{+}3{-}Ia}\;\,\quad\Bar{\delta}_\mathrm{C}&=0.0983(6)\;[0.249(157)]
    \\
    \mathrm{NV2{+}3{-}Ia}^\star\quad\Bar{\delta}_\mathrm{C}&=0.1747(4)\;[0.152(99)]
\end{align}
where the values are given for the VMC [GFMC]. Combining these factors with the two- and three-nucleon $\Bar{\delta}_\mathrm{NS}$ values of this work and Ref. \cite{King_2025}, we arrive at $V_{ud}$ determined from ${}^{10}\mathrm{C}$:
\begin{align}
    \mathrm{NV2{+}3{-}Ia}\;\,\quad V_{ud}|_{{}^{10}\mathrm{C}}&=0.97389(66)_\mathrm{exp}(12)_{g_V}(77)_\mathrm{\Bar{\delta}_\mathrm{C}}(38)_{g_V^\mathrm{NN}}
    \\
    \mathrm{NV2{+}3{-}Ia}^\star \quad V_{ud}|_{{}^{10}\mathrm{C}}&=0.97335(66)_\mathrm{exp}(12)_{g_V}(48)_{\Bar{\delta}_\mathrm{C}}(32)_{g_{V}^\mathrm{NN}},
\end{align}
where GFMC values are used except where only VMC is available. As in Refs. \cite{King_2025},
we consider the contribution from the unknown LECs $g^{\text{NN}}_{V, 1}$ and $g^{\text{NN}}_{V, 2}$ as an error.
The impact of the three-nucleon contributions to $\Bar{\delta}_\mathrm{NS}$ are shown in the left of Fig. \ref{fig:VudVus} for model $\mathrm{NV2{+}3{-}Ia}^\star$. The gray band shows the uncertainty of $V_{ud}$ with only two-nucleon contributions to $\Bar{\delta}_\mathrm{NS}$. The purple band shows the change in the central value of $V_{ud}$ due to the inclusion of the three-nucleon contributions of this work, shifting the $V_{ud}$ band to a larger value. The right plot in Fig. \ref{fig:VudVus} shows the breakdown of uncertainties entering the $V_{ud}$ constraint of the left plot. The theoretical uncertainties are shown in blue, experimental in red, and combined in purple. In this model, the largest contribution remains experimental uncertainty on the decay rate. On the theory side, $\Bar{\delta}_\mathrm{C}$ is the largest uncertainty in the EFT and QMC approach, followed by the LEC uncertainty.

\section{Conclusion}\label{sec:conclusion}

Testing the unitarity of the first row of the CKM matrix provides a precise test of the Standard Model and means to constrain possible BSM scenarios. The determination of $V_{ud}$ from superallowed $\beta$ decays requires theory input across a range of scales, from lattice QCD, chiral effective field theory,
and many-body nuclear calculations. One of the biggest theory uncertainties is the nuclear-structure-dependent correction $\Bar{\delta}_\mathrm{NS}$, which receives contributions from potential regions. The leading contributions arise from two-nucleon transition operators, and three-nucleon contributions appear at $\mathcal{O}(\alpha^2)$ and $\mathcal{O}(\alpha\epsilon_\chi^2)$. 

In this work, we derived the three-nucleon potentials at $\mathcal{O}(\alpha\epsilon_\chi^2)$ and computed matrix elements of three-nucleon contributions up to this order with quantum Monte Carlo methods. Within the dispersive approach, there are no vector-vector contributions to $\delta_\mathrm{NS}$ at order $\mathcal{O}(\alpha)$. We show how the VV contributions in the three-nucleon potentials of this work cancel with three-nucleon contributions built from the iteration of either ($i$) the LO two-nucleon potential with the one-pion-exchange potential or ($ii$) the LO one-body weak vertex with a three-nucleon ISB potential. 

We compute matrix elements of the remaining VA contributions, finding a small correction to $V_{ud}$, currently beyond the precision capabilities of unitarity tests. However, matrix elements of the $\mathcal{O}(\alpha^2)$ potentials reveal an important contribution to $V_{ud}$ determinations. Three-nucleon contributions of this order have a sizable effect on $\Bar{\delta}_\mathrm{NS}$ and the extraction of $V_{ud}$ from the superallowed decay of ${}^{10}\mathrm{C}\rightarrow{}^{10}\mathrm{B}$. Due to the structure of these two- and three-nucleon potentials, their contributions scale like $Z$ and $Z^2$, respectively. Therefore, the contribution of these three-nucleon potentials should be sizable for heavier systems.

We conclude by computing an updated value of $V_{ud}$ using quantum Monte Carlo calculations of $\Bar{\delta}_\mathrm{NS}$ and $\Bar{\delta}_\mathrm{C}$. The effect of the NLO three-nucleon contributions is to shift $V_{ud}$ to a larger value, closer to the neutron decay determination. In the case of $^{10}\mathrm{C}$, the experimental uncertainty remains the largest source of error. On the theory side, $\Bar{\delta}_\mathrm{C}$ is the largest uncertainty.

We conclude with two takeaways. First, the three-nucleon contributions appearing at $\mathcal{O}(\alpha^2)$ in the EFT expansion provide an important contribution to $\Bar{\delta}_\mathrm{NS}$, especially for larger nuclei, and should not be neglected. However, due to the simplicity in the operator structure, simple approximations using only two-nucleon calculations, such as the one presented in this work, may be useful to constrain the three-nucleon matrix element when explicit calculation presents a challenge. Second, for $^{10}$C,
the three-nucleon potentials at $\mathcal{O}(\alpha\epsilon_\chi^2)$ appear to provide a small impact on $\Bar{\delta}_\mathrm{NS}$, currently outside the precision needs of $V_{ud}$ determinations. This is a desirable outcome from the viewpoint of the convergence of the EFT expansion.
It should be checked whether three-nucleon contributions remain negligible in medium mass nuclei, where the two-body magnetic operators corrections to $\delta_{\text{NS}}$ are suppressed. In addition, for a full assessment of the EFT convergence,  there remain other contributions at $\mathcal O(\alpha \epsilon_\chi^2)$, namely two-nucleon potentials arising from pion-photon loops, which have not yet been evaluated. Matrix elements of these remaining operators should be studied in order to robustly determine the uncertainty on $\Bar{\delta}_\mathrm{NS}$.

\section*{Acknowledgments}

We are thankful for useful discussions with Wouter Dekens, Garrett King, Chien-Yeah Seng, and Bob Wiringa. 
This project is supported in part by the US Department of Energy under Contracts No. DE-SC0021027 (G.C. and S.P.) and the Office of Science Graduate Student Research (SCGSR) program under contract number DE-SC0014664 (G.C.). E.M. is supported by the U.S. Department of Energy Office and by the Laboratory Directed Research and Development (LDRD) program of Los Alamos National Laboratory under project numbers 20250164ER and 20260246ER. Los Alamos National Laboratory is operated by Triad National Security, LLC, for the National Nuclear Security Administration of the U.S. Department of Energy (Contract No. 89233218CNA000001). M.P. is supported by U.S. Department of Energy under the 2021 Early Career Award No. DE-SC0022002. 
We acknowledge support from the DOE Topical Collaboration “Nuclear Theory for New Physics,”
award No. DE-SC0023663. The QMC calculations were performed by the National Energy Research Scientific Computing Center, a DOE Office of Science User Facility supported by the Office of Science of the U.S. Department of Energy under Contract No. DE-AC02-05CH11231 using NERSC award NP-ERCAP0036336 and NP-ERCAP0032885.

\appendix

\section{Coordinate-space potentials}\label{app:VA}

We denote the Fourier transform of potential $\mathcal{V}(\boldsymbol{q}_i,\boldsymbol{q}_k)$ as

\begin{align}
    \tilde{\mathcal{V}}(\boldsymbol{r}_{ij},\boldsymbol{r}_{jk})&=\int\frac{\mathrm{d}^3\boldsymbol{q}_i}{(2\pi)^3}\int\frac{\mathrm{d}^3\boldsymbol{q}_k}{(2\pi)^3}\mathcal{V}(\boldsymbol{q}_i,\boldsymbol{q}_k)\exp(i\boldsymbol{q}_i\cdot\boldsymbol{r}_{ij}-i\boldsymbol{q}_k\cdot\boldsymbol{r}_{jk}).
\end{align}

We provide the coordinate-space expression of the VA part of potentials $\mathcal{V}_{0}^{(j)}$ and $\mathcal{V}_0^{(g)}$. The expression for $\mathcal{V}_{0,\mathrm{VA}}^{(j)}$ is given below, where we have factorized the three-nucleon spin operators into anticommutators of purely two-nucleon Tensor and Gamow-Teller operators. Identities translating from this factorized form to the three-nucleon form are given explicitly in Appendix B.

\begin{align}
    \begin{split}
     \tilde{\mathcal{V}}_{0,\mathrm{VA}}^{(j)}&=\frac{g_A^3}{F_\pi^2}\sum_{i\neq j\neq k}\frac{m_\pi^2}{16\pi}Y_\pi(r_{jk})\left(1+\frac{1}{m_\pi r_{jk}}\right)
     \\
     &\left(\frac{1}{36}\frac{1}{\hat{\boldsymbol{r}}_{ij}\cdot\hat{\boldsymbol{r}}_{jk}}\{\boldsymbol{\sigma}_i\cdot\boldsymbol{\sigma}_k,\boldsymbol{\sigma}_j\cdot\boldsymbol{\sigma}_k\}\{(S_{ij}+\boldsymbol{\sigma}_i\cdot\boldsymbol{\sigma}_j),(S_{jk}+\boldsymbol{\sigma}_j\cdot\boldsymbol{\sigma}_k)\}\right.
\\
&\left.-\frac{1}{18}\frac{r_{ik}}{r_{jk}}\frac{1}{\hat{\boldsymbol{r}}_{ij}\cdot\hat{\boldsymbol{r}}_{ik}}\{(S_{ij}+\boldsymbol{\sigma}_i\cdot\boldsymbol{\sigma}_j),(S_{ik}+\boldsymbol{\sigma}_i\cdot\boldsymbol{\sigma}_k)\}\right.
\\
&\left.+\frac{1}{6}\frac{r_{ij}}{r_{jk}}\{(S_{ij}+\boldsymbol{\sigma}_i\cdot\boldsymbol{\sigma}_j),\boldsymbol{\sigma}_i\cdot\boldsymbol{\sigma}_k\}\right)\tau_i^+(\tau_k^+\tau_j^--\tau_k^-\tau_j^+)   
    \end{split}
\end{align}

\noindent where the Yukawa and coordinate-space tensor operators are

\begin{align}
    Y_\pi(r)&=\frac{e^{-m_\pi r}}{m_\pi r},
    \\
    S_{ij}&=3(\boldsymbol{\sigma}_i\cdot\hat{\boldsymbol{r}}_{ij})(\boldsymbol{\sigma}_j\cdot\hat{\boldsymbol{r}}_{ij})-\boldsymbol{\sigma}_i\cdot\boldsymbol{\sigma}_j.
\end{align}

The Fourier transform of  $\mathcal{V}_{0,\mathrm{VA}}^{(g)}$ does not have a simple analytic form and must be evaluated numerically. The  expression is decomposed into different spin operators with derivatives acting on the scalar part of the potential, which we write as

\begin{align}
    \tilde{\mathcal{V}}_{0,\mathrm{VA}}^{(g)}&=i2\pi\frac{g_A^3}{F_\pi^2}\sum_{i\neq j\neq k}\int_0^\infty\mathrm{d}q\int_0^1\mathrm{d}x\;\mathcal{O}_{ijk}^{(g)}f^\prime(r_{ij},r_{jk},c_\theta)\tau_j^+(\tau_i^+\tau_k^--\tau_i^-\tau_k^+)\label{eq:pot_g}
\end{align}

The factor of $i$ appearing in \eqref{eq:pot_g} is a consequence of the form of the spin operators containing a cross product. When rewritten in a factorized form, as detailed below, an additional factor of $i$ will appear. The scalar function $f^\prime(r_{ij},r_{jk},c_\theta)$ is given by

\begin{align}
    f^\prime(r_{ij},r_{jk},c_\theta)&=\left(\frac{\partial^2}{\partial r_{ij}r_{jk}}-\frac{c_\theta}{r_{ij}}\frac{\partial^2}{\partial r_{jk}\partial c_\theta}-\frac{c_\theta}{r_{jk}}\frac{\partial^2}{\partial r_{ij}\partial c_\theta}\right.
    \\
    &\left.+\frac{1}{r_{ij}r_{jk}}\left((c_\theta^2-1)\frac{\partial^2}{\partial c_\theta^2}+c_\theta\frac{\partial}{\partial c_\theta}\right)\right)f(r_{ij},r_{jk},c_\theta)
    \\
    f(r_{ij},r_{jk},c_\theta)&=\frac{1}{16\pi^3}\frac{1}{q^3}\frac{\sin(q\abs{x\boldsymbol{r}_{ij}-(1-x)\boldsymbol{r}_{jk}})}{\abs{x\boldsymbol{r}_{ij}-(1-x)\boldsymbol{r}_{jk}}}\frac{e^{-r_{ik}\sqrt{q^2(x-x^2)+m_\pi^2}}}{\sqrt{q^2(x-x^2)+m_\pi^2}}
\end{align}

\noindent where $c_\theta=\hat{\boldsymbol{r}}_{ij}\cdot\hat{\boldsymbol{r}}_{jk}$. The spin operators and derivatives are given by

\begin{align}
\begin{split}
    \mathcal{O}_{ijk}^{(g)}&=(\boldsymbol{\sigma}_j\cdot\boldsymbol{\sigma}_k\times\boldsymbol{\sigma}_i)\mathcal{D}_0+(\boldsymbol{\sigma}_j\cdot\boldsymbol{\sigma}_k\times\hat{\boldsymbol{r}}_{ij})(\boldsymbol{\sigma}_i\cdot\hat{\boldsymbol{r}}_{ij})\mathcal{D}_1+(\boldsymbol{\sigma}_j\cdot\boldsymbol{\sigma}_k\times\hat{\boldsymbol{r}}_{ij})(\boldsymbol{\sigma}_i\cdot\hat{\boldsymbol{r}}_{jk})\mathcal{D}_3
    \\&+(\boldsymbol{\sigma}_j\cdot\hat{\boldsymbol{r}}_{jk}\times\boldsymbol{\sigma}_i)(\boldsymbol{\sigma}_k\cdot\hat{\boldsymbol{r}}_{jk})\mathcal{D}_2
    +(\boldsymbol{\sigma}_j\cdot\hat{\boldsymbol{r}}_{jk}\times\hat{\boldsymbol{r}}_{ij})(\boldsymbol{\sigma}_i\cdot\hat{\boldsymbol{r}}_{ij})(\boldsymbol{\sigma}_k\cdot\hat{\boldsymbol{r}}_{jk})\mathcal{D}_5
    \\
    &+(\boldsymbol{\sigma}_j\cdot\hat{\boldsymbol{r}}_{jk}\times\hat{\boldsymbol{r}}_{ij})(\boldsymbol{\sigma}_i\cdot\hat{\boldsymbol{r}}_{jk})(\boldsymbol{\sigma}_k\cdot\hat{\boldsymbol{r}}_{jk})\mathcal{D}_7+(\boldsymbol{\sigma}_j\cdot\hat{\boldsymbol{r}}_{jk}\times\boldsymbol{\sigma}_i)(\boldsymbol{\sigma}_k\cdot\hat{\boldsymbol{r}}_{ij})\mathcal{D}_3
    \\
    &+(\boldsymbol{\sigma}_j\cdot\hat{\boldsymbol{r}}_{jk}\times\hat{\boldsymbol{r}}_{ij})(\boldsymbol{\sigma}_i\cdot\boldsymbol{\sigma}_k)\mathcal{D}_3+(\boldsymbol{\sigma}_j\cdot\hat{\boldsymbol{r}}_{jk}\times\hat{\boldsymbol{r}}_{ij})(\boldsymbol{\sigma}_i\cdot\hat{\boldsymbol{r}}_{ij})(\boldsymbol{\sigma}_k\cdot\hat{\boldsymbol{r}}_{ij})\mathcal{D}_6
    \\
    &+(\boldsymbol{\sigma}_j\cdot\hat{\boldsymbol{r}}_{jk}\times\hat{\boldsymbol{r}}_{ij})(\boldsymbol{\sigma}_i\cdot\hat{\boldsymbol{r}}_{jk})(\boldsymbol{\sigma}_k\cdot\hat{\boldsymbol{r}}_{ij})\mathcal{D}_4
\end{split}
\end{align}

\begin{align}
    \mathcal{D}_0&=\frac{1}{r_{ij}r_{jk}}
    \\
    \mathcal{D}_1&=\frac{1}{r_{jk}}\left[\frac{\partial}{\partial r_{ij}}-\frac{c_\theta}{r_{ij}}\frac{\partial}{\partial c_\theta}-\frac{1}{r_{ij}}\right]
    \\
    \mathcal{D}_2&=\frac{1}{r_{ij}}\left[\frac{\partial}{\partial r_{jk}}-\frac{c_\theta}{r_{jk}}\frac{\partial}{\partial c_\theta}-\frac{1}{r_{jk}}\right]
    \\
    \mathcal{D}_3&=\frac{1}{r_{ij}r_{jk}}\frac{\partial}{\partial c_\theta}
    \\
    \mathcal{D}_4&=\frac{1}{r_{ij}r_{jk}}\frac{\partial^2}{\partial c_\theta^2}
    \\
    \mathcal{D}_5&=\left[\frac{\partial}{\partial r_{ij}}-\frac{c_\theta}{r_{ij}}\frac{\partial}{\partial c_\theta}\right]\left[\frac{\partial}{\partial r_{jk}}-\frac{c_\theta}{r_{jk}}\frac{\partial}{\partial c_\theta}-\frac{1}{r_{jk}}\right]-\frac{1}{r_{ij}}\left[\frac{\partial}{\partial r_{jk}}-\frac{c_\theta}{r_{jk}}\frac{\partial}{\partial c_\theta}-\frac{1}{r_{jk}}\right]
    \\
    \mathcal{D}_6&=\left[\frac{\partial}{\partial r_{ij}}-\frac{c_\theta}{r_{ij}}\frac{\partial}{\partial c_\theta}\right]\left[\frac{1}{r_{jk}}\frac{\partial}{\partial c_\theta}\right]-\frac{2}{r_{ij}r_{jk}}\frac{\partial}{\partial c_\theta}
    \\
    \mathcal{D}_7&=\left[\frac{\partial}{\partial r_{jk}}-\frac{c_\theta}{r_{jk}}\frac{\partial}{\partial c_\theta}\right]\left[\frac{1}{r_{ij}}\frac{\partial}{\partial c_\theta}\right]-\frac{2}{r_{ij}r_{jk}}\frac{\partial}{\partial c_\theta}
\end{align}

\noindent To implement the radial-dependence of $\tilde{\mathcal{V}}_{0,\mathrm{VA}}^{(g)}$ in the QMC, we construct a table of the eight double integrals evaluated numerically and perform a trilinear interpolation between grid points.

\section{Spin-operator identities}

In this section we provide identities which decompose the three-body spin operators into a factorization of purely two-body terms. Basic three-body identities, which can be found in Ref. \cite{Carlson_1983}, are given by

\begin{align}
    \boldsymbol{\sigma}_j\cdot\boldsymbol{\sigma}_k&=\frac{1}{2}\{\boldsymbol{\sigma}_i\cdot\boldsymbol{\sigma}_j,\boldsymbol{\sigma}_i\cdot\boldsymbol{\sigma}_k\}
    \\
    (\boldsymbol{\sigma}_j\cdot \hat{\boldsymbol{r}}_{ij})(\boldsymbol{\sigma}_k\cdot\hat{\boldsymbol{r}}_{ij})&=\frac{1}{6}\{(S_{ij}+\boldsymbol{\sigma}_i\cdot\boldsymbol{\sigma}_j),\boldsymbol{\sigma}_i\cdot\boldsymbol{\sigma}_k\}
    \\
    (\boldsymbol{\sigma}_j\cdot\hat{\boldsymbol{r}}_{ij})(\boldsymbol{\sigma}_k\cdot\hat{\boldsymbol{r}}_{ik})(\hat{\boldsymbol{r}}_{ij}\cdot\hat{\boldsymbol{r}}_{ik})&=\frac{1}{18}\{(S_{ij}+\boldsymbol{\sigma}_i\cdot\boldsymbol{\sigma}_j),(S_{ik}+\boldsymbol{\sigma}_i\cdot\boldsymbol{\sigma}_k)\}
    \\
    \boldsymbol{\sigma}_i\cdot\boldsymbol{\sigma}_j\times\boldsymbol{\sigma}_k&=\frac{1}{2i}[\boldsymbol{\sigma}_i\cdot\boldsymbol{\sigma}_j,\boldsymbol{\sigma}_i\cdot\boldsymbol{\sigma}_k]
    \\
    (\boldsymbol{\sigma}_j\cdot\hat{\boldsymbol{r}}_{ij})(\boldsymbol{\sigma}_k\cdot\boldsymbol{\sigma}_i\times\hat{\boldsymbol{r}}_{ij})&=\frac{1}{6i}[(S_{ij}+\boldsymbol{\sigma}_i\cdot\boldsymbol{\sigma}_j),\boldsymbol{\sigma}_i\cdot\boldsymbol{\sigma}_k]
    \\    (\boldsymbol{\sigma}_j\cdot\hat{\boldsymbol{r}}_{ij})(\boldsymbol{\sigma}_k\cdot\hat{\boldsymbol{r}}_{ik})(\boldsymbol{\sigma}_i\cdot\hat{\boldsymbol{r}}_{ij}\times\hat{\boldsymbol{r}}_{ik})&=\frac{1}{18i}[(S_{ij}+\boldsymbol{\sigma}_i\cdot\boldsymbol{\sigma}_j),(S_{ik}+\boldsymbol{\sigma}_i\cdot\boldsymbol{\sigma}_k)]
\end{align}

\noindent Additional identities are required for potentials $\tilde{\mathcal{V}}_{0,\mathrm{VA}}^{(g)}$:

\begin{align}
\begin{split}
(\boldsymbol{\sigma}_j\cdot\hat{\boldsymbol{r}}_{jk}\times\hat{\boldsymbol{r}}_{ij})(\boldsymbol{\sigma}_i\cdot\boldsymbol{\sigma}_k)&=-\frac{1}{324i}\frac{1}{\hat{\boldsymbol{r}}_{ij}\cdot\hat{\boldsymbol{r}}_{jk}}\{S_{ij}+\boldsymbol{\sigma}_i\cdot\boldsymbol{\sigma}_j,S_{jk}+\boldsymbol{\sigma}_j\cdot\boldsymbol{\sigma}_k\}
    \\
    &\times[S_{ij}+\boldsymbol{\sigma}_i\cdot\boldsymbol{\sigma}_j,S_{jk}+\boldsymbol{\sigma}_j\cdot\boldsymbol{\sigma}_k]\boldsymbol{\sigma}_i\cdot\boldsymbol{\sigma}_k
    \end{split}
\\
\begin{split}    (\boldsymbol{\sigma}_i\cdot\hat{\boldsymbol{r}}_{ij})(\boldsymbol{\sigma}_k\cdot\hat{\boldsymbol{r}}_{ij})(\boldsymbol{\sigma}_j\cdot\hat{\boldsymbol{r}}_{ij}\times\hat{\boldsymbol{r}}_{jk})
    &=\frac{1}{1944i}\frac{1}{\hat{\boldsymbol{r}}_{ij}\cdot\hat{\boldsymbol{r}}_{jk}}\{(S_{ij}+\boldsymbol{\sigma}_i\cdot\boldsymbol{\sigma}_j),\boldsymbol{\sigma}_j\cdot\boldsymbol{\sigma}_k\}
    \\
    &\times\{(S_{ij}+\boldsymbol{\sigma}_{i}\cdot\boldsymbol{\sigma}_j),(S_{jk}+\boldsymbol{\sigma}_j\cdot\boldsymbol{\sigma}_k)\}[(S_{ij}+\boldsymbol{\sigma}_i\cdot\boldsymbol{\sigma}_j),(S_{jk}+\boldsymbol{\sigma}_j\cdot\boldsymbol{\sigma}_k)]
\end{split}
\\
\begin{split}
    (\boldsymbol{\sigma}_i\cdot\hat{\boldsymbol{r}}_{jk})(\boldsymbol{\sigma}_k\cdot\hat{\boldsymbol{r}}_{jk})(\boldsymbol{\sigma}_j\cdot\hat{\boldsymbol{r}}_{ij}\times\hat{\boldsymbol{r}}_{jk})  
    &=\frac{1}{1944i}\frac{1}{\hat{\boldsymbol{r}}_{ij}\cdot\hat{\boldsymbol{r}}_{jk}}\{(S_{jk}+\boldsymbol{\sigma}_j\cdot\boldsymbol{\sigma}_k),\boldsymbol{\sigma}_i\cdot\boldsymbol{\sigma}_j\}
    \\
    &\times\{(S_{ij}+\boldsymbol{\sigma}_i\cdot\boldsymbol{\sigma}_j),(S_{jk}+\boldsymbol{\sigma}_j\cdot\boldsymbol{\sigma}_k)\}[(S_{ij}+\boldsymbol{\sigma}_i\cdot\boldsymbol{\sigma}_j),(S_{jk}+\boldsymbol{\sigma}_j\cdot\boldsymbol{\sigma}_k)]
\end{split}
\\
(\boldsymbol{\sigma}_j\cdot\boldsymbol{\sigma}_k\times\hat{\boldsymbol{r}}_{ij})(\boldsymbol{\sigma}_i\cdot\hat{\boldsymbol{r}}_{jk})&=-\frac{1}{648i}\frac{1}{\hat{\boldsymbol{r}}_{ij}\cdot\hat{\boldsymbol{r}}_{jk}}[(S_{ij}+\boldsymbol{\sigma}_i\cdot\boldsymbol{\sigma}_j),\boldsymbol{\sigma}_j\cdot\boldsymbol{\sigma}_k]
\\
&\times\{(S_{ij}+\boldsymbol{\sigma}_i\cdot\boldsymbol{\sigma}_j),(S_{jk}+\boldsymbol{\sigma}_j\cdot\boldsymbol{\sigma}_k)\}\{(S_{jk}+\boldsymbol{\sigma}_j\cdot\boldsymbol{\sigma}_k),\boldsymbol{\sigma}_i\cdot\boldsymbol{\sigma}_j\}
\\
(\boldsymbol{\sigma}_j\cdot\hat{\boldsymbol{r}}_{jk}\times\hat{\boldsymbol{r}}_{ij})(\boldsymbol{\sigma}_i\cdot\hat{\boldsymbol{r}}_{jk})(\boldsymbol{\sigma}_k\cdot\hat{\boldsymbol{r}}_{ij})&=\frac{1}{648i}\{(S_{jk}+\boldsymbol{\sigma}_j\cdot\boldsymbol{\sigma}_k),\boldsymbol{\sigma}_i\cdot\boldsymbol{\sigma}_j\}\\
&\times[(S_{jk}+\boldsymbol{\sigma}_j\cdot\boldsymbol{\sigma}_k),(S_{ij}+\boldsymbol{\sigma}_i\cdot\boldsymbol{\sigma}_j)]\{(S_{ij}+\boldsymbol{\sigma}_i\cdot\boldsymbol{\sigma}_j),\boldsymbol{\sigma}_j\cdot\boldsymbol{\sigma}_k\}
\end{align}

\bibliography{bibliography}

\end{document}